\documentclass[aps,pra,notitlepage,twocolumn,amsmath,amssymb,showpacs,nofootinbib,a4paper,accepted=2025-01-10]{quantumarticle}
\pdfoutput=1
\usepackage{graphicx}
\usepackage{amsmath}
\usepackage{bm}
\usepackage{xcolor}
\usepackage{comment}
\usepackage{dsfont} 

\usepackage[utf8]{inputenc}
\usepackage[english]{babel}
\usepackage[T1]{fontenc}
\usepackage{hyperref}

\usepackage{tikz}
\usepackage{lipsum}

\newcommand{\be}{\begin{equation}}  
\newcommand{\ee}{\end{equation}}
\newcommand{\ba}{\begin{array}}
\newcommand{\ea}{\end{array}}
\newcommand{\bea}{\begin{eqnarray}}
\newcommand{\eea}{\end{eqnarray}}

\newcommand{\bra}{\langle}
\newcommand{\ket}{\rangle}
\newcommand{\moy}[1]{\left\langle #1\right\rangle}

\newcommand{\nn}{\nonumber}
\newcommand{\idop}{\mathds{1}} 
\newcommand{\Cyril}[1]{{\color[rgb]{0,0.2,1}{#1}}}
\newcommand{\Cami}[1]{{\color[rgb]{1,0.15,0}{#1}}}

\begin{document}

\title{A thermodynamically consistent approach to the energy costs of quantum measurements}

\author{Camille Lombard Latune}
\email{cami-address@hotmail.com}

\affiliation{Laboratoire Interdisciplinaire Carnot de Bourgogne, CNRS UMR 6303,
Université Bourgogne Europe, BP 47870, F-21078 Dijon, France}
\author{Cyril Elouard}
\email{cyril.elouard@gmail.com} 
\affiliation{Université de Lorraine, CNRS, LPCT, F-54000 Nancy, France}

\date{20 Jan 2025}
\begin{abstract}
Considering a general microscopic model for a quantum measuring apparatus comprising a quantum probe coupled to a thermal bath, we analyze the energetic resources necessary for the realization of a quantum measurement, which includes the creation of system-apparatus correlations, the irreversible transition to a statistical mixture of definite outcomes, and the apparatus resetting. Crucially, we do not resort to another quantum measurement to capture the emergence of objective measurement results, but rather exploit the properties of the thermal bath which redundantly records the measurement result in its degrees of freedom, naturally implementing the paradigm of quantum Darwinism. In practice, this model allows us to perform a quantitative thermodynamic analysis of the measurement process. From the expression of the second law, we show how the minimal required work depends on the energy variation of the system being measured plus information-theoretic quantities characterizing the performance of the measurement -- efficiency and completeness. Additionally, we show that it is possible to perform a thermodynamically reversible measurement, thus reaching the minimal work expenditure, and provide the corresponding protocol. Finally, for finite-time measurement protocols, we illustrate the increasing work cost induced by rising entropy production inherent in finite-time thermodynamic processes. This highlights an emerging trade-off between velocity of the measurement and work cost, on top of a trade-off between efficiency of the measurement and work cost. We apply those findings to bring new insights in the thermodynamic balance of the measurement-powered quantum engines.
\end{abstract}

\maketitle

\section{Introduction}

Quantum measurement, along with the stochastic evolution it triggers on the measured quantum system, stands as one of the most perplexing phenomena in quantum mechanics. On the one hand, the dynamics induced by quantum measurements is well-modeled, even when accounting for non-ideal, realistic measurement scenarios \cite{WisemanBook}, and exhibit remarkable consistency with experimental observations. On the other hand, the underlying emergence of these dynamics from first principles \cite{Wigner1984} continues to be a subject of ongoing investigations \cite{Zurek09,Korbicz17}.

Moreover, during a measurement, both the entropy and energy of the measured system may undergo changes, rendering the measurement-induced dynamics akin to a thermodynamic transformation \cite{Elouard17Role}. This notion has led to the development of engines and refrigerators fueled by measurement-induced energy variations \cite{Yi17,Ding18,Elouard17,Manikandan22, Bresque21}. In a broader sense, quantum measurement represents a form of thermodynamic resource extending beyond the mere acquisition of information, which itself can be converted into work through protocols akin to Maxwell's demons harnessing information from classical measurements.

Despite these thermodynamic analyses of quantum measurement, little is known about the actual energetic cost of realizing quantum measurements and how this relates with the energy received by the system. How closely do realistic quantum measurements approach fundamental limits? How can their energetic costs be optimized? These latter questions gain newfound significance as the energetic cost of implementing quantum algorithms, which typically necessitate numerous measurements for error correction, comes under scrutiny \cite{Fellous-Asiani23}.

Pioneering studies have focused on the cost associated with information manipulation during the measurement \cite{Jacobs12,Deffner16}, paving the road towards the understanding of the thermodynamic description of feedback protocols and autonomous Maxwell demons \cite{Sagawa09,Jacobs09,Funo13,Watanabe14,Manikandan19, Belenchia20}, in particular by deriving Fluctuation Theorems, leading to subsequent experimental tests \cite{Mancino18,Linpeng22}. Another contribution to the cost of quantum measurement is the cost to generate system-apparatus correlations, which can in turn explain some of the energy exchanges received by the system \cite{Elouard18}. Analyzing this contribution had led to show that ideal projective measurements require infinite resource cost \cite{Guryanova20}, and can therefore only be asymptotically approached. However, the quantum measurement process cannot be reduced down to a reversible unitary interaction between the system and the meter storing the result without bringing paradoxes \cite{Elouard21} associated with (generalized) Wigner's friend scenarios \cite{Sokolovski21}. 
In contrast, one must include the step of ``objectification'' \cite{Korbicz17,Mohammady23} during which the apparatus undergoes a transition to classical behavior and the measurement outcome becomes an objective fact verifiable by independent observers, encoded into a statistical mixture of perfectly distinguishable states of the apparatus. Taking the objectification into account is necessary to fully grasp the thermodynamic balance of quantum measurement -- in particular, its irreversible nature. When it is not ignored, this step is usually described by resorting to another quantum measurement process, whose resource cost remains unevaluated (see also Appendix \ref{app:issueqm} for an analysis of some consequences of this approach).\\
Here, we circumvent this problem by considering microscopic models of measuring apparatus including a thermal bath at equilibrium whose role is to implement the objectification. This strategy was already used to analyze measurement-induced dynamics from specific microscopic models of measuring apparatuses \cite{Allahverdyan13,Goan01,BreuerPetruccione}, inspired by real-world setups. In contrast, we use it to perform a general thermodynamic analysis of a large class of measurement processes, including nonideal quantum measurements, and deduce fundamental constraints on the total measurement cost. In particular, we derive a lower bound on the total work cost of the measurement which is composed of the energy absorbed by the measured system during the measurement, as well as of information-theoretic quantities that quantify the measurement quality through information acquired by the apparatus.

Through the analysis of a specific measurement protocol, we demonstrate the attainability of this lower bound. Key conditions for such attainability include quasi-static manipulation of the system-apparatus interaction (very slow relative to the thermalization timescale of the reservoir), as well as the full utilization of information acquired during the protocol.
Moreover, we establish that a measurement of an observable commuting with the system's Hamiltonian can be performed without expending any driving work in the quasi-static limit. Finally, for measurements conducted at finite speeds, the energy cost escalates due to increased entropy production (predominantly ``classical friction"). This gives rise to a dual trade-off between measurement duration, measurement efficiency, and energy expenditure.

\begin{figure}
    \centering
   \raisebox{.73cm}{\includegraphics[width=0.45\textwidth]{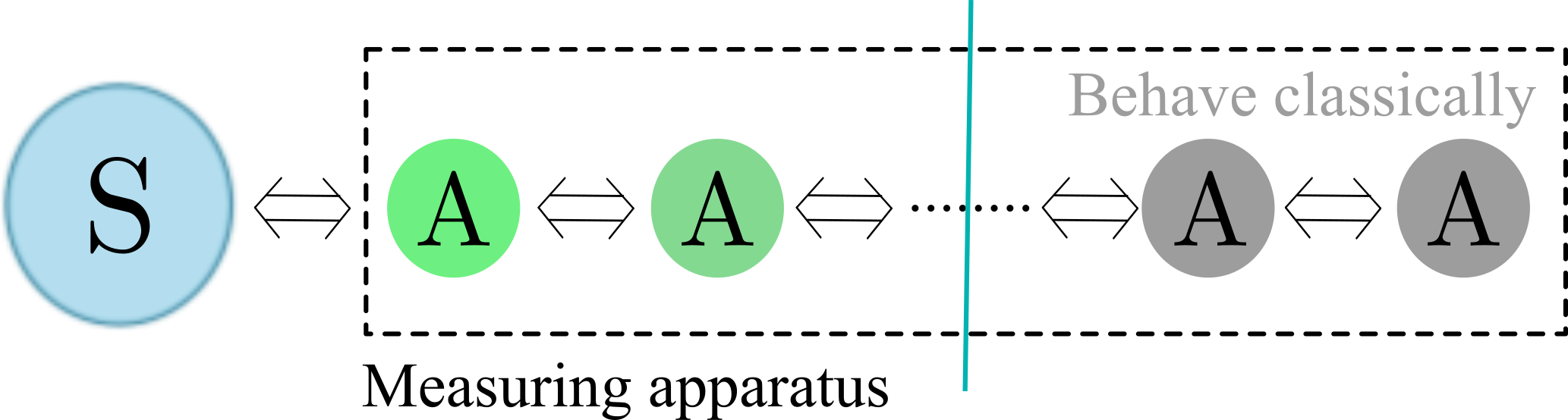}}
    \hspace{0.4cm}
    \includegraphics[width=0.38\textwidth]{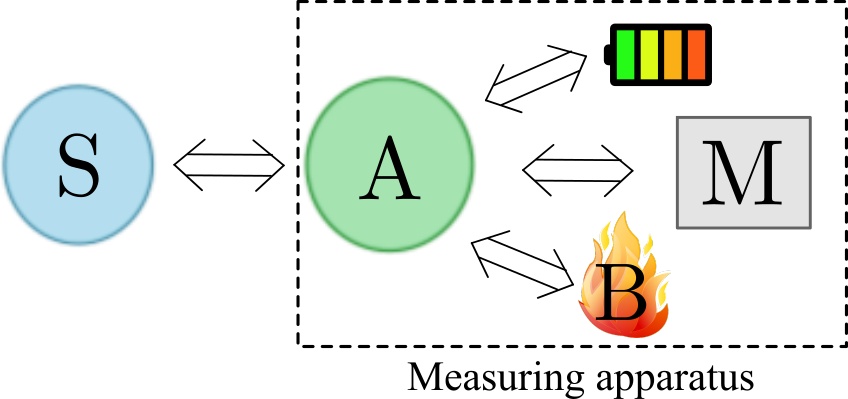}
    \caption{Up: Von Neumann chain model of the objectification process. Bottom: The model of measuring apparatus considered in this paper.}
    \label{fig:MeasModel}
\end{figure}

\section{Energetically closed model of quantum measurement}

\subsection{Modeling the objectification}\label{sec:meas_protocol}

In this section, we introduce and motivate our general model of measuring apparatus, used to draw conclusion about quantum measurement energetics. The measured system $S$, is assumed to be initially isolated in a state $\rho_S(0)$ with local Hamiltonian $H_S$. The aim of the operation is to measure an observable $Q_S$. Since $S$ is not directly accessible to the observer -- the value of $Q_S$ cannot be read by just ``looking" at it -- standard measurement protocols \cite{WisemanBook} consider an auxiliary quantum system $A$ (which is a subpart of the total measuring apparatus) which is set to interact with $S$. As a consequence of the correlations established between $S$ and $A$ during their interaction, the state of $A$ contains information about $Q_S$. In a realistic measurement setup, this information is, at least partially, amplified and transferred to many other degrees of freedom until it reaches a macroscopic pointer, which can be directly read by the observer.

The redundant encoding of the measurement outcome in many degrees of freedom, a large part of which are inaccessible to observer is responsible for the irreversible nature of the measurement process, as pointed out in particular by the paradigm of quantum Darwinism \cite{Zurek03}. The redundancy also makes the measurement outcome an objective reality, accessible by different observers \cite{Horodecki13}, and is also crucial to prevent well-known paradoxes associated with the Wigner's friend scenarios \cite{Elouard21}, which all incur from assuming an observer to have complete access to any information extracted from the system. 

This process is sometimes called \emph{objectification} \cite{Korbicz17, Mohammady23}, and was first modeled by von Neumann by considering a second auxiliary system $A^{(1)}$ interacting with $A$ so as to acquire information about $Q_S$, followed by a third auxiliary system $A^{(2)}$ interacting with $A^{(1)}$, and so on, forming a macroscopic chain $A^{(1)}...A^{(N)}$ \cite{WisemanBook}. The chain can be stopped when the total system $SA^{(1)}...A^{(N)}$ behaves essentially classically, 
and the measurement outcome can be simply accessed by looking at the apparatus to find out its macroscopic state -- a classical observation.

To bridge the gap with thermodynamics, we analyze here a slightly different model, where the role of the macroscopic chain of systems is played by a single quantum pointer $A$, together with a reservoir at thermal equilibrium $B$ and a classical memory $M$. 
In addition, a classical battery provides the work needed for the realization of the measurement (see Fig.~\ref{fig:MeasModel}b), and manifests itself in the form of time-dependent Hamiltonians.

The macroscopic reservoir $B$ ensures the objectification by inducing decoherence on the pointer: Namely, after interacting with the bath, if the decoherence process is complete, the state of the system $S$ and of the pointer $A$ is a fully incoherent mixture of perfectly distinguishable states associated with the different measurement outcomes. The distinguishable states corresponds to macrostates of the apparatus, in which the measurement outcome can be read by a mere classical information acquisition. There is therefore no need to invoke a quantum measurement on the apparatus itself. In addition, the measurement outcome can be independently verified by different observers, without showing any contradictions.
In other words, once $B$ is explicitly included in the description, positioning the Heisenberg cut after the system $A$ is fully justified.
In particular, reading the outcome in $A$ (or equivalently writing in down in classical memory $M$) does not trigger any additional exchange of energy and does not incur any energy cost, besides the cost to further process classical information \cite{Landauer}. This is a crucial property that makes our model energetically close, and allows us to perform an energy balance beyond previous analyses of quantum measurement energetics \cite{Guryanova20,Mohammady23,Sagawa09,Jacobs09, Abdelkhalek16,Funo13}. We finish by mentioning that our framework is inspired by specific dynamical models involving an open-system measuring apparatus to capture the emergence of the measurement-induced dynamics (without energetic analyses) \cite{Goan01,Allahverdyan13,BreuerPetruccione}.\\

\subsection{Measurement protocol} \label{sec:measprot}

We now present the protocol associated with a quantum measurement of $Q_S$. Before the measurement protocol starts, the global system $SAB$ starts in state $\rho_S(0)\otimes\rho_{AB}(0)$, that is, $A$ and $B$ are uncorrelated from $S$. To fully take into account any resource spent in the measurement process, we assume that $A$ and $B$ are initially at thermal equilibrium and weakly coupled such that $\rho_{AB}(0) = e^{-\beta(H_A+V_{AB}+H_B)}/Z_{AB} \simeq e^{-\beta H_A}/Z_{A}\otimes e^{-\beta H_B}/Z_{B}$, where we have introduced Hamiltonians $H_A$ and $H_B$ of $A$ and $B$ and (weak) coupling $V_{AB}$. Therefore, any non-equilibriumness in $A$ exploited during the measurement process is assumed to be prepared during next stages of the protocol.


The general protocol considered throughout the paper starts with a first step allowing the correlation of $S$ and $A$. During this step, the coupling between $S$ and $A$ is switched on, in presence of the decoherence induced by $B$. This is described by a global unitary evolution on $SAB$ for $t$ from $t_0$ to $t_M^-$, 
\be
U_\text{on} := {\cal T}\exp\left\{-i \int_{t_0}^{t_M^-} du H_{SAB}(u)\right\},
\ee
where the Hamiltonian of systems $S$, $A$, and $B$ is of the most general form 
\be 
H_{SAB}(t) = H_S(t) + V_{SA}(t) + H_A(t) + H_B + V_{AB}(t),
\ee 
(assuming no control over the bath $B$) and where ${\cal T}\exp$ denotes the time-ordered exponential. 

As mentioned in previous section \ref{sec:meas_protocol}, in order to model a complete measurement setup, up to a level where the Heisenberg cut is justified, we impose that, at the end of this step (denoted time $t_M$), the state of $SA$ is an incoherent mixture of orthogonal states that will be associated with different measurement outcomes. We stress that this decoherence is the result of the interaction with the bath $B$ (see Section \ref{sec:example} for illustrative example). We therefore write:
\be\label{eq:rhoSAr}
\rho_{SA}(t_M) :={\rm Tr}_B[U_\text{on}\rho_S(0)\rho_{AB}(0)U_\text{on}^\dag]  = \sum_r p_r \rho_{SA\vert_r}(t_M),
\ee
with 
\be\label{eq:ortho}
\text{Tr}\{\rho_{A\vert_r}(t_M) \rho_{A\vert_r'}(t_M)\} = \delta_{rr'} \text{Tr}\{\rho_{A\vert_r}^2(t_M)\},
\ee
where $\rho_{A\vert_r}(t_M) = \text{Tr}_S\{\rho_{SA|r}\}$ and $r$ labels the measurement results that can be obtained during the classical observation of $A$, each with probability $p_r$.
The orthogonality condition Eq.~\eqref{eq:ortho} ensures that the states $\rho_{A\vert_r}(t_M)$ of $A$ are fully distinguishable and therefore behave as macroscopically different classical states: this can be understood as a condition for classicality of the final state of the apparatus, or objectivity of the measurement outcome \cite{Horodecki13,Korbicz14,Engineer24}, which is a direct consequence of the ``quantum Darwinism'' assumptions \cite{Le2019Jan,Feller21}. 


We formally represent the operation of reading the classical measurement outcome in the apparatus via a reversible operation correlating each orthogonal state $\rho_{A\vert_r}(t_M)$ into a distinct state of a classical memory system $M$, that we represent in terms of density operators $\vert r\ket_M\bra{r}\vert$ for simplicity. The memory is assumed to be initially in a pure reference state $|{\text{ref}}\ket_M\bra{\text{ref}}\vert$, and its free dynamics is assumed to be negligible over the duration of the measurement protocol (for instance, fully degenerate Hamiltonian). After the reading process (time $t_M^+$), the correlated state of $SA$ and $M$ is therefore:
\be\label{eq:genstatetm}
\rho_{SAM}(t_M^+) = \sum_r p_r \rho_{SA\vert_r}(t_M) \otimes\vert r\rangle_M \langle r\vert.
\ee
This correlation can be done perfectly in principle owing to the orthogonality of the $\rho_{A\vert_r}(t_M)$ states, and at no work cost for a degenerate memory Hamiltonian. While (classical)  errors coming from imperfect encoding in the memory could be analyzed, we choose to focus on limitations of the measurement quality coming from the interaction between $S$, $A$ and $B$. 

To reinitialize the apparatus, two additional steps are needed. First, if the $S-A$ coupling was not off at the end of the first driving phase, it must be switched off, which is described by another global unitary on $SAB$ for $t$ from $t_M^+$ to $t_F$, 
\be
U_\text{off} :={\cal T}\exp\left\{-i \int_{t_M^+}^{t_F} du H_{SAB}(u)\right\}.
\ee
We also assume that $B$ has returned to equilibrium before this switching off process begins (or equivalently that the switching off process involves another thermal reservoir). At the end of this step, the correlated state of the system, apparatus and memory reads:
\be\label{eq:genstatetf}
\rho_{SAM}(t_F) = \sum_r p_r \rho_{SA|r}(t_F)\otimes\vert r\rangle_M\langle r\vert 
\ee
with, $\rho_{SA|r}(t_F) = {\rm Tr}_B[U_\text{off}\rho_{SA|r}(t_M)\otimes\rho_B(0)U_\text{off}^\dag]$. We also define the reduced final state of $X=S$, $A$, $SA$ as $\rho_X(t_F) := \sum_r p_r \rho_{X|r}(t_F)$.

Second, the system $AM$ must be reset to its initial state.
In particular, the memory $M$ must be erased (potentially after its content has been used e.g. to design feedback operation). 



When outcome $r$ is obtained, the conditional final state of the system can generically be cast under the form (see Appendix \ref{app:ISA}):

\be \label{eq:rhoSrtF}
\rho_{S|r}(t_F) = \frac{1}{p_r}\sum_{s} M_{r,s} \rho_S(0) M_{r,s}^\dagger,
\ee
where we have introduced the Kraus operators $M_{r,s} \propto {}_B\langle i'| {}_A \langle m'| U_\text{on}\Pi_r^A U_\text{off} |m\rangle_A|i\rangle_B$ verifying $\sum_{r,s} M_{r,s}^\dagger M_{r,s} = \idop$. Note that $s$ is a collective index gathering all sets $(n,i,n',i')$ of initial and final states of $A$ and $B$ yielding system operators ${}_B\langle n'| {}_A \langle i'| U_\text{on}\Pi_r^A U_\text{off} |n\rangle_A|i\rangle_B$ which are proportional to each other -- where $\{|n\rangle_A\}$ and $\{|i\rangle_B\}$ are bases of the Hilbert spaces of $A$ and $B$, respectively. Additionally, the projective operator $\Pi_r^A$ is the projector onto the support of $\rho_{A|r}(t_M)$. The quality of the measurement process can be related to the properties of the states $\{\rho_{S|r}(t_F)\}_r$ \cite{WisemanBook}. The textbook case of a projecting measurement, such that $\rho_{S|r}(t_F)\propto \pi_r^S\rho_S(t_0)\pi_r^S$, with $\pi_r^S$ some projective operator of $S$, is a limiting case expected to be reached only at infinite resource cost \cite{Guryanova20}. By contrast, {\it realistic measurement} models can yield to measurement that are {\it invasive} (the statistics of the measurement observable in the average post-measurement state $\sum_r p_r \rho_{S|r}(t_F))$ is different from the statistics prior to the measurement), {\it partially inefficient} (a fraction of the information leaking in the environment is not available in the memory, such that the post-measurement state can be a mixed state) and {\it incomplete} (such as weak measurements \cite{Jacobs06} which only partially evolve the system's state towards an eigenstate of the measured observable).





In the remainder of this paper, we derive a lower bound on the work required to perform such {\it realistic measurement} protocols. Such lower bound depends on the quality of the measurement. 
It is then followed by some illustrative situations and protocols, where it is also possible to study additional energetic cost induced by irreversible finite-time protocols.

\section{General thermodynamic argument}\label{sec:genargument}

\subsection{A lower bound on the work cost of the measurement process}\label{sec:lowerb}

In this section, we use quantum formulations of thermodynamic laws to derive a lower bound on the work expenditures during the measurement process. The lower bound, which depends on the properties of the measurement, takes the form:
\bea 
W_\text{dr}+W_\text{reset} \geq \Delta E_S + \frac{1}{\beta}\left(J_S + \moy{I_{S:A}}\right) \label{eq:genbound}.
\eea

On the left-hand side, $W_\text{dr}$ is the work cost to drive the system and apparatus through the measurement process, while $W_\text{reset}$ is the work cost to reset the apparatus and memory to their initial state. The right-hand side is the sum of three terms which depends on the measurement properties. First, $\Delta E_S =\text{Tr}\{H_S\left(\rho_S(t_F)-\rho_S(0)\right)\}$ is the average energy variation of the measured system. This term vanishes for measurements of an observable $Q_S$ which commutes with the Hamiltonian of the system. Such measurements can be made noninvasive and verify $[M_{r,s},H_S]=0$.

Second,
\be 
J_S = \sum_r p_r \left(S[\rho_S(t_0)]-S[\rho_{S|r}(t_F)]\right),\label{eq:JS}
\ee
with $S[\rho] = -\text{Tr}\{\rho\log\rho\}$ the Von Neumann entropy of state $\rho$, quantifies the average information obtained on the system state during each single run of the measurement protocols (each corresponding to a different measurement outcomes $r$). This quantity verifies $J_S\in\big[S[\rho_S(t_0)]-\ln d_s; S[\rho_S(t_0)] \big]$. As we detail below, it may be negative for inefficient measurements (where only a fraction of the information leaking from $S$ is accessible to the observer).  

Finally,
\be 
\moy{I_{S:A}} = \sum_r p_r\left( S[\rho_{S|r}(t_F)]+S[\rho_{A|r}(t_F)]-S[\rho_{SA|r}(t_F)]\right),\label{eq:Isa}
\ee
is the average residual mutual information at $t_F$ between $S$ and $A$. 
It quantifies information about $S$ transferred to $A$ but \emph{not} to the memory for it is encoded in inaccessible degrees of freedoms of the pointer. It is therefore one source of measurement inefficiency.\\

Eq. \eqref{eq:genbound} is the main result of this section. It relates the measurement work cost with the energy change of the system and information-theoretic quantities which depend on the measurement performances (in particular its efficiency and its strength as detailed below). Eq.~\eqref{eq:genbound} becomes an equality when the whole protocol is performed in a reversible way, which requires quasi-static drives allowing equilibration between $SA$ and $B$ at all times, but also to fully exploit all the information obtained about the states of $S$ and $A$. We will come back with more details on this point in Section~\ref{sec:revmeas}.

Additionally, Eq. \eqref{eq:genbound} can be re-expressed in the form of the entropy production generated during the quantum
measurement and the reset of the apparatus and memory,
\be
\sigma := \beta[W_\text{dr} + W_\text{reset} - \Delta E_S] - J_S - \bra I_{S:A}\ket \geq 0.
\ee
In Section \ref{sec:revmeas}, the entropy production is further analyzed and decomposed in three contributions whose physical origin is clearly identified (see Eq.~\eqref{eq:entropyprod}).

\subsection{Proof of the lower bound}

In this section, we derive Eq.~\eqref{eq:genbound}. To do so, we first gather the evolution of the total system $SAMB$ from time $t_0$ to $t_F$ in a single unitary $U = U_\text{off} U_\text{reg} U_\text{on}$ which is all-in-all generated by an Hamiltonian of the form $H(t) = H_{SA}(t)+H_B+V_{AB}+V_{AM}(t)$, where $V_{AM}(t)$ describes without loss of generality the encoding of the measurement outcome in $M$ mentioned in section \ref{sec:measprot}, which is also denoted by unitary $U_\text{reg}$. We then use the formalism from Ref.~\cite{Esposito10} to express the entropy production associated with the measurement protocol.

From \cite{Esposito10}, the first law implies that work performed on the total system $SAMB$ when varying in time Hamiltonian $H(t)$ is equal to its total energy variation:
\bea 
W_\text{dr} &=& \moy{H(t_F)}-\moy{H(t_0)}\nonumber\\
&=& \Delta E_{SA} + \Delta 
\moy{V_{AB}}+\Delta E_B,
\eea
where $\Delta X = X(t_F)-X(t_0)$ refers to a variation of the average quantity $X$ between times $t_F$ and $t_0$, while $E_{\alpha}(t) = \moy{H_{\alpha}(t)}$ denotes the internal energy of system $\alpha$ at time $t$. The associated expression of the second law for the transformation of system $SAM$ in contact with $B$ reads:
\be
\sigma = \Delta S_{SAM}-\beta Q_B \geq 0
\ee
with $\sigma$ the total entropy produced, and $Q_B = -\Delta E_B$ the average heat flow exchanged with the bath $B$. In all the article, $S_\alpha(t) = S[\rho_\alpha(t)]$ denotes the Von Neumanm entropy of system $\alpha$ at time $t$.
Combining these two laws leads to:
\bea
W_\text{dr} &\geq& \Delta E_{SA} + \Delta \moy{V_{AB}} -\frac{1}{\beta} \Delta S_{SAM}.
\eea
By injecting the specific form Eq.~\eqref{eq:genstatetf} of the conditional final states, noting that $\Delta S_{SAM} = \sum_r p_r S[\rho_{SA|r}(t_F)]+{\cal H}(\{p_r\})-S_A(t_0)-S_S(t_0)$, we obtain 
\bea
W_\text{dr} \geq \Delta F_S + \Delta F_A + \frac{1}{\beta}\left[\xi_S + \xi_A + \moy{I_{S:A}} - {\cal H}(\{p_r\})\right],\nn\\
\eea
 with 
\be
{\cal H}(\{p_r\}) = -\sum_r p_r \log p_r,
\ee
the Shannon entropy of the measurement outcome distribution $p_r$, and $\xi_S := J_S + \Delta S_S = S[\rho_S(t_F)] - \sum_r p_r S[\rho_{S|r}(t_F)]$ is the Holevo information \cite{Nielsen10} about the outcome $r$ in the state $\rho_S(t_F) = \sum_r p_r \rho_{S|r}(t_F)$. $\xi_A$ is defined analogously.

On the other hand, the work cost to reset systems $A$ and $M$ is lower-bounded by the associated variation of free energy:
\bea
W_\text{reset} &\geq& F_{AM}(t_0) -F_{AM}(t_F) = -\Delta E_{A} + \frac{1}{\beta}\Delta S_{AM},\nonumber\\ 
&=& -\Delta F_A + \beta^ {-1}[{\cal H}(\{p_r\}) - \xi_A],\label{eq:wreset}
\eea
remembering that we considered a memory register $M$ with a fully degenerate Hamiltonian.
We now sum up both inequalities and use the assumption $V_{SA}(t_F)=V_{SA}(t_0)=0$, such that $\Delta E_{SA} = \Delta E_S+\Delta E_A$. We obtain
\be
W_\text{dr}+W_\text{reset} \geq \Delta E_S + \frac{1}{\beta}\left(J_S + \moy{I_{S:A}}\right) + \Delta\bra V_{AB}\ket,
\ee
which leads to Eq.~\eqref{eq:genbound} when neglecting the variation apparatus-bath coupling energy $\Delta \moy{V_{AB}}$ (either because the initial value of the apparatus-bath coupling is restored due to return to equilibrium after $\chi(t)$ is switched off, or simply because the coupling is weak).

\subsection{Efficient measurements}

To compare with earlier results \cite{Sagawa09,Erratum-Sagawa09} and further interpret 
Eq.~\eqref{eq:genbound}, we first focus on the case of \emph{efficient} measurements, which are such that all the information acquired about the system can be transferred to the memory $M$. Mathematically, efficient measurements lead to a conditional state of $S$ associated with measurement outcome $r$ obtained from the application of a single Kraus operator $M_r$ on the initial system state:
\be 
\rho_{S|r}(t_F) = \frac{M_r \rho_S(t_0) M_r^\dagger}{p_r},
\ee 
with $\sum_r M_r^\dagger M_r = \idop$ and $p_r = \text{Tr}\{M_r^\dagger M_r\rho_S(t_0)\}$. Such measurement therefore prepare pure conditional states from pure initial states. Depending on the interaction strength between the $S$ and $A$, the measurement can be \emph{complete} \cite{WisemanBook} -- that is, $\rho_{S|r}(t_F)$ is independent on the initial state, as it is the case for a projective/strong measurement where $M_r$ is a projector onto an eigenstate of the measured observable -- or \emph{incomplete}, such as weak measurements \cite{Jacobs06}. 

As we show in Appendix \ref{app:ISA}, the condition of efficient measurement implies that $\moy{I_{S:A}}=0$. Indeed, a non-zero mutual information between $S$ and $A$ in the conditional states $\rho_{S|r}(t_F)$ reveals the existence of information on $S$ stored in A but not read and therefore not transferred to M, hence a loss of efficiency of the measurement. 


In addition, for an efficient measurement, $J_S$ becomes equal to the so-called quantum-classical mutual information between $S$ and the measurement outcomes \cite{Sagawa12}, defined as:
\be 
I_\text{QC} = S_S(t_0) - \sum_r p_r S\left[\frac{\sqrt{E_r}\rho_S(t_0)\sqrt{E_r}}{p_r}\right]
\ee
with $E_r$ the effect operator defined from $p_r = \text{Tr}\{E_r\rho_S(t_0)\}$. 
In the particular situation where $\Delta E_S = 0$ (for instance for the measurement of an observable $Q_S$ commuting with $H_S$, as it will be further detailed in the next section), our bound Eq.~\eqref{eq:genbound} recovers the result of \cite{Sagawa09,Erratum-Sagawa09} in the regime where the latter was derived, i.e. for efficient measurements. However, in the general case, when $\Delta E_S \ne 0$, our lower bound Eq.~\eqref{eq:genbound} takes into account the energy variation of $S$, as thermodynamic laws require, while the lower bound obtained in \cite{Sagawa09,Erratum-Sagawa09} does not. In particular, we point at situations in Appendix \ref{app:issueqm} where the lower bound obtained in \cite{Sagawa09,Erratum-Sagawa09} seems to violated the second law of thermodynamics, while our lower bound always satisfies the second law since it is derived from it. We believe that this issue is in part due to the use in \cite{Sagawa09,Erratum-Sagawa09} of a microscopic model resorting to quantum measurements. \\

The quantity $I_\text{QC}$ verifies \cite{Sagawa12,Sagawa08} 
\be
0\leq I_\text{QC} \leq {\cal H}(\{p_r\}).\label{eq:I_QCbound}
\ee
 The limit $I_\text{QC}\to 0$ corresponds to a weak measurement acquiring an infinitesimal amount of information \cite{Jacobs06}. 

An important consequence is that efficient measurements fulfill $J_S \geq 0$, i.e. always lead to an increased knowledge of the system's state. The bound Eq.~\eqref{eq:genbound} implies that work must be spent beyond the energy received by the system to perform the measurement.

A  projective measurement associated with Kraus operators which are rank-1 projectors fulfils $I_\text{QC}=S_S(t_0)$ and therefore saturates the upper bound ${\cal H}(\{p_r\})$ of Eq.~\eqref{eq:I_QCbound} when the initial state is a mixture of the eigenstates of the measured observable. 
Note that according to our lower bound Eq.~\eqref{eq:genbound}, it is in principle possible to realize a projective measurement at finite energetic cost, apparently conflicting the results in \cite{Guryanova20}. We will see in Section \ref{sec:example} that the key to the resolution of this apparent contradiction is to take into account finite-time dynamics which unavoidably brings additional costs.

\subsection{General case: possibly inefficient measurements}\label{subsec:generalcase}

Our results also apply to the broader case of measurement with finite detection efficiency, i.e. the more realistic case where a fraction of the information extracted from the system leaks into inaccessible degrees of freedoms in the environment. In this case, the conditional state of the system can generically be cast under the form:
\be 
\rho_{S|r}(t_F) = \frac{1}{p_r}\sum_{s} M_{rs}\rho_S(t_0) M_{rs}^\dagger,
\ee
with $\sum_{r,s}M_{rs}^\dagger M_{rs} = \idop$. In this general regime, the different contributions $J_S$, $\moy{I_{S:A}}$ and $\Delta E_S$ are related to qualitatively different properties of the measurement:
\begin{itemize}
    \item $J_S$ quantifies the net  average information gain during the measurement. In the case of inefficient measurement, this quantity does not coincide with the quantum-classical mutual information, and can be negative (which means that the measurement is so inefficient that the state's entropy of each conditional state is on average larger than the initial entropy \cite{Naghiloo18}). While $J_S<0$ implies a lower work cost for the measurement process, it corresponds to cases where information is lost, so a measurement process consuming information about the system. A particular case of protocol verifying $J_S<0$ is the combination of an efficient measurement with a protocol converting information about the initial state into extracted work (i.e. a Maxwell Demon/Szilard engine protocol), which indeed enables one to reach lower net work costs (see also Appendix \ref{app:refined}).  
     \item The Holevo information $\xi_S =  J_S + \Delta S_S  \in [0;{\cal H}(\{p_r\})]$ quantifies how well the states $\rho_{S|r}(t_F)$ can be distinguished from each other, and therefore the average gain of information about the final system state obtained when reading the measurement outcome versus not reading it. It therefore quantifies the efficiency of the measurement (and goes to zero with the efficiency see Section \ref{sec:measqual}). Additionally, $\xi_S$ quantifies the information about $S$ stored in the memory at final time $t_F$, $\xi_S = I_{S:M}(t_F)$, which is also the quantity determining the performance of Bayesian Metrology \cite{Lecamwasam24}. We therefore see that the efficiency of the measurement is associated with larger work costs. Finally, it is insightful to rewrite the lower bound Eq.~\eqref{eq:genbound} in a form which makes explicit the link the work cost and the performance in Bayesian Metrology,
 \be
 W_\text{dr}  + W_\text{reset} \geq \Delta F_S + \frac{1}{\beta}(\xi_S + \bra I_{S:A}\ket) \nn
 \ee 
recalling that $F_S$ denotes the non-equilibrium free energy of $S$.
     
    \item $\log d_S \leq-\Delta S \leq 0$ is the average information loss about the system state when measurement outcome is ignored, and therefore only provides information about the strength of the measurement. Note that it depends on the initial state of the system, and can be zero (for instance for a system initially in a mixture of eigenstates of the measured obervable). In contrast, it reaches maximum value $\log d_S$, with $d_S$ the dimension of the Hilbert space of $S$, for a pure initial state transformed into a mixture of measurement observable eigenstates by the measurement process (when the outcomes are not read).
    
    \item $\moy{I_{S:A}}$ is associated to one source of inefficiency for the measurement (as already mentioned in section \ref{sec:lowerb}): the loss information inside the measuring apparatus due to the coarseness of the reading procedure (i.e. the rank of the projectors $\Pi_{A,r}$). We see that this coarse-graining tends to reduce the amount of information available from the measurement outcome quantified by $J_S$, but also tends to increases the term $\moy{I_{S:A}}$, leaving the measurement cost constant. In other words, the latter is determined by the total amount of extracted information, whether it is stored in accessible degrees of freedom or not.
    Note that the switching off process can also erase remaining correlations between $S$ and $A$, so that one can have $\moy{I_{S:A}} = 0$ even though some information remained in $A$ after the observation. This irreversible  information loss in the environment then leads to an increased entropy production, and therefore still increases the work cost (see Section \ref{sec:revmeas} for an illustration of this phenomenon).
    
    \item $\Delta E_S$ corresponds to energy exchanged directly between the system and the apparatus. This quantity, which was neglected in \cite{Sagawa09,Erratum-Sagawa09}, is non-zero when the measured observable does not commute with the system's Hamiltonian. In such cases, $\Delta E_S$ can be positive or negative depending on the initial system state. When the system receives energy from the measuring apparatus ($\Delta E_S\geq 0$), the lower bound Eq.~\eqref{eq:genbound} is increased by the same amount. This mechanism lies at the basis of measurement-driven engines \cite{Elouard17,Ding18,Buffoni19,Bresque20} which convert into work this same energy gained by the system via the measurement. One of the insights brought by our results is therefore to identify precisely the source of this energy gained by the system in a large class of models of measuring apparatuses. The lower bound on the measurement work cost Eq.~\eqref{eq:genbound} can be used to upper bound the efficiency of such engines when seen as work-to-work transducers \cite{Perna23}. We come back to these aspects in more details in Section \ref{sec:MPE}.
\end{itemize}

Our inequality Eq.~\eqref{eq:genbound} therefore provides a general lower bound which can be applied to diverse situations of interest. In the remainder of the paper, we consider more specific measurement protocols, allowing us to test the validity of this lower bound, show its tightness and investigate additional work expenditure related to finite-time measurements. 

\section{Reaching the minimum work cost: case of a measurement of an observable which commutes with the Hamiltonian}\label{sec:example}

In this section, we analyze how a class of protocols for the measurements of observables which commute with the system's Hamiltonian (i.e. quantum non-demolition measurements \cite{Braginsky1980}) can saturate the bound Eq.~\eqref{eq:genbound}. 

\subsection{Measurement from the variation of the system-meter coupling}\label{sec:settings}

We specify the general protocol introduced in section \ref{sec:meas_protocol} in two ways: First, we consider that only the coupling term $V_{SA}(t)$ in $H_{SAB}(t)$ is time dependent, and that it is of the form
\bea\label{gencoupling}
V_{SA}(t) = \chi(t) Q_S\,R_A,
\eea
where $Q_S$ is an observable of the system (the one being measured), $R_A$ an observable of the apparatus and $\chi(t)$ a time dependent coupling strength. We consider that $\chi(0) = \chi(t_F) = 0$, and denote $\chi(t_M) = \chi_M$, the coupling strength at the time where the meter is read. The system Hamiltonian is time-independent $H_S = \sum_{k=1}^{d_S}e_k \pi_k^S$, with $\pi_k^S$ the projector onto the $k$th energy eigenstate. 
In this section, we focus on observables $Q_S$ fulfilling $[Q_S,H_S]=0$. We further assume $[R_A,H_A]=0$, a condition which is fulfilled by some (but not all) experimental protocols and enables to draw simple analytical conclusions. 

Second,  we consider a bosonic bath of Hamiltonian $H_B = \sum_j \omega_j b_j^\dagger b_j$, coupled to $A$ via Hamiltonian $V_{AB} =  AB$ where $A$ is an arbitrary operator of $A$ which does not commute with $R_A$ nor $H_A$, and $B=\sum_j g_j (b_j^{\dag} + b_j)$ is a bosonic bath operator. In the weak system-bath coupling limit, we can describe the dynamics of $S$ and $A$ during the switching processes via a Bloch-Redfield master equation (that we derive for arbitrary systems $S$ and $A$ in Appendix \ref{app:MEderivation}). We consider for simplicity that $\chi(t)$ varies on a time-scale much longer than the correlation time $\tau_c$ of the bath, leading to time-dependent system transition frequencies, and therefore time-dependent dissipation rates. 
For a given value of the coupling constant $\chi_M$, its steady state $\rho_{SA}^\text{ss}$ is diagonal in the energy eigenbasis of system $S$, i.e. it takes the form (see Appendix \ref{app:steadystate}):

\be\label{genss}
   \rho_{SA}(t_M) := \rho_{SA}^{\rm ss}  = \sum_{k=1}^{d_S} p_k(0) \pi_k^S\otimes\rho_{A,k}^{\rm th},
 \ee
  where 
  \bea\label{ssak}
  \rho_{A,k}^{\rm th} = \frac{1}{Z_{A,k}}e^{-\beta H_{A,k}},
  \eea
  is the thermal equilibrium state of $A$ when $S$ is initially in the $k$th energy eigenstate, with $Z_{A,k} = {\rm Tr}[e^{-\beta H_{A,k}}]$, $H_{A,k} = \sum_n E_{n,k}\pi_n^A$ and  $E_{n,k}:= E_n + \chi_M q_k r_n$. Additionally, $\pi_n^A$ is the projector onto the $n$th energy level of $H_A$, associated with energy $E_n$, and $r_n$ is the $n$th eigenvalue of $R_A$.\\
  The state Eq.~\eqref{genss} is not yet in the form of Eq.~\eqref{eq:rhoSAr} (since the states $\rho_{A,k}^\text{th}$ are not orthogonal). 
  However, as $A$ is a statistical mixture of states of well defined energy, namely the eigenstates $|n\ket_A$, 
  it can be rewritten in such form. For that, we introduce the orthogonal subspaces in which $A$ is assumed to be read. More precisely, the measurement outcome $r$ is associated to the subspace of energy $E_A$ belonging to the interval $[E_{n_r};E_{m_r}]$, where $[n_r;m_r]$ is a subset of $\mathbb{Z}$. In more operational terms, this means that the energy of $A$ is assumed to be read with a potentially finite resolution. Introducing the projectors $\Pi_r^A = \sum_{n=n_r}^{m_r} \pi^A_n$, the state $\rho_{SA}(t_M)$ has the following form $\rho_{SA}(t_M)=\sum_rp_r\rho_{SA|r}(t_M)$, with:  
\be  
  \rho_{SA|r}(t_M) = \frac{1}{p_r}\sum_{k=1}^{d_S} p_k(0) \pi_k^S\otimes\Pi^A_r\rho_{A,k}^\text{th}\Pi_r^A,
 \ee 
 having $p_r=\sum_{k=1}^{d_s}p_{r|k}p_k(0)$ and $p_{r|k} := {\rm Tr}_A[\rho_{A,k}^\text{th}\Pi_r^A]$.

 Thus, the steady-state of $SA$ in presence of a non-zero coupling satisfies our assumption \eqref{eq:rhoSAr} for a measurement process. We can therefore consider the following protocol: (i) $A$ is initially in the equilibrium state $\rho_A(0)=e^{-\beta H_A}/Z_A$; (ii) The coupling $\chi(t)$ is increased from $0$ to $\chi_M$, and then kept constant for a time long enough to reach the steady state and ensure that $SA$ is in a classical mixture of the different $\rho_{SA|r}$; (iii) At time $t_M$, the result $r$ is read and encoded in the memory $M$ which was initially in a reference pure state; (iv) The coupling is decreased up to $\chi(t)=0$ and kept constant until $A$ thermalizes back to $\rho_A(0)$; (v) After the result has been potentially used for feedback operations, the memory is reset to a reference pure state.

At the final time $t_F$ of the protocol, the system and meter are in the state:
\be 
\rho_{SA\vert r} = \rho_{S|r}(t_F)\otimes \rho_A(0).
\ee
 with
 \be 
 \rho_{S|r}(t_F) = \frac{1}{p_r}\sum_{k=1}^{d_S} p_{r|k}p_k(0) \pi_k^S.\label{eq:rhoSrF}
 \ee

\subsection{Measurement quality and lower bound}\label{sec:measqual}

Due to the form of the final state, we have $\moy{I(S:A)}=0$ (the final state has a factorized form). In addition, $\Delta E_S =0$ as the measured observable commutes with the system Hamiltonian. For this class of protocols, we therefore have:
\be  
W_\text{drive}+W_\text{reset} \geq \frac{1}{\beta}J_S.
\ee

We can see  Eq.~\eqref{eq:rhoSrF} that the average system state at time $t_F$ is $\rho_S(t_F) = \sum_{k=1}^{d_S} p_k(0)\pi_k^S$ for any value of $\chi_M>0$.  That is, the measurement always fully projects the system in the eigenbasis of the observable (it has maximal strength). In contrast, $\chi_M$ controls the efficiency of the measurement which can be quantified by $\eta = 1+(J_S-S_S(0))/\ln d_S$ which reaches $1$ ($0$) when the states $\rho_{S|r}$ are pure (have maximal entropy $\ln d_S$). From Eq.~\eqref{eq:rhoSrF}, we simplify the information gain $J_S$ to
\be 
J_S = S_S(0) + \mathcal{H}[\{p_r\}] - \mathcal{H}[\{p_{k,r}\}],
\ee
where $\mathcal{H}[\{p_i\}]= -\sum_ip_i\ln p_i$ is Shannon's entropy of distribution $\{p_i\}$. It is also insightful to express the information gain as $J_S = I[p_k(0):p_r] - \Delta S_S$, where $I[p_k(0):p_r]$ is the mutual information between the distribution $\{p_k(0)\}$ 
and the distribution $p_r$, which is the measured distribution. Note also that  $I[p_k(0):p_r]$ is equal to the final mutual information between $S$ and the memory $M$, $I[p_k(0):p_r] = I_{S:M}(t_F) = \xi_S$. 

\begin{figure*}
    \centering
    \includegraphics[width=0.9\textwidth]{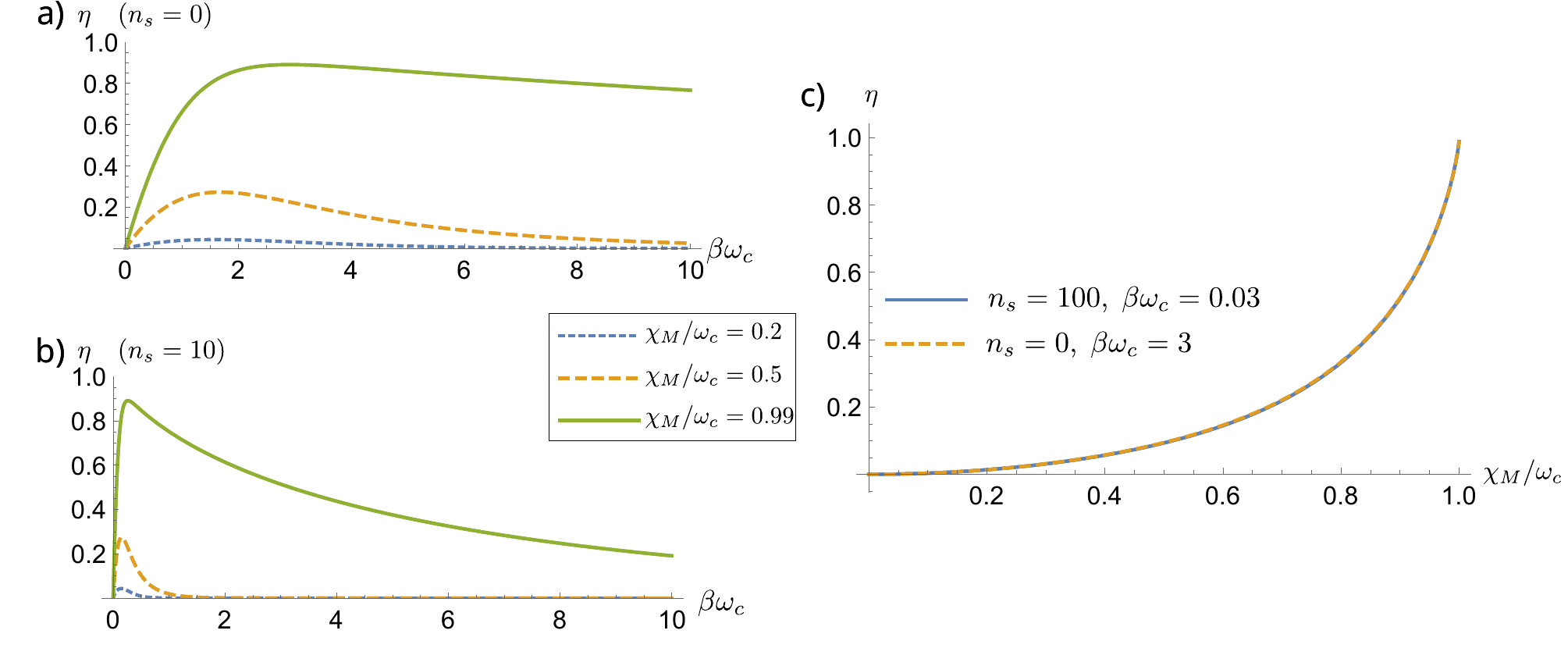}
    \caption{Efficiency $\eta = 1 + (J_S-S_S(0))/\log 2$ of the measurement for the illustrative example of a qubit $S$ measured owing to a harmonic oscillator (cavity mode) $A$ of frequency $\omega_c$ as a function of the bath inverse temperature $\beta$ for the energy threshold $n_s=0$ (a) and $n_s=10$ (b). The blue solid, red dashed and green dotted curves correspond to $\chi_M/\omega_c=0.3$, $0.7$ and $0.99$, respectively. (c): Efficiency $\eta$ as a function of $\chi_M/\omega_c$ for $(n_s,\beta\omega_c)=(100,0.03)$ (solid blue) and $(n_s,\beta\omega_c)=(0,3)$ (dashed orange). For all plots, the qubit is assumed to be initially in a pure state with $p_e(0) = p_g(0) = 1/2$.}\label{fig:JS}
\end{figure*}

To relate the efficiency of the measurement to the parameters of the model, we analyze the case where the system is a qubit of Hamiltonian $\frac{\omega_S}{2}\sigma_z$, with $\sigma_z=|e\rangle\langle e|-|g\rangle\langle g|$ and the system $A$ is an oscillator of Hamiltonian $\omega_c a^\dagger a$ ($\omega_c$ in reference to ``cavity", having in mind experimental realizations) with coupling observable $R_A = a^\dag a$. Moreover, we split the oscillator space in two by introducing the energy threshold $n_s$ and by defining the projectors $\Pi_{r=\pm}^A$ as $\Pi_{+}^A = \sum_{n\leq n_s} |n\rangle_{A}\langle n|$ and $\Pi_{-}^A = \sum_{n > n_s} |n\rangle_{A}\langle n|$. 
This leads to 
\bea
\rho_{SA}(t_M) &=& p_e(0)|e\ket\bra e|\rho_{A,e}^\text{th} +p_g(0)|g\ket\bra g|\rho_{A,g}^\text{th}\\
&=& p_+ \rho_{SA|+}(t_M) + p_- \rho_{SA|-}(t_M)
\eea
with $H_{A,e} = \sum_{n=0}^\infty (\omega_c +\chi_M)n \pi_n^A$ and $H_{A,g} = \sum_{n=0}^\infty (\omega_c -\chi_M)n \pi_n^A$. Additionally, \bea
&&\rho_{SA|+} = \nn\\&&\frac{p_e(0)}{p_+}|e\ket\bra e| \sum_{n\leq n_s} e^{-\beta n(\omega_c+\chi_M)}(1-e^{-\beta (\omega_c+\chi_M)})\pi_n^A \nn\\
&+& \frac{p_g(0)}{p_+}|g\ket\bra g| \sum_{n\leq n_s} e^{-\beta n(\omega_c-\chi_M)}(1-e^{-\beta (\omega_c-\chi_M)})\pi_n^A\nn\\
\eea
and 
\bea
&&\rho_{SA|-} = \nn\\&&\frac{p_e(0)}{p_-}|e\ket\bra e| \sum_{n> n_s} e^{-\beta n(\omega_c+\chi_M)}(1-e^{-\beta (\omega_c+\chi_M)})\pi_n^A \nn\\
&+& \frac{p_g(0)}{p_-}|g\ket\bra g| \sum_{n> n_s} e^{-\beta n(\omega_c-\chi_M)}(1-e^{-\beta (\omega_c-\chi_M)})\pi_n^A\nn\\
\eea
with 
\be
p_\pm = p_e(0)p_{\pm|e} + p_g(0) p_{\pm|g}
\ee
and $p_{+|e} = 1 - e^{-\beta (\omega_c+\chi_M)(n_s+1)}$, $p_{+|g} = 1 - e^{-\beta (\omega_c-\chi_M)(n_s+1)}$, $p_{+|e} =  e^{-\beta (\omega_c+\chi_M)(n_s+1)}$, $p_{-|g} = e^{-\beta (\omega_c-\chi_M)(n_s+1)}$. The resulting efficiency is therefore $\eta = 1 -\frac{1}{\ln 2}\Big\{\mathcal{H}[\{p_r\}] - \mathcal{H}[\{p_{k,r}\}]  \Big\}$, with $p_{k,r} = p_k(0)p_{r|k}$, for $k=e, g$ and $r= +,-$, and it is plotted in Fig.~\ref{fig:JS} for a pure qubit initial state with $p_e(0) = p_g(0) = 1/2$. From panels \ref{fig:JS}a) and b), it is clear that $J_S$ is maximized for a finite value of the inverse temperature $\beta$ which depends on the choice of energy threshold $n_s$. When $n_s$ and $\beta$ are scaled accordingly, the curves collapse to a unique function of $\chi_M$ as shown in Fig.\ref{fig:JS}c). Then, the limit of an efficient measurement is reached asymptotically for $\chi_M\to\omega_c$.

\subsection{Quasi-static limit and finite-time-induced costs}\label{sec:QSL}

We first analyze the work $W_\text{drive}$ associated with the variation of the system's Hamiltonian over the entire protocol (where the $S-A$ coupling is switched on and off). Neglecting the work to record the measurement outcome in the classical memory (e.g. assumed to have a degenerate Hamiltonian), the total work spent to switch on and off the coupling between the system and the apparatus can be computed from:
\bea
W_\text{dr} &=& \int_0^{t_F} dt {\rm Tr}\left[\rho_{SAB}(t)\frac{d}{dt}H_{SAB}(t)\right]\nonumber\\
&=& \int_0^{t_F} dt \dot \chi(t) {\rm Tr}_{SA}\left[\rho_{SA}(t)Q_SR_A\right].
\eea

In appendix \ref{app:quasisdrive}, we show that in the quasi-static limit 
the work cost $W_\text{drive}$ vanishes, that is:
\bea\label{eq:qswork}
\underset{t_f\to\infty}{\lim} W_\text{dr} = 0.
\eea

We further analyze the behavior in finite time and how the adiabatic limit is reached. For illustrative example introduced in the previous section, namely the qubit measured by the harmonic oscillator, the adiabatic limit corresponds to $ \dot\chi(t)/\kappa \ll \chi_M$ (equivalent to $\kappa t_M \gg1$), where $\kappa$ corresponds to the dissipation rate induced by the bath. In Appendix \ref{app:finitetime}, we compute the first non-zero correction to the adiabatic work cost, which is a function of the ratio $\dot\chi/\kappa$ and of the initial qubit population $p_e(0)=\langle e|\rho_S(0)|e\rangle$. We plot the two extreme values $W_\text{dr}^\pm$ of $W_\text{dr}$ (associated with $p_e(0)=1,0$) in Fig.~\ref{fig:3} using the explicit expressions provided in Appendix \ref{app:finitetime}. We observe that the adiabatic limit is more easily reached at small temperature and that both the average work cost and its fluctuations (characterized by the discrepancy between $W_\text{dr}^+$ and $W_\text{dr}^-$) can diverge for large values of $\chi_M\lesssim \omega_c$.
Put in the perspective of Fig.~ \ref{fig:JS}, efficient measurements are obtained in the limit $\chi_M \rightarrow \omega_c$, which therefore leads to a diverging work cost. As projective measurements constitute a subfamily of efficient measurement, we recover the diverging work cost pointed out in \cite{Guryanova20}.


In summary, we observe here a double trade-off between the efficiency of the measurement, the duration of the measurement and the work cost: Both decreasing the duration and increasing the efficiency require more work.

\begin{figure}
    \centering
    \includegraphics[width=0.4\textwidth]{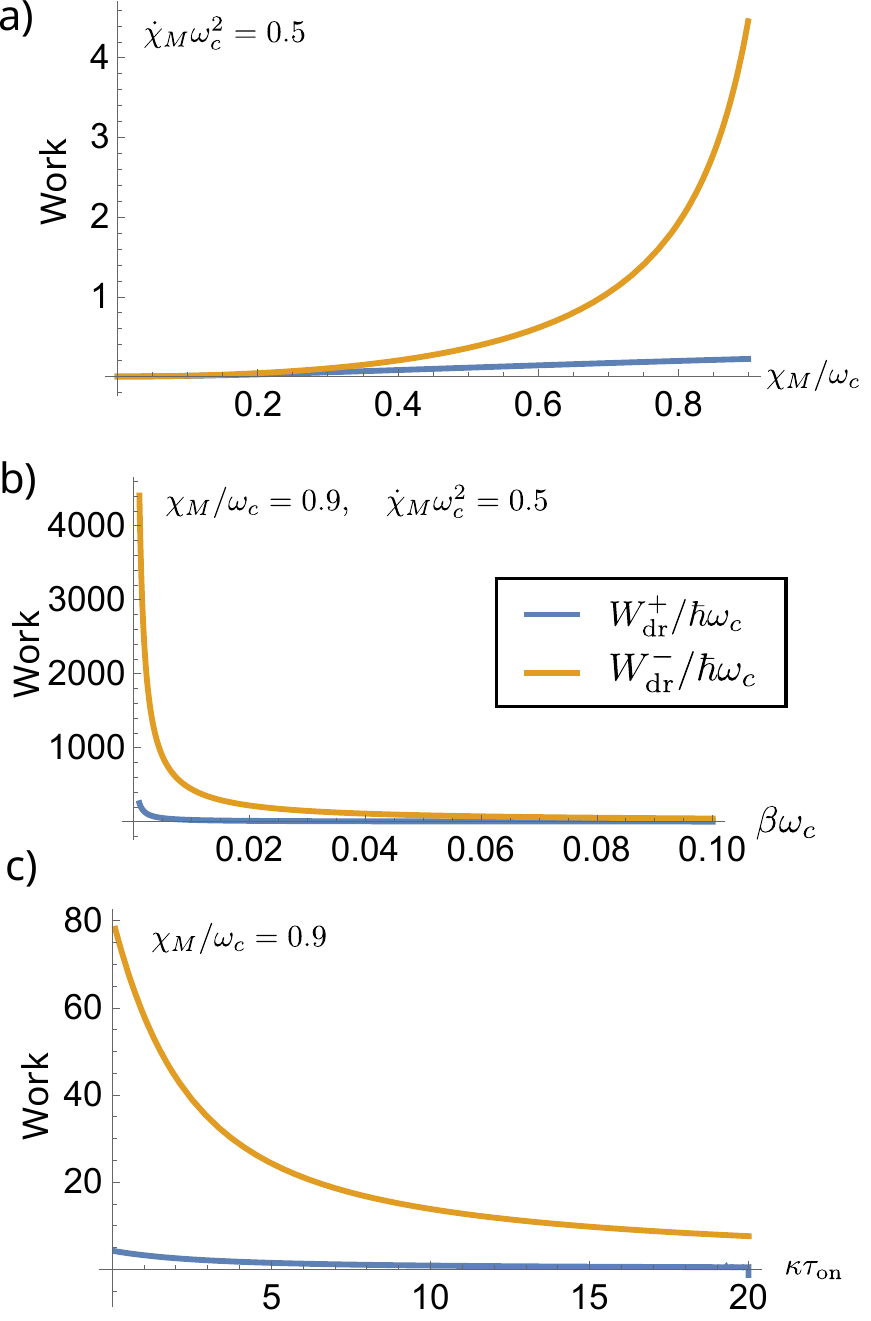}
    \caption{First nonadiabatic correction to the driving work cost $W_\text{drive}^{(1)}$ in the case where $S$ is a qubit and $A$ an harmonic oscillator (cavity mode) of frequency $\omega_c$. Blue (Orange) curves correspond to the case where the qubit is initially state $|e\rangle$ ($|g\rangle)$. a) As a function of the maximum coupling strength $\chi_M$ (at fixed coupling switching rate $\dot\chi$), b) As a function of the inverse bath temperature $\beta$, c) As a function of the time length $\tau_M=\chi_M/\dot\chi$ taken to switch on the coupling (at fixed final coupling strength $\chi_M$). The constant $\kappa$ stands for the dissipation rate induced by the bath.} 
    \label{fig:3}
\end{figure}

\subsection{Reversible measurement protocol}\label{sec:revmeas}

In this section, we explain how the quantum measurement can be performed as a reversible thermodynamic process, thereby reaching the minimum work cost.
We have shown that in the adiabatic limit, that driving work $W_\text{drive}$ vanishes while the total work cost $W_\text{drive} + W_\text{reset}$ does not saturate the lower bound Eq.~\eqref{eq:genbound}. This behavior signals additional sources of entropy production which are present even for quasi-static variations of the coupling. We identified two origins for this extra irreversibility (see Appendix \ref{app:refined}).
The first one, $\sigma_S$, corresponds to the entropy increase of the average qubit state due to the projection in the measurement basis, and is equal to 
\be
\sigma_S = {\cal H}(\{p_k(0)\})- S_S(0).
\ee
This term fulfils a fluctuation theorem and was identified as the entropy production associated with the measurement in previous works \cite{Elouard17Role}. 
The second contribution is the entropy production associated with the thermalization of the meter $A$ after the measurement result has been read. It is equal to 
\bea
\sigma_A &=& D[\rho_{SA\vert r}(t_M^+)|\rho_{SA|r}^\text{th}]\nonumber\\
&=& {\cal H}(\{p_{k,r}\}) - {\cal H}(\{p_k(0)\}).
\eea
and is due to the nonequilibrium nature of the state $\rho_{SA|r}(t_M^+) = \frac{1}{p_r}\sum_{k=1}^{d_S} p_k(0)\pi_k^S\otimes\Pi_r^A\rho_{A,k}^\text{th}\Pi_r^A$. In the quasi-static limit, such out-of-equilibrium state irreversibly relaxes to $\rho_{SA|r}^\text{th} := \frac{1}{p_r}\sum_{k=1}^{d_S} p_k(0)p_{k|r} \pi_k^S\otimes \rho_{A,k}^\text{th}$ before the coupling is slowly switched off, yielding a non-zero value of $\sigma_A$.

\begin{figure}
    \centering
    \includegraphics[width=\columnwidth]{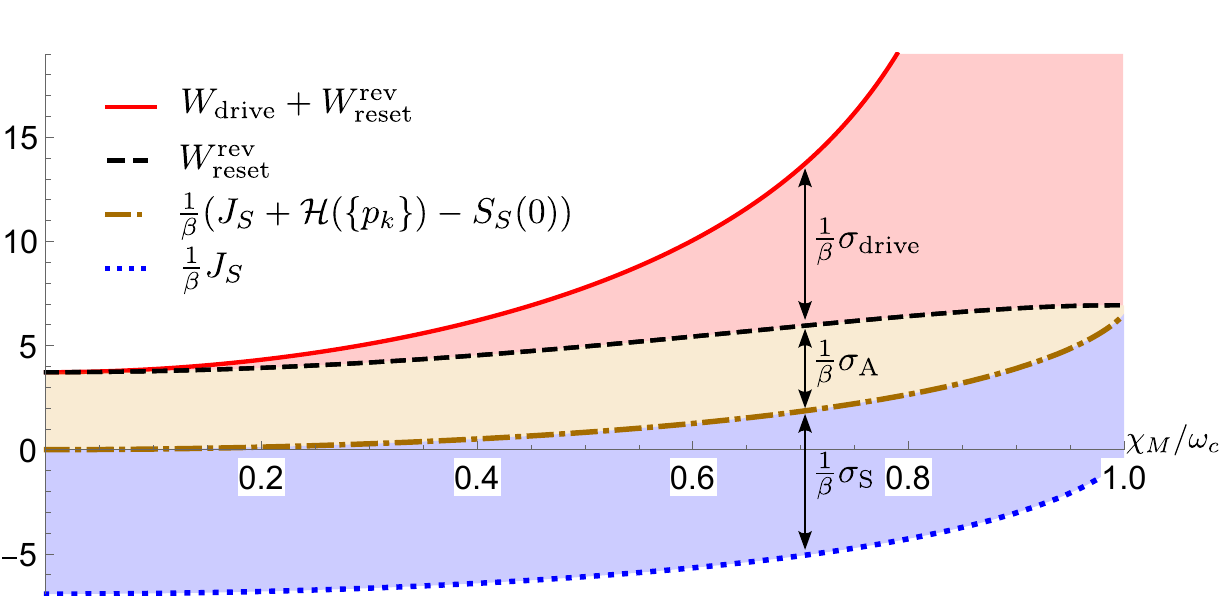}
    \caption{The three sources of entropy production inflating the work cost beyond the lower bound Eq.~\eqref{eq:genbound} for the case of a qubit $S$ measured owing to a harmonic oscillator $A$. The total average work $W_\text{drive}+W_\text{reset}^\text{rev}$ (red solid) reduces to $W_\text{reset}^\text{rev}$ (dashed black) in the quasi-static limit (see Eq.~\eqref{eq:qswork}). The distance between the red solid and black dashed curves is proportional to the entropy production $\sigma_\text{drive}$ associated with the nonadiabatic variation of the coupling $\chi(t)$. The brown dash-dotted line corresponds to the work cost when the information gained about $A$ during the readout is optimally used. It is separated from the black-dashed curve by $\sigma_A/\beta$, which vanishes for $\chi_M \rightarrow \omega_c$ (see main text). Finally, the quantity $J_S/\beta$ is plotted (blue dotted line), separated from the brown dash-dotted line by an amount $\sigma_S/\beta$. We have taken $p_e(0) = p_g(0) = 1/2$ and a pure initial qubit state.}
    \label{fig:RevMeas}
\end{figure}

All in all, the total work cost is equal to the lower bound $J_S/\beta$ plus three sources of entropy production, $\sigma = \sigma_\text{drive} + \sigma_S + \sigma_A$,
\be\label{eq:entropyprod}
W_\text{drive} + W_\text{reset}^\text{rev} = \frac{1}{\beta}(\sigma_\text{drive} + \sigma_S + \sigma_A) + \frac{1}{\beta}J_S,
\ee
where we introduced $W_\text{reset}^\text{rev} := \frac{1}{\beta}{\cal H}(\{p_r\})$ as the reversible reset work (the minimal amount of work needed to reset the memory $M$, obtained from Eq.~\eqref{eq:wreset}). 
These three contributions to the measurement irreversibility, $\sigma_\text{drive}$, $\sigma_S$, and $\sigma_A$ are illustrated in Fig.~\ref{fig:RevMeas} together with the total work cost $W_\text{drive}+W_\text{reset}^\text{rev}$ for the example of the qubit measured by the oscillator. Note also that the heat provided by the bath $B$, $Q_B = -\Delta E_B$, is directly obtained from $W_\text{drive} = - Q_B$ since we are in the situation where $\Delta E_{SA} =0$.\\

Identifying those two irreversible contributions beyond the quasi-static limit allowed us to propose a protocol reaching reversibility, at least in principle. The core idea behind this protocol originates from information thermodynamics, which demonstrated, first in a classical context, that information about a system can be consumed to extract work \cite{Parrondo15}. Such engine was first introduced by Szilard and can be seen as the reverse process of Landauer's reset protocol which consumes work to decrease the entropy of a memory. Here, reaching the minimum work cost imposed by the second law requires to fully exploit all the information about the systems $A$ and $S$ available to the observer to extract work before it is irreversibly erased during the measurement process.\\


First, with respect to $\sigma_S$, the entropy production due to the decoherence in the eigenbasis of $Q_S$, it can be avoided by extracting work from initial coherences before switching on $\chi(t)$ without affecting the measurement process. This operation can be done reversibly in principle via the following protocol \cite{Kammerlander16}: (i) Sudden variation (quench) of the system Hamiltonian to $H_S^{(1)} =-\log \rho_S(0)\beta$; (ii) While in contact with a thermal bath at temperature $\beta$, quasi-static variation of the system Hamiltonian to $H_S^{(2)} = -\log \tilde\rho_S(0)/\beta$, where $\tilde\rho_S(0)=\sum_k \pi_k^S\rho_S(0)\pi_k^S$ is the incoherent mixture resulting from the projection onto the the eigenstates of $Q_S$; (iii) Sudden variation of the system's Hamiltonian back to $H_S$. This protocol leads to work extraction $W_S = \beta^{-1} (S_S(0)-{\cal H}(\{p_k(0)\})$, implying a reduction of the overall work cost (see illustrative representation in Fig.~\ref{fig:revprot}). This reduction compensates exactly the term $\frac{1}{\beta}\sigma_S$ in Eq.~\eqref{eq:entropyprod}.\\
Secondly, with respect to $\sigma_A$, work can be extracted in a similar way by exploiting the knowledge of state $\rho_{SA|r}$ right after reading the measurement outcome, and before starting to switch off the coupling. More precisely, work can be extracted while transforming the conditional state $(1/p_r)\Pi_r^A\rho_{A,k}(t_M^+)\Pi_r^A$ of the apparatus into a thermal state $\rho_{A,k}^\text{th}$. The optimal reversible protocol involves a sudden variation of the Hamiltonian of $A$ to the Hamiltonian $H_{A,k}$, followed by a quasi-static restoration of the initial value $H_A$ in presence of a bath at inverse temperature $\beta$ \cite{Bera19,Elouard23}. We show in Appendix \ref{app:refined} that such protocol extracts precisely a quantity of work equal to $\frac{1}{\beta}\sigma_A$.

Remarkably, this entropy production $\sigma_A$ naturally vanishes in the case of efficient measurement. This can be seen as follows. For efficient measurements, the states $\rho_{A,k}^\text{th}$ tend to be orthogonal between each other, so that there exists a choice of projectors $\{\Pi_r^A\}_r$ such that $(1/p_r)\Pi_r^A\rho_{A,k}^\text{th}\Pi_r^A=\delta_{r,k}\rho_{A,k}^\text{th}$. This implies that the state post-observation $\rho_{SA|_r}(t_M^+)$ is already an equilibrium state, and therefore there is no entropy production associated with relaxation to equilibrium, hence $\sigma_A=0$. 
In the implementation where $S$ is a qubit measured by an harmonic oscillator, this is achieved in the limit $\chi_M\to\omega_c$ together with the choice $n_s=0$.
This is illustrated in Fig.~\ref{fig:RevMeas} where one can see that the brown area (corresponding to $\sigma_A$) vanishes for $\chi_M\to\omega_c$. Fig.~\ref{fig:RevMeas} was plotted using the explicit expression of $\sigma_S$ and $\sigma_A$ provided in Appendix \ref{app:refined}, and using $\sigma_\text{drive} = \frac{1}{\beta}W_\text{drive}$ with the explicit expression of $W_\text{drive}$ provided in Appendix \ref{app:finitetime} for an initial pure qubit state such that $p_e(0)=p_g(0)=1/2$.

\section{Measurements of an observable not commuting with the Hamiltonian}

To prove the achievability of the bound in the general case, we analyze in this section a class of protocols mapping measurements of observable not commuting with $H_S$ onto measurements of observables which do commute with the system's Hamiltonian, and show they saturate the bound.\\

For a general observable $Q_S = \sum_{k=1}^{d_S} q_k |q_k\rangle_S\langle q_k|$, one can always define a unitary $U_S$ acting on the system only such that $[\tilde Q_S,H_S]=0$ with $\tilde Q_S=U_S^\dagger Q_S U_S$. One can for instance pick $U_S= \sum_{k=1}^{d_S} |e(k)\rangle_S\langle q_k|$ for some choice of ordering $e(k)$ of the energy eigenstates $|k\rangle_S$ of $S$. Then, one can use the following protocol to implement the measurement of $Q_S$:
\begin{enumerate}
    \item Apply unitary $U_S$ to the initial state of the system. After this step, the state of $SAB$ is then $\tilde\rho_S(0)\rho_{AB}(0)$ with $\tilde\rho_S(0)=U_S\rho_S(0)U_S^\dagger$.
    \item Perform the measurement of observable $\tilde Q_S$. After this step (including the apparatus and memory reset), the total system $SAB$ is in a state $\tilde\rho_S(t_F)\otimes\rho_{AB}(0)$.
    \item Apply unitary $U_S^\dagger$, transforming $\tilde\rho_S(t_F)$ into $\rho_S(t_F)$.
\end{enumerate}

It is straightforward to check that this protocols implements a quantum measurement characterized by $\rho_{S|r}(t_F)=\frac{1}{p_r} \tilde{M}_{r,s} \rho_S(0) \tilde{M}_{r,s}$ with $\tilde M_{r,s}=U_S^\dagger  M_{r,s}  U_S$. We are interested in particular in the case where the measurement of $\tilde Q_S$ performed in step 2 is noninvasive,
such that $M_{r=k,s}\propto |e(k)\rangle_S\langle e(k)|$, which implies $\tilde M_{r=k,s} \propto |q_k\rangle_S\langle q_k|$ and therefore a noninvasive measurement of $Q_S$.\\

We now analyze the work cost associated with the protocol. Steps 1 and 3 are unitary and therefore are associated with minimum work costs equal to the energy variations of system S, namely
\bea 
W_\text{S}^{(1)} &=& \text{Tr}\{(\tilde\rho_S(0)-\rho_S(0))H_S\}\nonumber\\
W_\text{S}^{(3)} &=& \text{Tr}\{(\rho_S(t_F)-\tilde\rho_S(t_F))H_S\}.
\eea
The work to perform the measurement in step 2, that is $ W^{(2)}_\text{drive}+W^{(2)}_\text{reset}$, obeys to bound Eq.~\eqref{eq:genbound} with $\Delta E_S^{(2)}=0$ since the observable $\tilde Q_S$ commutes with $H_S$. We have:
\be 
W^{(2)}_\text{drive}+W^{(2)}_\text{reset} \geq \frac{1}{\beta}(J_S^{(2)}+\moy{I_{S:A}}^{(2)}),
\ee
with an equality reached for the reversible protocol presented in section \ref{sec:revmeas}.\\

We now use the fact that the measurement of $\tilde Q_S$ in step 2 preserves the system's energy, that is $\text{Tr}\{\tilde\rho_S(t_F)H_S\}=\text{Tr}\{\tilde\rho_S(0)H_S\}$ to find that:
\be
W^{(1)}_S+W^{(3)}_S = \text{Tr}\{(\rho_S(t_F)-\rho_S(0))H_S\} = \Delta E_S.
\ee
Finally, we see that the total work cost $W_\text{tot} = W^{(1)}_S+W^{(3)}_S+W^{(2)}_\text{drive}+W^{(2)}_\text{reset}$ is increased with respect to the case of an observable commuting with the Hamiltonian by the minimum amount predicted by the bound \eqref{eq:genbound}, that is $\Delta E_S$. Moreover, if the measurement in step 2 is performed according to the thermodynamically reversible protocol presented in section \ref{sec:example}, $W_\text{tot}$ is saturating the lower bound Eq.~\eqref{eq:genbound}. In conclusion, we show that the general lower bound Eq.~\eqref{eq:genbound} is tight and we exhibit an explicit protocol reaching it.

\section{Consequences for measurement-powered engines}\label{sec:MPE}

Measurement-powered engines (MPEs) use the thermodynamic consequences of the dynamics induced by a quantum measurement to enable work extraction. Various examples of such machines have been proposed, based on a feedback loop \cite{Elouard17,Elouard18,Bresque20,Manikandan22,Fadler23}, or on an unread measurement process \cite{Yi17,Ding18}. Although refrigerators were also proposed \cite{Buffoni19}, we will focus in this section on motors.

To draw the consequences of our results for MPEs, we consider a cyclic engine including a quantum measurement that injects an energy $E_\text{meas}$ into the working medium, as well as strokes in contact with a single thermal bath at temperature $T_\text{bath}$ resulting in a work extraction equal to $-W_\text{extr}$. For the whole cycle, we can write: 
\bea
  \text{First law}&\text{:}&\quad W_\text{extr}+ Q_\text{bath} + E_\text{meas}= 0 \label{eq:1stlawMPE}\\
  \text{Second law}&\text{:}&\quad 
  \frac{Q_\text{bath}}{T_\text{bath}} + \frac{Q_\text{meas}}{T_\text{meas}} \leq  0. \label{eq:2ndlawMPE}
\eea
where $E_\text{meas} = W_\text{meas}+Q_\text{meas}$ is the sum of the work $W_\text{meas} = W_\text{dr}+W_\text{reset}$ paid to perform the measurement and reset the measuring apparatus, as analyzed in previous sections of the present article, and the heat provided by the bath used for the measurement, which we denote in this section $Q_\text{meas}$ to distinguish it from the heat $Q_\text{bath}$ exchanged during the other strokes of the engine.


Applying our result Eq.~\eqref{eq:genbound}, we deduce (focusing on a measurement with $\moy{I_{S:A}}=0$ to simplify the discussion):
\bea\label{eq:2lawMPE}
 W_\text{meas} \geq E_\text{meas} + \frac{J_S}{T_\text{meas}} \Rightarrow Q_\text{meas} \leq -\frac{J_S}{T_\text{meas}}.
\eea
We also assume $-W_\text{extr}\geq 0$, which means that the machine behaves as an engine. These thermodynamic constraints give rise to three different regimes, that we analyze below, each associated with a different origin of the extracted work $-W_\text{extr}$, namely $W_\text{meas}$, $Q_\text{bath}$, and $Q_\text{meas}$. 

First, when $J_S \geq 0$, Eq.~\eqref{eq:2lawMPE} implies that $Q_\text{meas} \leq 0$. If furthermore, $W_\text{meas} + W_\text{extr} = -Q_\text{meas}-Q_\text{bath}\geq 0$, the origin of the extracted work is $W_\text{meas}$: the machine is a work-to-work transducer, which converts part of the work spent during the measurement $W_\text{meas}$ into extracted work. The efficiency of this conversion process is quantified by  $\eta_\text{transd.} = - W_\text{extr}/W_\text{meas} \leq 1$, and is degraded by the presence of non-zero heat flows $Q_\text{meas}+Q_\text{bath} \leq 0$. This is the operation regime of most measurement-driven engines, where the measurement replaces the hot bath of a conventional thermal engine, while allowing some information extraction, as in \cite{Elouard17,Elouard18,Bresque21}.\\
Another regime is obtained when $W_\text{meas} + W_\text{extr} \leq 0$ while still having $J_S \geq 0$, which implies that $Q_\text{bath}\geq 0$. Then, in this case, the fuel of the engine is the heat provided by the engine's bath, converted into work owing to the information brought by the measurement: the machine then implements a Maxwell demon. The feasibility of such machines in the case of arbitrary quantum measurements, including with finite detection efficiency, has been demonstrated experimentally in \cite{Cottet17,Naghiloo18}. The efficiency of the engine is then quantified by $\eta_\text{Demon} = -(W_\text{extr}+W_\text{meas})/Q_\text{bath} \leq 1-T_\text{meas}/T_\text{bath}$, such that this regime of operation is only possible if the bath involved in the measurement is colder than the engine's bath. We also stress that in this case, the energy fueling the engine does not come from the measurement, and only the informational role of the measurement is crucial (hence the existence of classical Maxwell demons, like the Szilard engine  \cite{Parrondo15}). Note that the efficiency is limited by the Carnot bound because the cost of information acquisition is already included in $W_\text{meas}$.\\

Finally, our analysis reveals the existence of a third working regime, which is obtained when $Q_\text{meas}\geq 0$ contributes to fuel the engine. This is possible only when $J_S \leq 0$ (the measurement is inefficient and causes a net loss of information about the system). If $W_\text{meas} \leq 0$, the engine is purely fueled by the heat from the measurement bath $T_\text{meas}$, which is allowed provided $T_\text{meas} \geq T_\text{bath}$. The engine efficiency is then limited by the corresponding Carnot efficiency, $1-T_\text{bath}/T_\text{meas}$. When $W_\text{meas} \geq 0$, the engine is fueled both by work and heat resources. By combining Eq.~\eqref{eq:1stlawMPE} and Eq.~\eqref{eq:2ndlawMPE}, we obtain the thermodynamic constraint on the machine:
\be
 - W_\text{extr} \leq W_\text{meas} + Q_\text{meas}(1-T_\text{bath}/T_\text{meas}).
\ee
While no textbook approach exists to quantify the efficiency of such hybrid machine, a framework has been proposed in \cite{Manzano20}. One can for instance use the ratio
\be
\epsilon_\text{hybrid} = \frac{- W_\text{extr}}{W_\text{meas} + Q_\text{meas}(1-T_\text{bath}/T_\text{meas})} \leq 1,
\ee
which is equal to $1$ in the case of optimal use of the resources. A first straightforward application of this regime corresponds to engines which include a quantum measurement in their cycle, but do not require the use of the measurement information (that is, involve no feedback step), as proposed for instance in \cite{Yi17,Ding18}. However, we stress that if the information brought by the measurement is acquired by some classical memory and simply not used, such machines actually falls under the first case described above (a work-to-work transducer, simply used under-optimally). To achieve this heat-powered regime, the measurement process must be engineered to make the information about the outcome unavailable \cite{MeasFridgePaper}. More generally, we predict the existence of more subtle thermodynamic cycles involving measurements in which information about the system is lost overall ($J_S \leq 0$), but gained about some specific observables of the system, allowing for feedback. In this context, existing ability to engineer a measurement's back-action would then translate into the ability to deliver specifically heat or work into quantum systems. The feasibility and capability of such engines require further studies.

\section{Conclusions}

We have analyzed a generic physical model of a measurement apparatus coupled to a system being measured. This model captures the two key stages of the measurement process: the pre-measurement, in which system-apparatus correlations are generated, and objectification, entailing the generation of classical apparatus states. The latter is ensured by the presence of a thermal reservoir inducing decoherence between different apparatus states associated with different measurement outcomes. 
By applying the second law of thermodynamics to this model, we have derived a lower bound for the work expended in measuring an arbitrary observable $Q_S$ of a quantum system $S$. This lower bound comprises the energy variation of $S$ along with additional contributions related to the information extracted from the system, which we link to the measurement's quality. 
Our lower bound extends the bound derived in \cite{Sagawa09}, taking into account the energy variation $\Delta E_S$ of the system as well as encompassing arbitrary measurement operations, including inefficient measurements. Unlike previous studies, our results were obtained without resorting to quantum measurements in the description of the microscopic model, and therefore avoid related issues (see Appendix \ref{app:issueqm}) and naturally follows the second law of thermodynamics.

Analyzing a protocol with a time-dependent system-apparatus coupling in the presence of a thermal bath, we have examined the behavior of entropy production, which increases the work cost beyond the lower bound. Entropy production consists of three distinct contributions: one arising from finite-time open dynamics (which vanishes in the quasi-static limit), a second one intrinsically tied to the measurement process, stemming from the projection of the initial state onto the measurement basis, and a third associated with the information gained about the apparatus itself during the measurement. 

We then demonstrate through an explicit protocol that quantum measurements can be performed in a thermodynamically reversible manner, allowing us to saturate our general lower bound, which is therefore tight, irrespective of whether the measured observable commutes with the system's Hamiltonian. Such reversible protocols must be run quasi-statically, must involve the extraction of work from initial coherences present in $\rho_S(0)$, and must fully exploit information gained about the apparatus state to tame sources of irreversibility.
Lastly, we identify a double trade-off for finite-time protocols between, on the one hand, the efficiency and the work cost, and on the other hand, the duration of the measurement and the work cost. Our methodology paves the road towards a systematic optimization of energy cost and heat dissipated during quantum measurements taking into account constraints on measurement quality. This is one of the necessary step towards the energy optimization of fault-tolerant quantum algorithms \cite{Fellous-Asiani23}.

Finally, we have drawn some consequences of our results concerning the thermodynamic analysis of Measurement-Powered Engines. We identify three distinct regimes for thermodynamic cycles based on a measuring apparatus and a single heat bath. They correspond respectively to work-to-work transducer, quantum Maxwell demon, and hybrid engine that may be powered by both work and heat, or solely by heat coming from the reservoir involved in the measurement. The latter regime is largely unexplored and suggests the existence of new types of measurement-driven machines -- yet to be explored -- which could exploit imperfect measurements to achieve better efficiency.

\acknowledgements
C.LL. acknowledges funding from the French National Research Agency (ANR) under grant ANR-23-CPJ1-0030-01.
C.E. acknowledges funding from the French National Research Agency (ANR) under grant ANR-22-CPJ1-0029-01.
We thank Lorena Ballesteros Ferraz, Nathan Shettell, Alexia Auffèves and Benjamin Huard for useful discussions.

\onecolumngrid
\appendix

\section{Issue related to resorting to quantum measurements in the microscopic model}\label{app:issueqm}
In this section we expose one issue related to the use of quantum measurements in the description of the microscopic model aimed precisely at analyzing energetic resources used during a quantum measurement. Indeed, previous studies \cite{Sagawa09, Jacobs09,Abdelkhalek16,Funo13}  use a projective measurement (or a generalized measurement \cite{Minagawa23}) on the measurement apparatus after its interaction with the system $S$ of interest. In the microscopic model used here, we avoid that by imposing that the bath interaction naturally leads to the emergence of statistical mixture \cite{Allahverdyan13, Korbicz14}. We expose in the following the consequences of this difference for energetic analyses.

\subsection{Projective measurement on the apparatus}\label{app:projmeas}
Following \cite{Sagawa09,Abdelkhalek16,Funo13}, the system of interest $S$ interacts unitarily with a measurement apparatus $A$ through $U_\text{int}$, assuming $S$ and $A$ are initially and finally uncoupled, and their initial and final local Hamiltonian $H_S$ and $H_A$ are unchanged. Right after their interaction, a projective measurement described by the projectors $\{\Pi_{A,k}\}_k$ is applied on $A$. The resulting state is
\be\label{eq:SAtau}
\rho_{SA}(\tau) = \sum_k \Pi_{A,k} U_\text{int} \rho_S(0)\rho_A(0)U_\text{int}^{\dag}\Pi_{A,k},
\ee
resulting in (generalized) measurement on $S$,
\be
\rho_S(\tau) = \sum_{k,l,m} M_{k,l,m}\rho_S(0)M_{k,l,m}^\dag
\ee
with $M_{k,l,m}:= \sqrt{r_m} \bra k,l|U_\text{int}|r_m\ket$, where $\Pi_k := \sum_l |k,l\ket\bra k,l|$ and $\rho_A(0) = \sum_m r_m |r_m\ket\bra r_m|$.
The energy cost associated with such operation is the sum of two contributions, the first being the work required for the application of $U_\text{int}$, and the second is the work required to perform the projective measurement of $A$. The first contribution is simply the energy variation of $SA$ during the application of $U_\text{int}$,
\be
W_\text{int} = {\rm Tr}\left[\left(U_\text{int}\rho_S(0)\rho_A(0)U_\text{int}^\dag - \rho_S(0)\rho_A(0)\right)(H_S + H_A)\right].
\ee
However, we do not know the second contribution associated with the projective measurement on $A$. One common hypothesis, see for instance in \cite{Sagawa08, Sagawa09, Jacobs09, Abdelkhalek16, Funo13, Mancino18, Minagawa23}, is to consider that the (average) work injected by the projective measurement is simply the energy difference between the pre and post-measurement states,
\be
W_{\Pi_{A,k}} =  {\rm Tr}\left[\left(\rho_{SA}(\tau) - U_\text{int}\rho_S(0)\rho_A(0)U_\text{int}^\dag \right)(H_S + H_A)\right].
\ee
All together, the work cost is simply the energy variation of $SA$,
\be\label{eq:ev}
W_\text{meas} := W_\text{int} + W_{\Pi_{A,k}} = \Delta E_{SA}:= {\rm Tr}\left[\left(\rho_{SA}(\tau) - \rho_S(0)\rho_A(0)\right)(H_S + H_A)\right].
\ee
One problem with such approach is to implicitly assume that the energy exchange during a projective quantum measurement is purely work, which seems arbitrary and problematic in the light of the inherent entropy change induced by projective measurements. A second issue is detail below.  

\subsection{Bath-induced projection}
Now, recapping and simplifying the approach followed in this paper, and inspired from \cite{Allahverdyan13,Korbicz14}, the system $S$ of interest interacts with a measurement apparatus $A$ itself interacting with a thermal bath $B$. The whole ensemble evolves unitarily throught  $U_{SAB}$, and the coupling between $S$ and $A$ is initially and finally null, and the local Hamiltonians of $S$ and $HA$ are initially and finally unchanged. The final state is of the form
\be
\rho_{SAB}(\tau) = U_{SAB} \rho_S(0)\rho_{AB}(0)U_{SAB}^\dag,
\ee
We now assume that the bath induces a dissipation reproducing the CPTP (quantum) map ${\cal E}_A[\cdot] = \sum_k \Pi_{A,k}\cdot\Pi_{A,k}$, with the same set of projectors $\{\Pi_{A,k}\}_k$ involved in the measurement of $A$ in previous section. We then obtain:
\be
\rho_{SA}(\tau) = \sum_k \Pi_{A,k} U_\text{int} \rho_S(0)\rho_A(0)U_\text{int}^{\dag}\Pi_{A,k},
\ee
which is precisely the state obtained in \eqref{eq:SAtau}.
The work required for such operation is the energy variation of $SAB$,
\be
 W_\text{dr} := \int_0^{t_f} du {\rm Tr}[\rho_{SAB}(u)\dot H_{SAB}(u)] =  \Delta E_{SAB} = \Delta E_S + \Delta E_A + \Delta E_B + \Delta V_{AB}.
\ee 
The entropy production associated with the interaction with $B$ is \cite{Esposito10},
\be
\Sigma := \Delta S_{SA} + \beta \Delta E_B \geq 0.
\ee 
Equivalently:
\be
\Delta E_B = -\beta^{-1} \Delta S_{SA} + \beta^{-1} \Sigma,
\ee
implying 
\be\label{eq:wdr}
W_\text{dr} = \Delta F_{SA} + \beta^{-1} \Sigma + \Delta V_{AB} \geq \Delta F_{SA} + \Delta V_{AB}.
\ee

\subsection{Conclusion}
Setting $\Delta V_{AB} = 0 $ or neglecting it (vanishing coupling),
and since $\Delta S_{SA}$ is always positive (because the final state of $SA$ can be equivalently described by projective measurement on A), we have
\be
W_\text{dr} \geq \Delta F_{SA},
\ee
while for microscopic model using projective measurements on $A$ we have
$W_\text{meas} = \Delta E_{SA} \geq \Delta F_{SA}$.
 Thus, $W_\text{meas} = \Delta E_{SA}$, generally overestimates the work cost associated with the overall measurement operation on $SA$. This results in estimations which are not tight, in addition to being based on the apparently arbitrary choice of considering that projective measurements only involve work exchanges. Additionally, from eq. \eqref{eq:wdr}, one can see that the work cost depends on the level of entropy production generated during the operation, and therefore depends on the protocol even at  fixed final state, which is not the case of $W_\text{meas} = \Delta E_{SA}$.

\subsection{Lower bound obtained in \cite{Sagawa09} and the second law of thermodynamics}

The authors of  \cite{Sagawa09} follow the above approach described in section \ref{app:projmeas}, with its inherent issues as just mentioned. On top of that, the authors neglect the energetic contribution from the system, which can lead to violation of the second law. This point is detail further in this section. Note that in our notation, the measurement apparatus is denoted by $A$, while in \cite{Sagawa09}, it is denoted by $M$. In the following, we will use the notation $M$ to ease the direct comparison with results from \cite{Sagawa09}.  \\

In \cite{Sagawa09}, the quantum measurement of a system $S$ is described by the interaction with a measurement apparatus $M$, followed by a projective measurement on $M$, and finally concluded by switching off the coupling between $S$ and $M$ followed by the interaction between $M$ and a bath $B$. The initial state of the ensemble is
\be
\rho_{SMB}(0) = \rho_S(0)\otimes\rho_M(0)\otimes\rho_B(0)
\ee
where $\rho_M(0)$ and $\rho_B(0)$ are thermal states at the same inverse temperature $\beta$. After the aforementioned protocol, the final state is of the form (after assumption Eq.(1) of \cite{Sagawa09} and assumption of efficient measurements acknowledged in \cite{Erratum-Sagawa09}),
\be\label{fstate}
\rho_{SMB}(t_f) = \sum_k M_k \rho_S(0) M_k^{\dag} \otimes U_f \rho_{MB,k}U_f^{\dag},
\ee
where $M_k$ is the resulting measurement operator on $S$ resulting from the unitary interaction with $M$ followed by its projective measurement described by projective operators $\{P_k\}_k$, $\rho_{MB,k}$ is the post-measurement state of $MB$ associated with the outcome $k$, and $U_f$ is the unitary interaction between $M$ and the bath $B$. \\

The work required for such operation, $W_\text{meas} = {\rm Tr}\{[\rho_{MB}(t_f)-\rho_{MB}(0)][H_M + H_B]\} $, is defined as the energy difference of $MB$ between the initial state $\rho_M(0)\otimes\rho_B(0)$ and $ \sum_k p_k \rho_{MB,k},$ the state just before their unitary interaction trhough $U_f$, with $p_k := {\rm Tr}[ \rho_S(0) M_k^{\dag}M_k]$. From there, the authors derive a lower bound for the energy variation undergone by $MB$, arriving at (eq.(2) of \cite{Sagawa09}),
\be\label{SUbound}
W_\text{meas} \geq \bra \Delta F_M^\text{eq}\ket + \beta^{-1}(J_S - H) := W_\text{S\&U}
\ee
where $\bra \Delta F_M^\text{eq}\ket := \sum_k p_k F^\text{eq}_{M,k} - F^\text{eq}_{M,0}$ is the average equilibrium free energy variation, with $F_{M,k}^\text{eq} = -\beta^{-1} \ln Z_{M,k}$ and $Z_{M,k} = {\rm Tr}\left(e^{-\beta P_k H_M P_k}\right)$. Note that the authors also assumed that $\rho_S(0) = Z_{M,0}^{-1}e^{-\beta H_{M,0}}$. Finally, $H =- \sum_k p_k \ln p_k$ is the Shanon entropy associated with the distribution $\{p_k\}_k$, and 
\be
J_S = S_S(0) - \sum_k p_k S(\rho_{S,k})
\ee
is the average information gain on $S$ obtained by the knowledge of the outcome $k$, with $\rho_{S,k} := p_k^{-1} M_k \rho_S(0)M_k^{\dag}$.\\

Now, as seen in the beginning of this appendix, if instead of using a quantum measurement on the apparatus we resort to an adequate interaction with a thermal bath, the resulting work cost is obtained by the second law of thermodynamics and fulfilled 
\be 
W_\text{dr} \geq \Delta F_{SM}
\ee
 Using Eq. \eqref{fstate} to compute $\Delta F_{SM}$, we obtain
\bea
\Delta F_{SM} &=& F_{SM}[\rho_{SM}(t_f)] - F_{SM}[\rho_{SM}(0)] \nn\\
&=&\Delta F_M + \Delta E_S + \beta^{-1}\left[  \xi_S - \Delta S_S + \xi_M - \xi_{SM} \right]\nn\\
&=& \Delta F_M + \beta^{-1}(J_S - H) + \Delta E_S + \beta^{-1}\left[ \xi_M + H-\xi_{SM}\right] \nn\\
&=& W_\text{S\&U} + \Delta E_S + \beta^{-1}\left[ \xi_M + H-\xi_{SM}\right] + \bra \Delta F^\text{eq}_M\ket - \Delta F_M\\
&=& W_\text{S\&U} + \Delta E_S + \beta^{-1}\left[  H-\xi_{SM} + \sum_k p_k D[\rho_{M,k}^f|| \rho_{M,k}^\text{th}]\right],
\eea
where $\xi_X = S[\sum_k p_k \rho_X] - 
\sum_k p_k S[\rho_X]$, $X=S, M, SM$.
Finally, the lower bound Eq. \eqref{SUbound} is saturated when \cite{Sagawa09} $S[\rho_{SMB}(0)] = S[\rho_{SMB}(t_f)]$ and when $\rho_k^{MB} = \rho_{M,k}^\text{th} \otimes \rho_B^\text{th}$, where $\rho_{M,k}^\text{th} := Z_{M,k}^{-1}e^{-\beta P_K H_M P_k}$. Then, under such circunstances, we have that the work cost introduced by Sagawa and Ueda satisfies,
\be
W_\text{meas} = W_\text{S\&U}
\ee
while the lower bound imposed by the second law becomes
\be
W_\text{dr} \geq  W_\text{S\&U} + \Delta E_S + \beta^{-1}\left[  H-\xi_{SM} \right]
\ee
Thus, it seems that the work cost introduced in \cite{Sagawa09} misses the energetic contribution from $S$ (as well as a residual correlations between $S$ and $M$, $H-\xi_{SM} \geq 0$), which can lead to violation of the second law when $\Delta E_S > 0$.

\section{Efficient measurement}\label{app:ISA}

The system state conditioned on a given measurement outcome $r$ has the form:

\bea  
p_r \rho_{S\vert_r} &=& \text{Tr}_{A,B}\left\{U_\text{on}\Pi_{r}^AU_\text{off}(\rho_S(0)\otimes\rho_A(0)\otimes\rho_B(0))U_\text{on}^\dagger\Pi_{r}^AU_\text{off}^\dagger\right\}\nonumber\\
&=& \sum_{n,n'} p_{A,n}(t_0)\sum_{i,i'} p_{B,i}(t_0) {}_{AB}\langle n' i'| U_\text{off}\Pi_{r}^AU_\text{on} | ni\rangle_{AB}\; \rho_S(0) \;{}_{AB}\langle ni| U_\text{on}^\dagger\Pi_{r}^AU_\text{off}^\dagger | n'i'\rangle_{AB}\nonumber\\
&=& \sum_s M_{r,s} \rho_S(0) M_{r,s}^\dagger.\label{eq:condstate}
\eea
The projective operator $\Pi_r^A$ is the projector onto the support of $\rho_{A|r}(t_M)$. Additionally, to go to the last line, we have gathered together all the tuplets $(n,n',i,i')$ leading to system operators which are proportional to each others.  Namely, denoting ${\cal S}^{r}_{s}$ each such set of tuplets, there exists an operator acting on $S$ denoted by $K_{r,s}$ such that  
\be  
\forall (n,n',i,i') \in {\cal S}^{r}_s, \;\exists  \lambda_{n,n',i,i'} \in\mathbb{C}, \;\;{}_{AB}\langle n' i'| U_\text{on}\Pi_{r}^AU_\text{off} | ni\rangle_{AB} = \lambda_{n,n',i,i'} K_{r,s},
\ee
and
\be
M_{r,s} = \sqrt{\sum_{(n,n',i,i')\in {\cal S}^{r}_s } p_{A,n}(t_0)p_{B,i}(t_0)\lambda_{n,n',i,i'}} K_{r,s}.
\ee

In the case of an efficient measurement, there must be a unique term in the sum over $s$ of Eq.~\eqref{eq:condstate}, i.e. only one of the $M_{r,s}$ is non-zero for each given $r$ (and there is a unique set ${\cal S}^{r}_s$ for each $r$). Keeping this property in mind, we can now examine the conditioned state of system $SA$ given an outcome $r$. The latter has a similar expression as Eq.~\eqref{eq:condstate} without the trace over $A$, i.e.:
\bea  
p_r \rho_{SA\vert_r} &=& \sum_{n} p_{A,n}(t_0)\sum_{i,i'} p_{B,i}(t_0) {}_{B}\langle i'| U_\text{off}\Pi_{r}^AU_\text{on} | ni\rangle_{AB}\; \rho_S(0) \;{}_{AB}\langle ni| U_\text{on}^\dagger\Pi_{r}^AU_\text{off}^\dagger | i'\rangle_{B}\nonumber\\
&=& \sum_{n,n',n''} p_{A,n}(t_0)\sum_{i,i'} p_{B,i}(t_0) {}_{AB}\langle n'i'| U_\text{on}\Pi_{r}^AU_\text{off} | ni\rangle_{AB}\; \rho_S(0) \;{}_{AB}\langle ni| U_\text{on}^\dagger\Pi_{r}^AU_\text{off}^\dagger | n''i'\rangle_{AB}  \;\; |n'\rangle_A\langle n''|\nn\\
\eea
Note that as we did not take the trace over the space of $A$, we have to sum over three indices $n,n',n''$ for the apparatus. Now, gathering has before the system operators which are proportional to each others will lead to terms proportional to $M_{r,s} \rho_S(0) M_{r,s'}$ as $(n,n',i,i')$ and $(n,n'',i,i')$ may belong to two different sets ${\cal S}^{r}_s$ and ${\cal S}^{r}_{s'}$. However, under the assumption that the measurement is efficient, ${\cal S}^{r}_s={\cal S}^{r}_{s'}$ and 
\be 
\forall (n,n',i,i') \in {\cal S}_s^r,\;\; \exists \lambda_{n,n',i,i'}\in \mathbb{C},\;\;{}_{AB}\langle n' i'| U_\text{off}\Pi_{r}^AU_\text{on} | ni\rangle_{AB} = \lambda_{n,n',i,i'} K_{r},
\ee
such that
\bea  
p_r \rho_{SA\vert_r} 
&=& K_r\rho_S(0)K_r^\dagger \,\otimes\,\sum_{n,n',n''}\sum_{i,i'} p_{A,n}(t_0) p_{B,i}(t_0) \lambda_{n,n',i,i'}\lambda^*_{n,n'',i,i''} |n'\rangle_A\langle n''|.
\eea
As this is a factorized state, it verifies $S[\rho_{SA|r}] = S[\rho_{S|r}] + S[\rho_{A|r}]$ and therefore:
\be 
\moy{I_{S:A}} = 0.
\ee

\section{Entropy variation during a weak measurement}

We consider a weak measurement of observable $Q_S$. In the continuous limit of measurement strength going to zero, the action of the measurement operator on the system can be written as a stochastic master equation \cite{Jacobs06}:

\be
\frac{1}{q_r}M_r \rho M_r = \rho + \Gamma_\text{meas}\Delta t {\cal D}[Q_S]\rho + \sqrt{\eta\Gamma_\text{meas}}\Delta W(r) {\cal H}[Q_S]\rho, 
\ee
where $\Gamma_\text{meas}$ is the measurement rate, $\eta \in [0,1]$ the detection efficiency and $\Delta W(r)$ a Wiener increment and we have introduced
\bea 
{\cal D}[X]\rho &=& X\rho X^\dagger-\tfrac{1}{2}(X^\dagger X \rho - \rho X^\dagger X)\\
{\cal H}[X]\rho &=& X\rho+\rho X^\dagger - \text{Tr}\{X\rho+\rho X^\dagger\}.
\eea
We now compute the variation of entropy during the weak measurement. For this we use:
\bea
\log(\rho_0+\epsilon X) &=& \log[\rho_0 (\idop+\epsilon\rho_0^{-1}X)]\nonumber\\
&\simeq& \log(\rho_0)+ \epsilon\rho_0^{-1}X-\tfrac{\epsilon^2}{2}\rho_0^{-1}X\rho_0^{-1}X+{\cal O}(\epsilon^3),
\eea
such that
\be 
-\text{Tr}\{(\rho_0+\epsilon X)\log(\rho_0+\epsilon X) \} +\text{Tr}\{\rho_0\log(\rho_0)\} \simeq -\epsilon(\text{Tr}\{X\} + \text{Tr}\{X\log\rho_0\})-\frac{\epsilon^2}{2}(\text{Tr}\{X^2\rho_0^{-1}\})+{\cal O}(\epsilon^3).
\ee
We then inject $\rho_0=\rho$, $\epsilon X = \Gamma_\text{meas}\Delta t {\cal D}[Q_S]\rho + \sqrt{\eta\Gamma_\text{meas}}\Delta W(r) {\cal H}[Q_S]\rho$, and expand to first order in $\Delta t$ (second order in $\Delta W(r)$, using $\Delta W^2(r)=\Delta t$):
\bea
 S\left[\frac{1}{q_r}M_r \rho M_r\right]
&=& S[\rho]-\sqrt{\eta\Gamma_\text{meas}}\Delta W(r) \text{Tr}\{({\cal H}[Q_S]\rho) \log \rho\}- \Gamma_\text{meas}\Delta t  \text{Tr}\{({\cal L}[Q_S]\rho) \log \rho\}\nonumber\\
&& -\frac{\eta\Gamma_\text{meas}\Delta t}{2}\left[3 \moy{Q_S^2}+\text{Tr}\{Q_S\rho^2Q_S\rho^{-1}\}-4\moy{Q_S}^2\right].
\eea
We now average over the measurement outcome probability distribution. The term proportional to $\Delta W(r)$ cancel out since $\Delta W(r)$ has zero mean, while the other terms are unchanged (they only contain deterministic terms). In addition, we use that $\sum_{r} M_r\rho M_r = \langle M_r\rho M_r/q_r\rangle$ such that:
\be
S\left[\sum_{r} M_r\rho M_r\right] = S[\rho]- \Gamma_\text{meas}\Delta t  \text{Tr}\{({\cal L}[Q_S]\rho) \log \rho\}.
\ee
Finally:
\bea 
\xi_S &=& S\left[\sum_{r} M_r\rho M_r\right]-\sum_{r} q_r S\left[\frac{1}{q_r}M_r\rho M_r\right]\nonumber\\
&=&\frac{\eta\Gamma_\text{meas}\Delta t}{2}\left[3 \moy{Q_S^2}+\text{Tr}\{Q_S\rho^2Q_S\rho^{-1}\}-4\moy{Q_S}^2\right],
\eea
and hence $\xi_S = {\cal O}(\eta\Gamma_\text{meas}\Delta t)$.

\section{Saturating the lower bound on work expenditure}\label{app:saturatingprotocol}
Firstly, we show in Sec. \ref{app:quasisdrive} that the protocol from  Sec. \ref{sec:settings} yields no work expenditure in the quasi-static limit. Then,
in Sec. \ref{app:refined}, we detail how the protocol introduced in Sec. \ref{sec:settings} can be refined to saturate the general lower bound Eq.\eqref{eq:genbound} derived from thermodynamic arguments. Finally, in Sec. \ref{app:finitetime}, we derive the expression of the work expenditure for finite-time protocols.

\subsection{Quasi-static limit}\label{app:quasisdrive}
The expression of the work invested in the overall protocol is given by the work performed by external drives \cite{Esposito10}, 
\be
W_\text{dr} = \int_0^{t_F} dt {\rm Tr}_{SAMB}\left[\rho_{SAMB}(t)\frac{d}{dt}H_{SAMB}(t)\right].
\ee
Since only $V_{SA}(t)$ varies in the switch on and off processes, and the result of the observation of $A$ is encoded in $M$ via a unitary operation on $SAM$ (assumed to occur on a timescale much shorter than the dissipation rate induced by the bath $B$), we have
\bea
W_\text{dr} = \int_0^{t_M^{-}} dt \dot \chi(t) {\rm Tr}_{SA}\left[\rho_{SA}(t)Q_SR_A\right] + \int_{t_M^{-}}^{t_M^{+}} dt {\rm Tr}_{SAM}\left[\rho_{SAM}(t) \dot H_{SAM}(t)\right] + \int_{t_M^{+}}^{t_F} dt \dot \chi(t) {\rm Tr}_{SA}\left[\rho_{SA}(t)Q_SR_A\right].
\eea
The encoding's contribution is
\bea
\int_{t_M^{-}}^{t_M^{+}} dt {\rm Tr}_{SAM}\left[\rho_{SAM}(t) \dot H_{SAM}(t)\right] &=& {\rm Tr}_{SAM}[\rho_{SAM}(t_M^{+})H_{SAM}(t_M^{+})] - {\rm Tr}_{SAM}[\rho_{SAM}(t_M^{-})H_{SAM}(t_M^{-})]  \nn\\
&=&  {\rm Tr}_{SAM}\left[ \sum_r p_r \rho_{SA|r}(t_M)\otimes |r\ket \bra r| H_{SA}(t_M)\right] - {\rm Tr}_{SA}\left[\sum_k p_k(0)\pi_k^S\rho_{A,k}^\text{th}H_{SA}(t_M)\right]\nn\\
&=&  {\rm Tr}_{SA}\left[ \sum_r p_r \rho_{SA|r}(t_M) H_{SA}(t_M)\right] - {\rm Tr}_{SA}\left[\sum_k p_k(0)\pi_k^S\rho_{A,k}^\text{th}H_{SA}(t_M)\right] \nn\\
&=& 0
\eea
since, by construction, the measurement of $A$ is a classical observation, implying that on average it does not affect the state of $SA$, $\sum_r p_r \rho_{SA|r}(t_M) = \rho_{SA}(t_M) = \sum_k p_k(0)\pi_k^S\rho_{A,k}^\text{th}$.

To proceed with the contribution from the switching on and off, we need to use the structure of the dynamics. We recall that during the switching on (off), the dynamics is given by a unitary operation of $SAB$, $U_\text{on}(t) := {\cal T}\exp{\Big[-i {\cal T}\int_{t_0}^{t} du H_{SAB}(u)\Big]}$, $t_0\leq t\leq t_M$, ($U_\text{off} := {\cal T}\exp{\Big[-i {\cal T}\int_{t_M}^{t} du H_{SAB}(u)\Big]}$, $t_M \leq t\leq t_F$) 
with, 
\be
H_{SAB}(t) = H_S + \chi(t)Q_S R_A + H_A +  V_{AB} + H_B.
\ee
Since $[Q_S,H_S]=0$, the global unitary $U_\text{on}$ can be decomposed as 
\be
U_\text{on}(t) = \sum_k \pi_k^S \otimes U_\text{on, k}(t)
\ee
with 
\be
U_\text{on, k}(t) := {\cal T}e^{ -i\int_{t_0}^{t} dt [e_k + \chi(t)q_kR_A + H_A + V_{AB} + H_B]},
\ee
and similarly for the switching off process. This implies that 
\bea
\int_0^{t_M^{-}} dt \dot \chi(t) {\rm Tr}_{SA}\left[\rho_{SA}(t)Q_SR_A\right] &=& \int_0^{t_M^{-}} dt \dot \chi(t) {\rm Tr}_{SAB}\left[U_\text{on}(t)\rho_{S}(0)\otimes\rho_A(0)\otimes\rho_B(0)U_\text{on}^\dag(t)Q_SR_A\right] \nn\\
&=& \int_0^{t_M^{-}} dt \dot \chi(t) \sum_k p_k(0)q_k {\rm Tr}_{A}\left[\rho_{A,k}(t)R_A\right],\label{eqapp:switchon}
\eea
with $\rho_{A,k}(t) := {\rm Tr}_B[U_\text{on,k}(t)\rho_A(0)\otimes\rho_B(0)U_\text{on,k}^{\dag}]$ for $t_0 \leq t \leq t_M$.
Similarly,
\bea
\int_{t_M^{-}}^{t_F} dt \dot \chi(t) {\rm Tr}_{SA}\left[\rho_{SA}(t)Q_SR_A\right] &=& \int_{t_M^{-}}^{t_F} dt \dot \chi(t) {\rm Tr}_{SAB}\left[U_\text{off}(t)\rho_{SAB}(t_M)U_\text{off}^\dag(t)Q_SR_A\right] \nn\\
&=& \int_0^{t_M^{-}} dt \dot \chi(t) \sum_r\sum_k q_k p_k(0)p(r|k) {\rm Tr}_{A}\left[\rho_{A,k|r}(t)R_A\right],\label{eqapp:switchoff}
\eea
with $\rho_{A,k|r}(t) := \frac{1}{p(r|k)} {\rm Tr}_B[U_\text{off,k}(t)\pi_r^A\rho_{A,k}^\text{th}\pi_r^A\otimes\rho_B(0)U_\text{off,k}^{\dag}]$ for $t_M \leq t \leq t_F$.

Then, for quasi-static drive, $A$ is at all times in the instantaneous equilibrium state $\rho_{A,k}^\text{th}(t) := Z_{A,k}^{-1}e^{-\beta H_{A,k}(t)}$ (see Appendix \ref{app:steadystate}), yielding 
\bea
W_\text{dr} &=& \int_0^{t_M^{-}} dt \dot \chi(t) \sum_k p_k(0)q_k {\rm Tr}_{A}\left[\rho_{A,k}^\text{th}(t)R_A\right] + \int_{t_M^{+}}^{t_F} dt \dot \chi(t) \sum_k p_k(0)q_k {\rm Tr}_{A}\left[\rho_{A,k}^\text{th}(t)R_A\right] \nn\\
&=& \int_0^{t_F} dt \dot \chi(t) \sum_k p_k(0)q_k {\rm Tr}_{A}\left[\rho_{A,k}^\text{th}(t)R_A\right].
\eea
Finally, since $\rho_{A,k}^\text{th}(t)$ depends only on the instantaneous value of $\chi(t)$, ${\rm Tr}_{A}\left[\rho_{A,k}^\text{th}(t)R_A\right] \equiv f_k[\chi(t)]$ is a real function of $\chi(t)$, and we can make the following change of variable
\bea
W_\text{dr} =  \sum_k p_k(0)q_k\int_0^{t_F} dt \dot \chi(t) f_k[\chi(t)] =  \sum_k p_k(0)q_k\int_{\chi(t_0)}^{\chi(t_F)} d\chi   f_k(\chi) = 0,
\eea
since $\chi(t_0) = \chi(t_F) = 0$. We therefore recover the result announced in the main text, namely that the protocol introduced in Sec. \ref{sec:settings} leads to no work expenditure in the quasi-static limit.

\subsection{Reversible protocol}\label{app:refined}
\begin{figure}
    \centering
\includegraphics[width=1.05\textwidth]{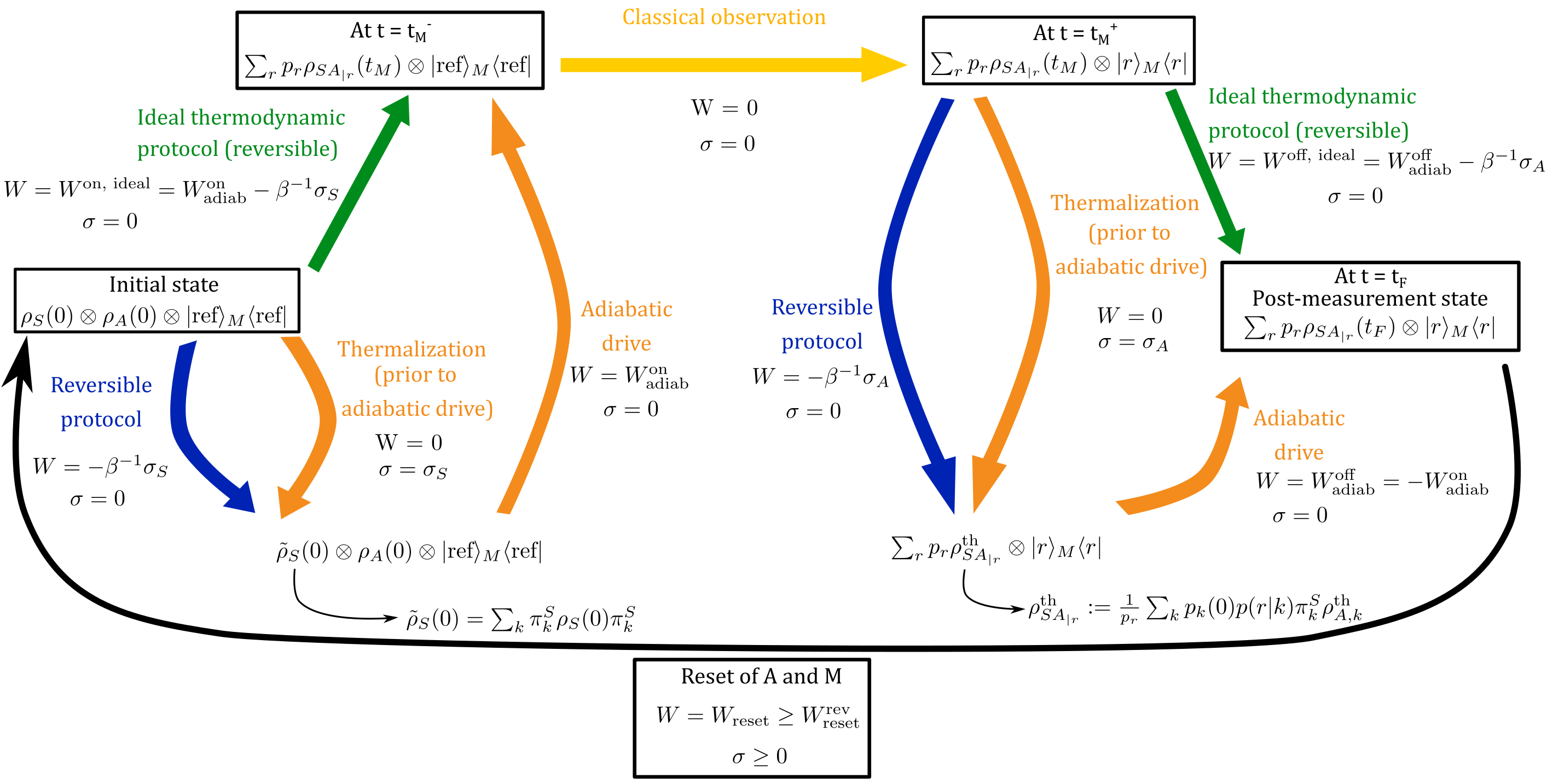}
\caption{Schematic representation of the reversible measurement protocol.}
    \label{fig:revprot}
\end{figure}

As commented in the main text, operating the driving quasi-statically is not enough to make the whole measurement process reversible because of the entropy produced in the system and in the apparatus. Here we detail how additional steps can be included to exploit the information gained during the measurement to extract additional work and reach the bound (that is, how to make the whole protocol reversible).  The graph in Fig. \ref{fig:revprot} summarizes the main steps detailed in the following.

To start with, the first source of irreversibility is introduced when the switching on process is performed quasi-statically. While $A$ is assumed to be initially in equilibrium, $S$ is in an arbitrary initial state, and in particular can contain initial coherences (in the eigenenergy basis of $H_S$). As soon as the interaction between $S$ and $A$ is switched on, with $\chi(t)$ much smaller than all energy transition of $S$ and $A$, $SA$ equilibrates with the bath due to the quasi-static nature of the protocol, reaching an equilibrium state (see Appendix \ref{app:steadystate}) infinitesimally close to $\sum_k \pi_k^S \rho_S(0)\pi_k^S\otimes\rho_A(0)$. The irreversible dissipation of coherences in the bath leads to an entropy production given by
\bea
\sigma_S &:=& S\left[\sum_k p_k(0)\pi_k^S\rho_A(0)\right] - S[\rho_{SA}(0)] - \beta {\rm Tr}\left[\sum_k p_k(0)\pi_k^S\rho_A(0) H_{SA}(t)\right] + \beta {\rm Tr}[\rho_{SA}(0)H_{SA}(0)]\nn\\
&=& {\cal H}(\{p_k(0)\}) - S[\rho_S(0)],
\eea
where the second line is taken in the limit of $\chi(t)$ going to zero so that the energy difference on the right-hand side is negligible, and ${\cal H}(\{p_k(0)\})$ stands for the Shanonn entropy associated with the distribution $\{p_k(0)\}_k$. Alternatively to the quasi-static switching on of $\chi(t)$, one could perform a protocol \cite{Kammerlander16} to extract an amount of work  equal to 
\be\label{eq:wini}
W_\text{ini} = \frac{1}{\beta}\left\{S\left[\sum_k p_k(0)\pi_k^S\right] - S[\rho_{S}(0)]\right\} = \frac{1}{\beta} \{H[p_k(0)] - S[\rho_{S}(0)]\} = \frac{1}{\beta} \sigma_S,
\ee
from the initial coherences potentially contained in $\rho_S(0)$. One should note that such protocol requires the previous knowledge of $\rho_S(0)$, which is not the case in typical situations of measurement.

Additionally, there is a second source of entropy production (irreversibility), which stems from the thermalisation of $SA$ just after the observation of an outcome $r$. More precisely, for an observation $r$, we know that just after the observation, $SA$ is in the state 
\be\label{eq:A11}
\rho_{SA|r}(t_M^+) = \frac{1}{p_r} \sum_k p_k(0) \pi_k^S \Pi_r^A \rho_{A,k}^\text{th}\Pi_r^A  =\frac{1}{p_r} \sum_k p_k(0)p(r|k) \pi_k^S \rho_{A,k|r}(t_M),
\ee
which is a non-equilibrium state. Then, the switching off protocol of \ref{sec:settings} operated in a quasi-static way leads to the equilibration of $SA$ (before $\chi_M$ changes significantly), namely to the state 
\be\label{eq:A13}
\rho_{SA|r}^\text{th} :=\frac{1}{p_r} \sum_k p_k(0) p(r|k)\pi_k^S \rho_{A,k}^\text{th},
\ee
which is the equilibrium state of $SA$ associated with the initial state $\rho_{SA|r}(t_M^+)$.
The entropy production associated to this dissipative evolution is 
\bea
\sigma_{A,r} &:=& \Delta S_{SA,r} - \beta Q_r = \Delta S_{SA,r} - \beta \Delta E_{SA, r} \nn\\
&=& S[\rho_{SA|r}^\text{th}] - S[\rho_{SA|r}(t_M^+)] - \beta {\rm Tr}\{[\rho_{SA,r}^\text{th} - \rho_{SA|r}(t_M^+)](H_{S} + \chi_MQ_SR_A +H_A)\}\nn\\
&=& S[\rho_{SA|r}^\text{th}] - S[\rho_{SA|r}(t_M^+)] + \frac{1}{p_r}\sum_k p_k(0)p(r|k){\rm Tr}\{[\rho_{A,k}^\text{th} - \rho_{A,k|r}(t_M^+)]\ln \rho_{A,k}^\text{th}\}\nn\\
&=& S[\rho_{SA|r}^\text{th}] - S[\rho_{SA|r}(t_M^+)] + {\rm Tr}\{[\rho_{SA|r}^\text{th} - \rho_{SA|r}(t_M^+)]\ln \rho_{SA|r}^\text{th}\}\nn\\
&=& D[\rho_{SA,r}(t_M^+)|\rho_{SA|r}^\text{th}].
\eea
where $D[\sigma|\rho]:= {\rm Tr}[\sigma (\ln \sigma -\rho)]$ is the relative entropy.

One can then design a specific protocol to extract work from the non-thermal features present in $\rho_{SA|r}(t_M)$. One such possible protocol is a reversible procedure consisting in a quench followed by a quasi-reversible driving, as presented in \cite{Elouard23, Bera19}. The amount of extracted work is  
\bea
W_\text{non-eq,r} &=& \beta^{-1}D[\rho_{SA|r}(t_M^+)|\rho_{SA|r}^\text{th}]\nn\\
&=& \beta^{-1} \sigma_{A,r}.
\eea 
One can show that the average extracted work $\beta\sum_r p_r W_\text{non-eq,r} = \sum_r p_r \sigma_{A,r} := \sigma_A$ is equal to $H(p_r) - I[p_k(0):p_r]$. Here, using the expression Eqs. \eqref{eq:A11} and \eqref{eq:A13} we briefly present the main steps of this derivation:

\bea
\sigma_A &:=&\beta\sum_r p_r W_\text{non-eq,r}\nn\\
&=& \sum_r p_r D[\rho_{SA|r}(t_M^+)|\rho_{SA|r}^\text{th}] \nn\\
&=& - \sum_r p_r S[\rho_{SA|r}(t_M^+)] - {\rm Tr}\left[ \sum_r\sum_k p_k(0)\pi_k^S\Pi_r^A\rho_{A,k}^\text{th}\Pi_r^A\ln\left(\sum_{k'} \frac{p_{k'}(0)p(r|k')}{p_r}\pi_{k'}^S\rho_{A,k'}^\text{th}\right) \right]\nn\\
&=& - \sum_r p_r S[\rho_{SA|r}(t_M^+)] - \sum_r\sum_k p_k(0){\rm Tr}\left[ \pi_k^S\Pi_r^A\rho_{A,k}^\text{th}\Pi_r^A\ln\left(\sum_{k'} \frac{p_{k'}(0)p(r|k')}{p_r}\pi_{k'}^S\sum_{r'}\Pi_{r'}^A\rho_{A,k'}^\text{th}\Pi_{r'}^A\right) \right]\nn\\
&=& - \sum_r p_r S[\rho_{SA|r}(t_M^+)] - \sum_r \sum_k p_k(0){\rm Tr}\left[\pi_k^S\Pi_r^A\rho_{A,k}^\text{th}\Pi_r^A\ln\left( \frac{p_k(0)p(r|k)}{p_r}\pi_k^S\Pi_r^A\rho_{A,k}^\text{th}\Pi_r^A\right) \right]\nn\\
&=& - \sum_r p_r S[\rho_{SA|r}(t_M^+)] - \sum_r \sum_k p_k(0)p(r|k){\rm Tr}\left[\pi_k^S\rho_{A,k|r}(t_M)\ln\left( \frac{p_k(0)p^2(r|k)}{p_r}\pi_k^S\rho_{A,k|r}(t_M)\right) \right]\nn\\
&=& - \sum_r p_r S[\rho_{SA|r}(t_M^+)] - \sum_r \sum_k p_k(0)p(r|k) \left\{ \ln\frac{p_k(0)p^2(r|k)}{p_r}   - S[\pi_k^S\rho_{A,k|r}(t_M)]\right\}\nn\\
&=& \sum_r p_r \sum_k \frac{p_k(0)p(r|k)}{p_r}\ln \frac{p_k(0)p(r|k)}{p_r} - \sum_r \sum_k p_k(0)p(r|k)  \ln\frac{p_k(0)p^2(r|k)}{p_r} \nn\\
&=& - \sum_r\sum_k p_k(0)p(r|k)\ln p(r|k)\nn\\
&=& {\cal H}(\{p_{k,r}\}) -  {\cal H}(\{p_{k}(0)\}),
\eea
where we used in the third line the property $\rho_{A,k}^\text{th} = \sum_p \pi_r^A \rho_{A,k}^\text{th}\pi_r^A$ and in the seventh line $S[\rho_{SA|r}(t_M^+)] = - \sum_k \frac{p_k(0)p(r|k)}{p_r} \ln \frac{p_k(0)p(r|k)}{p_r} + \sum_k \frac{p_k(0)p(r|k)}{p_r}S[\pi_k^S\rho_{A,k|r}(t_M)]$.

Finally, taking into account the two above refinements, allowing one to avoid the two sources of irreversibility, the work expenditure becomes
\bea
W_\text{dr}^\text{refined} : = W_\text{dr} - W_\text{ini} - \sum_r p_r W_\text{non-eq,r} &=& W_\text{dr} - \beta^{-1}\sigma_S -\beta^{-1}\sigma_A \nn\\
&=& \frac{1}{\beta}\left[ S[\rho_S(0)] - H[p_k(0)] - H(p_r) + I[p_k(0):p_r] \right]\nn\\
&=& \frac{1}{\beta}[J-H(p_r)],
\eea
which, together with $W_\text{reset}$, saturates the lower bound \eqref{eq:genbound}, as mentioned in the main text.

\subsection{Finite-time drive}\label{app:finitetime}
For finite-time drive, the difference from Sec. \ref{app:quasisdrive} is that the contributions to the work expenditure, Eqs. \eqref{eqapp:switchon} and \eqref{eqapp:switchoff}, have to be computed by solving the dynamics of $A$. More precisely, based on the beginning of Sec. \ref{app:quasisdrive}, the work expenditure is 
\be
W_\text{dr} = \int_0^{t_M^{-}} dt \dot \chi(t) \sum_k p_k(0)q_k {\rm Tr}_{A}\left[\rho_{A,k}(t)R_A\right] + \int_{t_M^{-}}^{t_F} dt \dot \chi(t) \sum_r\sum_k q_k p_k(0)p(r|k) {\rm Tr}_{A}\left[\rho_{A,k|r}(t)R_A\right],
\ee
with 
\be
\rho_{A,k}(t) := {\rm Tr}_{SB}[U_\text{on}(t)\pi_k^S\otimes\rho_A(0)\otimes\rho_B(0)U_\text{on}^{\dag}] = {\rm Tr}_B[U_\text{on,k}(t)\rho_A(0)\otimes\rho_B(0)U_\text{on,k}^{\dag}],
\ee
for $t_0 \leq t \leq t_M$ and
\be
\rho_{A,k|r}(t) := \frac{1}{p(r|k)} {\rm Tr}_B[U_\text{off}(t)\pi_k^S\otimes\pi_r^A\rho_{A,k}^\text{th}\pi_r^A\otimes\rho_B(0)U_\text{off}^{\dag}]=  \frac{1}{p(r|k)} {\rm Tr}_B[U_\text{off,k}(t)\pi_r^A\rho_{A,k}^\text{th}\pi_r^A\otimes\rho_B(0)U_\text{off,k}^{\dag}]
\ee
for $t_M \leq t \leq t_F$.
 In the limit of weak coupling between the apparatus and the bath $B$, the reduced dynamics of $A$ can be described by a master equation depending on the state of $S$ (see Appendix \ref{app:reduceddynA}),   
\be
\frac{d}{dt}\rho_{A,k}(t) = {\cal L}_k(t)\rho_{A,k}(t),
\ee
with
\bea\label{Lk}
  {\cal L}_{k}(t) \sigma_A = -i[H_{A,k}+ H_{LS}^{k},\sigma_A]  + 
  \sum_{l,l'\in {\cal E}_k}\gamma_{\nu_l^t,\nu_{l'}^{t}}\left[a_{l} \sigma_{A} a_{l'}^{\dag} -\frac{1}{2}\{a_{l'}^{\dag}a_{l},\sigma_{A}\}\right],
\eea
 where 
\bea
\nu_l^t &=&   \nu_{k,n,m}^t = E_{m}-E_n + \chi(t)q_k(r_m - r_n),\nn\\
  l &\equiv& (k,n,m) \nn\\
a_l &=& a_{n,m} := \pi_n^A A \pi_{m}^A, \quad\text{(indep. of k),}\nn\\
 \gamma_{\nu_l^t,\nu_{l'}^{t}} &=&\Gamma(\nu_l^t) + \Gamma^{ *}(\nu_{l'}^{t}),\nn\\
 \gamma(\nu_l^t) &=& \gamma_{\nu_l^t,\nu_l^t}=  2{\rm Re}[\Gamma(\nu_l^t)],\nn\\
 S(\nu_l^t,\nu_{l'}^t) &=& \frac{\Gamma(\nu_l^t)-\Gamma^{ *}(\nu_{l'}^t)}{2i}\nn\\
  \Gamma(\nu_l^t) &=&\int_0^\infty d\tau {\rm Tr}[\rho_B B(\tau)B] e^{i\nu_l^t\tau},\nn\\
  H_{LS}^{k} &=& \sum_{l,l' \in {\cal E}_k}S(\nu_l^t,\nu_{l'}^t) a_{l'}^{\dag}a_{l}
  \eea
and ${\cal E}_k$ is the ensemble of triple indices $(k',n,m)$ with fixed $k'=k$. The derivation of the master equation is detailed in the following section \ref{app:MEderivation} .

In order to compute the dynamics of $A$, we now specify the systems $S$ and $A$ using an example taken from standard experimental setups consisting in a qubit measured by an electromagnetic cavity mode of frequency $\omega_c$ (see for instance \cite{Linpeng22}).  The corresponding total Hamiltonian for a measurement of $\sigma_z$ can be described by
\be\label{Hexample}
H_{\rm tot} = \frac{\omega_S}{2}\sigma_z + \chi(t)\sigma_z a^{\dag}a + \omega_c a^{\dag}a + (a^{\dag} + a) B + H_B,
\ee
where $a$ and $a^{\dag}$ are the bosonic creation and annihilation operators. In particular, the steady state when $\chi(t)$ reaches the final value $\chi_M$ is (see Appendix \ref{app:steadystate}),
\be
\rho_{SA}^\text{th}(t_M^{-}) = p_e(0)\pi_e \rho_{A,e}^{\rm th} +  p_g(0)\pi_g \rho_{A,g}^{\rm th},
\ee
where $p_e(0)$, $p_g(0)$ are the initial excited and ground state populations, $\rho_{A,e}^{\rm th}$ and $\rho_{A,g}^{\rm th}$ are thermal state  given by \eqref{appssak}, associated with the Hamiltonians $H_{A,e} = (\omega_c + \chi_M)a^{\dag} a$ and $H_{A,g} = (\omega_c - \chi_M)a^{\dag} a$, respectively. 

Thus, in order to compute $W$, we have to derive the dynamics of $\bra a^{\dag}a\ket_k  := {\rm Tr}[\rho_{A,k}(t) a^\dag a]$ and $\bra a^{\dag}a\ket_{k|r}  := {\rm Tr}[\rho_{A,k|r}(t) a^\dag a]$, $k= e,g$. According to Eq. \eqref{Lk}, we have,
\be
\frac{d}{dt}\bra a^{\dag}a\ket_{k=e}  = - [\gamma(\nu_e^t) - \gamma(-\nu_e^t)]\bra a^{\dag}a\ket_e +\gamma(-\nu_e^t) + 2i S(\nu_e^t,-\nu_e^t) \bra a^2\ket_e -2i S(-\nu_e^t,\nu_e^t)\bra a^{\dag 2}\ket_e
\ee
and
\bea
\frac{d}{dt}\bra a^2\ket_{k=e}  &=& -[\gamma(\nu_e^t)-\gamma(-\nu_e^t) + 2i\nu_e^t +2i S(\nu_e^t,\nu_e^t) + 2iS(-\nu_e^t,-\nu_e^t)] \bra a^2\ket_e \nn\\
&&- 4iS(-\nu_e^t,\nu_e^t)\bra a^{\dag}a\ket_e -2i S(-\nu_e^t,\nu_e^t) - \gamma_{-\nu_e^t,\nu_e^t}
\eea
 Due to the time dependence of the frequency $\nu_e^t := \omega_c +  \chi(t)$, all coefficients are time dependent and there is no general analytical solutions for these coupled equations. We therefore integrate formally, inject one expression into the other, and after some manipulations (see detail in Appendix \ref{Appdetailn}), we arrive at
%
%
\bea\label{eq:A27}
\bra a^{\dag}a \ket_e(t) &=& n_{\nu_e^t}  -\int_0^t du e^{-\int_u^t ds \kappa_e}\frac{\partial n_{\nu_e^u}}{\partial u}+ e^{-\int_0^t du \kappa_e}[\bra a^{\dag}a\ket_e (0)-n_{\nu_e^0}] + T^{(2)}(t),
\eea
with $\kappa_k := \gamma(\nu_k^t) - \gamma(-\nu_k^t) = 2\pi J(\nu_k^t)$, $k=e,g$.
 The first term corresponds to the adiabatic contribution, $n_{\nu_k^t}=(e^{\beta\nu_k^t}-1)^{-1}$, the second term is a non-adiabatic correction (see more detail below), 
 the third term is an initial non-equilibrium contribution, and the last term $T^{(2)}(t)$ contains higher order corrections.
 For the switching on process, the initial state of $A$ is the thermal equilibrium state at the bath temperature so that $\bra a^{\dag}a\ket_e (0) = n_{\nu_e^0} =  n_{\omega_c}$, implying, 
 \bea
\bra a^{\dag}a \ket_e(t) &=& n_{\nu_e^t}-\int_0^t du e^{-\int_u^t ds \kappa_e}\frac{\partial n_{\nu_e^t}}{\partial u} + T^{(2)}(t).
\eea

For the switching off process associated with the observation $r$, the dynamics is the same, but the initial state is $\rho_{A,k|r}(t_M)=\frac{1}{p(r|k)}\pi_r^A \rho_{A,k}^\text{th}\pi_r^A$. We therefore obtain, for $t \in [t_M; t_F]$,
\bea
\bra a^{\dag}a \ket_{e|r}(t)=  n_{\nu_e^t}  -\int_{t_M}^t du e^{-\int_u^t ds \kappa_e}\frac{\partial n_{\nu_e^t}}{\partial u}+ e^{-\int_{t_M}^t du \kappa_e}\left[{\rm Tr}[ \rho_{A,e|r}(t_M)a^\dag a] -n_{\nu_k^{t_M}}\right] + T^{(2)}_{|r}(t).
\eea

The expressions of $\bra a^{\dag}a \ket_{k=g}(t)$ and $\bra a^{\dag}a \ket_{k=g|r}(t)$ 
can be obtained in the same way, resulting in similar expressions, namely, 
\bea
\bra a^{\dag}a \ket_{k=g}(t) &=& n_{\nu_g^t}-\int_0^t du e^{-\int_u^t ds \kappa_g}\frac{\partial n_{\nu_g^u}}{\partial u} + T^{(2)}(t),
\eea
with $\nu_g^t := \omega_c - \chi(t)$,
and 
\bea
\bra a^{\dag}a \ket_{k=g|r}(t)=  n_{\nu_g^t}  -\int_{t_M}^t du e^{-\int_u^t ds \kappa_g}\frac{\partial n_{\nu_g^u}}{\partial u}+ e^{-\int_{t_M}^t du \kappa_g}\left[{\rm Tr}[\rho_{A,g|r}(t_M) a^\dag a] -n_{\nu_g^{t_M}}\right]+ T^{(2)}_{|r}(t).
\eea
Altogether we obtain,
%
\bea\label{eqapp:work}
W_\text{dr} &=& \int_0^{t_M^{-}} dt \dot \chi(t) \sum_k p_k(0)q_k n_{\nu_k^t} + \int_{t_M^{-}}^{t_F} dt \dot \chi(t) \sum_r\sum_k q_k p_k(0)p(r|k)n_{\nu_k^t}\nn\\
&-& \int_0^{t_M^{-}} dt \dot \chi(t) \sum_k p_k(0)q_k \int_0^t du e^{-\int_u^t ds \kappa_k}\frac{\partial n_{\nu_k^u}}{\partial u} - \int_{t_M^{-}}^{t_F} dt \dot \chi(t) \sum_r\sum_k q_k p_k(0)p(r|k)\int_{t_M}^t du e^{-\int_u^t ds \kappa_k}\frac{\partial n_{\nu_k^u}}{\partial u}\nn\\
&+&\int_{t_M^{-}}^{t_F} dt \dot \chi(t) \sum_r\sum_k q_k p_k(0)p(r|k)e^{-\int_{t_M}^t du \kappa_k}\left[{\rm Tr}[\rho_{A,k|r}(t_M) a^\dag a] -n_{\nu_k^{t_M}}\right]\nn\\\label{eqapp:lastline}
&+& \int_0^{t_M} dt \dot \chi(t) T^{(2)}(t) + \int_{t_M}^{t_F} dt \dot \chi(t) T^{(2)}_{|r}(t) 
\eea
The first line corresponds to the adiabatic contributions and cancels out on average, as already discussed in Sec. \ref{app:quasisdrive}. The third line is equal to,
\be
\int_{t_M^{-}}^{t_F} dt \dot \chi(t) \sum_k q_k p_k(0)e^{-\int_{t_M}^t du \kappa_k}\left[{\rm Tr}[\sum_r\pi_r^A\rho_{A,k}^\text{th}\pi_r^A a^\dag a] -n_{\nu_k^{t_M}}\right] = 0,\nn\\
\ee
since ${\rm Tr}[\sum_r\pi_r^A\rho_{A,k}^\text{th}\pi_r^A a^\dag a] = {\rm Tr}[\rho_{A,k}^\text{th}a^\dag a]= n_{\nu_k^{t_M}}$.
The last line of \eqref{eqapp:lastline} corresponds to terms of order order ${\cal O}(g^2_k)$ and will be neglected in the following.
Note that although we do not consider explicitly the higher order contributions, the above expression already provides a relevant estimation of the required work and furthermore is insightful regarding the trade-offs at play emerging from finite-time protocols. Indeed, the second line gives always positive contributions, and grow with the velocity of the driving, represented by the time derivative $\partial n_{\nu_k^u}/\partial u$. This yields,
\bea
W_\text{dr} &=& - \sum_{k=e,g} p_k(0)q_k \left[\int_0^{t_M^{-}} dt \dot \chi(t) \int_0^t du e^{-\int_u^t ds \kappa_k}\frac{\partial n_{\nu_k^u}}{\partial u} + \int_{t_M^{-}}^{t_F} dt \dot \chi(t) \int_{t_M}^t du e^{-\int_u^t ds \kappa_k}\frac{\partial n_{\nu_k^u}}{\partial u}\right].
\eea

For a linear ramp $\chi(t) = R t$, for $t \in [0;t_M]$, $R = \chi_M/t_M$, and $\chi(t) =\chi_M - R (t-t_M)$, for $t \in [t_M;t_F]$, $t_F=2t_M$, one obtains (see detail in Appendix \ref{applinearramp}), 
\be\label{Willustration}
W_\text{dr} = \sum_{k=e,g} p_k(0)\frac{ R^2\beta}{\kappa}\Bigg\{ \int_0^{t_M}du\frac{e^{(\omega_c+q_kRu)\beta}}{(e^{(\omega_c+q_kRu)\beta}-1)^2} \left[2-e^{-\kappa(t_M-u)} -e^{-\kappa u}\right]\Bigg\} .  
\ee
remembering that $q_e=1$ and $q_g=-1$, and assuming that the bath spectral density $J(\omega) \simeq \kappa/2\pi$ is approximately constant over the range $\omega \in [\omega_c-\chi_M;\omega_c+\chi_M]$.
Interestingly, one can rewrite expression \eqref{Willustration} in the following form, 
\bea\label{appeq:finalW}
W_\text{dr} &=&  \sum_{k=e,g} p_k(0)(R/\kappa)^2\beta\Bigg\{ \int_0^{\chi_M\kappa/R}du\frac{e^{(\omega_c+q_k(R/\kappa)u)\beta}}{(e^{(\omega_c+q_k(R/\kappa)u)\beta}-1)^2} \left[2-e^{u- \chi_M\kappa/R} -e^{-u}\right]\Bigg\},
\eea
making explicit that multiplying the ramp $R$ and the damping rate $\kappa$ by the same factor does not change the average work. One can verify that Eq.\eqref{appeq:finalW} tends to zero when $R/\kappa $ tends to zero, recovering the adiabatic limit of Section \ref{app:quasisdrive}. 
This emphasizes that what matters is the slope of the ramp (or more generally the ``velocity'' of the driving) with respect to the equilibration velocity of $SA$. This is indeed the core idea of adiabaticity (in the classical sense of quasi-static evolution): what matters is the driving timescale compared to the equilibration timescale.     \\

The quantities plotted in Fig.~\ref{fig:3} corresponds to the above expression $W_\text{dr}$ when the qubits $S$ is initially in the excited state and ground state respectively, namely
\be
W_\text{dr}^+ =   p_e(0)(R/\kappa)^2\beta\Bigg\{ \int_0^{\chi_M\kappa/R}du\frac{e^{(\omega_c+(R/\kappa)u)\beta}}{(e^{(\omega_c+(R/\kappa)u)\beta}-1)^2} \left[2-e^{u- \chi_M\kappa/R} -e^{-u}\right]\Bigg\},
\ee
and 
\be
W_\text{dr}^- =   p_g(0)(R/\kappa)^2\beta\Bigg\{ \int_0^{\chi_M\kappa/R}du\frac{e^{(\omega_c-(R/\kappa)u)\beta}}{(e^{(\omega_c-(R/\kappa)u)\beta}-1)^2} \left[2-e^{u- \chi_M\kappa/R} -e^{-u}\right]\Bigg\}.
\ee

\section{Derivation of the master equation with time dependent Bohr frequencies}\label{app:MEderivation}
In this section we detail the derivation of the master equation describing the reduced dynamics of $SA$ in contact with the bosonic thermal bath $B$ while switching on and off the coupling $\chi(t)$. We remind that the Hamiltonian of $SA$ is chosen to be of the form
\bea
H_{SA}(t) = H_S + \chi(t) Q_S R_A + H_A &=& \sum_{k,n} (e_k + \chi(t) q_k r_n + E_n)\pi_k^S\pi_n^A,
\eea
considering $[Q_S,H_S]=0$ and we recall the notations $\pi_k^S := |e_k\ket\bra e_k|$, $\pi_n^A:=|n\ket\bra n|$, with $|e_k\ket$, $|n\ket$ are respectively the eigenstates of $H_S$ and $H_A$, associated to the eigenvalues $e_k$ and $E_n$, and $q_k$ and $r_n$ are the eigenvalues of $Q_S$ and $R_A$, respectively. Note that $H_{SA}(t)$ can be expressed as
\be\label{1stfacto}
H_{SA}(t) =\sum_k \pi_k^S (e_k + H_{A,k}),
\ee
with $H_{A,k} = \sum_n E_{n,k}\pi_n^A$ and $E_{n,k}:= E_n + \chi(t) q_k r_n$.
 Then, the free unitary evolution of $SA$ (when $V_{AB}$ is switched off) is 
\be
U_{SA}(t) = e^{-i{\cal T}\int_0^t du H_{SA}(u) }= \sum_{k,n} e^{-i (e_k + \bar\chi_t q_k r_n + E_n)t} \pi_k^S\pi_n^A,
\ee
with $\bar \chi_t = \frac{1}{t}\int_0^t du \chi(u)$. The coupling between $A$ and $B$ is considered to be of the standard form $V_{AB} =  AB$, where $A$ is an arbitrary operator of $A$, and $B=\sum_k g_k (b_k^{\dag} + b_k)$ is the usual bosonic bath operator. The jump operators that will appear in the master equation can be obtained from the interaction picture of the coupling operator $A$,
\be
U_{SA}^{\dag}(t) A U_{SA}(t) = \sum_{l} e^{-i\omega_l^t t} A_{l}.
\ee
with the collective index $l\equiv (k,n,m)$, $A_l = A_{k,n,m} := \pi_k^S  \pi_n^A A \pi_{m}^A$, and
\bea
\omega_l^t =\omega_{k,n,m}^t:=\frac{1}{t}\int_0^t du [E_{m,k}(u) - E_{n,k}(u)] = E_m-E_n + \bar \chi_t q_k (r_m-r_n),
\eea
 Note that we added a superscript $t$ to highlight the fact that the Bohr frequencies are time dependent since $\bar \chi_t$ is time dependent. However, the associated jump operator $A_{l}$ is time-independent because the energy eigenstates of $H_{SA}(t)$ are time-independent; only the energy transition $\omega_l^t$ is time dependent.

Then, following a standard derivation of master equation without coarse-graining (see for instance \cite{BreuerPetruccione}), we arrive at the following reduced dynamics for $SA$
\bea
\dot \rho_{SA} &=& \int_0^{t} d\tau  c_B(\tau) \sum_{l,l'} e^{  i \omega_{l'}^t t - i\omega_l^{t-\tau}(t-\tau)}\left(A_{l}\rho_{SA}A^{\dag}_{l'} - A^{\dag}_{l'}A_{l}\rho_{SA}\right) + {\rm h.c}\nn\\
\eea
where we introduced the bath correlation function  $c_B(\tau) := {\rm Tr}_B[\rho_BB(\tau)B]$. We re-write $\omega_l ^{t-\tau}(t-\tau)$, as
\bea
\omega_l^{t-\tau}(t-\tau) &=&  (E_{m}-E_n)(t-\tau) + \bar\chi_{t-\tau}(t-\tau)q_k(r_{m}-r_n) \nn\\
&=&  (E_{m}-E_n)(t-\tau) +  q_k(r_{m}-r_n) \int_0^{t-\tau} du \chi(u)\nn\\
&=&  (E_{m}-E_n)(t-\tau) +  q_k(r_{m}-r_n) \left(\frac{1}{t}\int_0^{t}du \chi(u)\right)t  -  q_k(r_{m}-r_n) \left(\frac{1}{\tau}\int_{t-\tau}^{t}du \chi(u)\right)\tau \nn\\
&=&  (E_{m}-E_n)(t-\tau) +  q_k(r_{m}-r_n) \bar\chi_t t  -  q_k(r_{m}-r_n) \left(\frac{1}{\tau}\int_{t-\tau}^{t}du [\chi(t) + (u-t)\dot\chi(t) + \frac{1}{2}(u-t)^2\ddot \chi(t) + ...]\right)\tau \nn\\
&=&  (E_{m}-E_n)(t-\tau) +  q_k(r_{m}-r_n) \bar\chi_t t  -  q_k(r_{m}-r_n) \left(\chi(t) - \frac{\tau}{2}\dot\chi(t) + \frac{\tau^2}{6}\ddot \chi(t) + ...\right)\tau \nn\\
&=&  [E_{m}-E_n +\bar\chi_t q_k(r_{m}-r_n) ]t  - [E_{m}-E_n +\chi(t) q_k(r_{m}-r_n)]\tau - q_k(r_{m}-r_n) \left( - \frac{\tau}{2}\dot\chi(t) + \frac{\tau^2}{6}\ddot \chi(t) + ...\right)\tau \nn\\
&\simeq&  [E_{m}-E_n +\bar\chi_t q_k(r_{m}-r_n) ]t  - [E_{m}-E_n +\chi(t) q_k(r_{m}-r_n)]\tau 
\eea
assuming that $\chi(u)$ does not change significantly from $t -\tau$ to $t$ (more precisely, that $ - \frac{\tau}{2}\dot\chi(t) + \frac{\tau^2}{6}\ddot \chi(t) + ...$ is much smaller that $\chi(t)$). Note that  thanks to the bath correlation function $c_B(\tau)$, this time interval $[t-\tau,t]$ is at most of the length of the bath correlation time $\tau_c$.  Thus, our approximation is valid as soon as $\chi(t)$ does not change significantly over a time interval of the order of the bath correlation time. Crucially, in the first term, it is the time average $\bar\chi_t$ which appears, while in the second term it is the instantaneous value $\chi(t)$. To highlight this important difference we denote 
 \be
 \nu_l^t = \nu_{k,n,m}^t :=  E_{m,k}(t) - E_{n,k}(t) = E_{m}-E_n +\chi(t) q_k(r_{m}-r_n),
 \ee
 while we recall $\omega_l^t = E_{m}-E_n +\bar\chi_t q_k(r_{m}-r_n) $.
Thus, according to the above derivation we have the following identity $\omega_l^{t-\tau} (t-\tau) =  \omega_l^t t - \nu_l^t \tau$, as long as $\chi(t)$ does not change significantly during the bath correlation time.
Then, with the above identity and the updated notations, we can re-write the above master equation as 
\bea
\dot \rho_{SA} &=&  \sum_{l,l'} e^{ i (\omega_{l'}^{t} -\omega_l^{t})t}\left(A_{l}\rho_{SA}A^{\dag}_{l'} - A^{\dag}_{l'}A_{l}\rho_{SA}\right)\int_0^{t} d\tau  c_B(\tau)e^{i\nu_l^t \tau} + {\rm h.c}\nn\\
&=&  \sum_{l,l'}\Gamma_t(\nu_l^t) e^{  i (\omega_{l'}^t - \omega_l^t)t}\left(A_{l}\rho_{SA}A^{\dag}_{l'} - A^{\dag}_{l'}A_{l}\rho_{SA}\right) + {\rm h.c}\nn\\
\eea
with $\Gamma_t(\nu_l^t) := \int_0^{t} d\tau  c_B(\tau)e^{i\nu_l^t \tau}$. Going back to the Schrodinger picture we have,
\bea
\dot \rho_{SA} &=&  -i[H_{SA}(t),\rho_{SA}] + \sum_{l,l'} \Gamma_t(\nu_l^t)\left(A_{l}\rho_{SA}A^{\dag}_{l'} - A^{\dag}_{l'}A_{l}\rho_{SA}\right) + {\rm h.c}.
\eea 
 Secondly, for time larger than the bath correlation time, and in particular when we are interested in the steady state, we can safely substitute $\Gamma_t(\nu_l^{t})$ by
\be
\Gamma(\nu_l^{t}) : = \int_0^{ \infty} d\tau c_B(\tau) e^{ i\nu_l^{t}\tau}.
\ee
Thus, according to this derivation, the only approximation beyond the standard Born and Markov ones is that $\chi(t)$ change on a timescale much larger than the bath correlation time $\tau_c$.

Finally, the above master equation can be recast into the standard form
\bea\label{eqapp:genme}
\dot \rho_{SA} = {\cal L} \rho_{SA} := -i[H_{SA}(t)+H_{LS}(t),\rho_{SA}] + \sum_{l,l'}\gamma_{\nu_l^t,\nu_{l'}^{t}}\left[A_{l} \rho_{SA} A_{l'}^{\dag} -\frac{1}{2}\{A_{l'}^{\dag}A_{l},\rho_{SA}\}\right]
\eea
with
\be
H_{LS}(t)= \sum_{l,l'}S(\nu_l^t,\nu_{l'}^{t}) A_{l'}^{\dag}A_{l},
\ee
and where
\bea
 \gamma_{\nu_l^t,\nu_{l'}^{t}} &=&\Gamma(\nu_l^t) + \Gamma^{ *}(\nu_{l'}^{t}),\nn\\
 \gamma(\nu_l^t) &=& \gamma_{\nu_l^t,\nu_l^t}=  2{\rm Re}[\Gamma(\nu_l^t)],\nn\\
 S(\nu_l^t,\nu_{l'}^t) &=& \frac{\Gamma(\nu_l^t)-\Gamma^{ *}(\nu_{l'}^t)}{2i}\nn\\
 s(\nu_l^t)&=&  S(\nu_l^t,\nu_l^t) = {\rm Im}\Gamma(\nu_l^t),\nn\\
  \Gamma(\nu_l^t) &=&\int_0^\infty d\tau {\rm Tr}[\rho_B B(\tau)B] e^{i\nu_l^t\tau},\nn\\
  \nu_l^t &=&   \nu_{k,n,m}^t = E_{m}-E_n + \chi(t)q_k(r_m - r_n).
  \eea
In particular, for $\omega>0$, $\gamma(\omega) = 2\pi J(\omega)( n_{\omega}+1)$, $\gamma(-\omega) = 2\pi J(\omega) n_{\omega}$, $n_\omega = (e^{\omega\beta}-1)^{-1}$, $\beta=1/k_bT_B$ is bath inverse temperature, and $J(\omega) = \sum_k g_k^2 \delta(\omega-\omega_k)$ is the bath spectral density.

\subsection{Illustrative example}
When applied to the practical illustrative example of a qubit $S$ measured by a cavity mode $A$ in contact with a bosonic thermal bath $B$, corresponding to the total Hamiltonian  
\be\label{app:htot}
H_{SAB}(t) = \frac{\omega_S}{2}\sigma_z + \chi(t)\sigma_z a^{\dag}a + \omega_c a^{\dag}a + (a^{\dag} + a) \sum_k g_k (b_k^\dag + b_k) + H_B,
\ee
the above master equation reduces to 
\bea
\dot \rho_{SA} = - i[H_{SA}(t) + H_{LS}(t),\rho_{SA}] &+& \sum_{k,k'=e,g} \gamma_{\nu_k^t,\nu_{k'}^t}\left[\pi_k^S a \rho_{SA}\pi_{k'}^Sa^\dag - \frac{1}{2}\delta_{k,k'}\{\pi_k^Sa^\dag a,\rho_{SA}\}\right]\nn\\
 &+& \sum_{k,k'=e,g} \gamma_{\nu_k^t,-\nu_{k'}^t}\left[\pi_k^S a \rho_{SA}\pi_{k'}^S a - \frac{1}{2}\delta_{k,k'}\{\pi_k^S a^2,\rho_{SA}\}\right]\nn\\
 &+& \sum_{k,k'=e,g} \gamma_{-\nu_k^t,\nu_{k'}^t}\left[\pi_k^S a^\dag \rho_{SA}\pi_{k'}^Sa^\dag - \frac{1}{2}\delta_{k,k'}\{\pi_k^Sa^{\dag 2} ,\rho_{SA}\}\right]\nn\\
  &+& \sum_{k,k' = e,g} \gamma_{-\nu_k^t,-\nu_{k'}^t}\left[\pi_k^S a^\dag \rho_{SA}\pi_{k'}^Sa - \frac{1}{2}\delta_{k,k'}\{\pi_k^Saa^\dag ,\rho_{SA}\}\right],
\eea
with $\nu_k^t : = \omega_c + q_k\chi(t)$, and 
\be
H_{LS}(t) = \sum_{k=e,g} \pi_k^S \left[ S(\nu_k^t,\nu_{k}^t) a^\dag a +  S(-\nu_k^t,\nu_{k}^t) a^{\dag 2} + S(\nu_k^t,-\nu_{k}^t) a^2 + S(-\nu_k^t,-\nu_{k}^t) aa^\dag\right].
\ee

\subsection{Reduced dynamics of $A$ depending on the state of $S$}\label{app:reduceddynA}
Furthermore, since $[Q_S,H_S]=0$, the eigenstates of $H_S$ remain invariant under the whole dynamics. In particular, if we consider that $S$ is initially in one of these eigenstates, $\rho_S(0) = \pi_k^S$, then the reduced state of $S$ remains unchanged throughout the evolution and the reduced dynamics of $A$, $\rho_{A,k}(t) := {\rm Tr}_{SB}[U_{SAB}(t)\pi_k^S\otimes\rho_A(0)\otimes\rho_B(0)U_{SAB}^\dag(t)]$, is given by
  \bea
 \dot \rho_{A,k}(t) = {\cal L}_{k} \rho_{A,k}(t) = -i[H_{A,k}+ H_{LS}^{k},\rho_{A,k}(t)]  + 
  \sum_{l,l'\in {\cal E}_k}\gamma_{\nu_l^t,\nu_{l'}^{t}}\left[a_{l} \rho_{A,k}(t) a_{l'}^{\dag} -\frac{1}{2}\{a_{l'}^{\dag}a_{l},\rho_{A,k}(t)\}\right],
\eea
where ${\cal E}_k$ is the ensemble of triple indices $(k',n,m)$ with fixed $k'=k$, and
\be
a_l = a_{n,m} := \pi_n^A A \pi_{m}^A,
\ee
\be
H_{A,k} = \sum_n [E_n + \chi(t) q_k r_n]\pi_n^A,
\ee
\be
H_{LS}^{k}= \sum_{l,l' \in {\cal E}_k}S(\nu_l^t,\nu_{l'}^t) a_{l'}^{\dag}a_{l}.
\ee


When applied to the practical example of a qubit $S$ measured by a cavity mode $A$ in contact with a bosonic thermal bath $B$, of total Hamiltonian \eqref{app:htot},
the above master equation reduces to 
\bea
\dot \rho_{A,k} = - i[H_{A,k}(t) + H_{LS}^k(t),\rho_{A,k}] &+&  \gamma_{\nu_k^t,\nu_{k}^t}\left[ a \rho_{A,k}a^\dag - \frac{1}{2}\{a^\dag a,\rho_{A,k}\}\right]\nn\\
 &+&  \gamma_{\nu_k^t,-\nu_{k}^t}\left[ a \rho_{A,k} a - \frac{1}{2}\{ a^2,\rho_{A,k}\}\right]\nn\\
 &+&  \gamma_{-\nu_k^t,\nu_{k}^t}\left[ a^\dag \rho_{A,k}a^\dag - \frac{1}{2}\{a^{\dag 2} ,\rho_{A,k}\}\right]\nn\\
  &+&  \gamma_{-\nu_k^t,-\nu_{k}^t}\left[ a^\dag \rho_{A,k} a - \frac{1}{2}\{aa^\dag ,\rho_{SA}\}\right],
\eea
with $\nu_k^t : = \omega_c + q_k\chi(t)$, and 
\be
H_{A,k} = \nu_k^t a^\dag a,
\ee
\be
H_{LS}^k(t) =  \left[ S(\nu_k^t,\nu_{k}^t) a^\dag a +  S(-\nu_k^t,\nu_{k}^t) a^{\dag 2} + S(\nu_k^t,-\nu_{k}^t) a^2 + S(-\nu_k^t,-\nu_{k}^t) aa^\dag\right].
\ee

\section{Steady states}\label{app:steadystate}
In this section we find the steady states of the Redfield master equation \eqref{eqapp:genme} derived in the previous section. More precisely, we are looking for the state of $SA$ such that ${\cal L} \rho_{SA}^\text{ss} = 0$. note that since the map ${\cal L}$ is time-dependent due to the time dependence of $\chi(t)$, the fixed point of the map ${\cal L}$ does depend on time as well (but for simplicity, we have omitted in the notation the explicit time-dependence). Additionally, we will see in the following that the fixed point is in fact not unique, and depends on the initial state $\rho_S(0)$.

As a preliminary observation, the master equation \eqref{eqapp:genme} contains terms of order 2 in the $A-B$ coupling strength, namely the dissipative part 
\be
{\cal D}\rho_{SA} := \sum_{l,l'}\gamma_{\nu_l^t,\nu_{l'}^t}\left[A_{l} \rho_{SA} A_{l'}^{\dag} -\frac{1}{2}\{A_{l'}^{\dag}A_{l},\rho_{SA}\}\right],
\ee
and the Lamb Shift part,
\be
 -i[H_{LS}(t),\rho_{SA}] ,
\ee
since both $\gamma_{\nu_l^t,\nu_{l'}^t}$ and $S(\nu_l^t,\nu_{l'}^t)$ are of order 2. By contrast, the term $-i[H_{SA}(t),\rho_{SA}]$ is of order 0 in the $A-B$ coupling strength. This suggests that the steady state $\rho_{SA}^{\rm ss}$ is also composed of terms of order 0, $\rho_{SA}^{\rm ss,0}$ and terms of order 2,  $\rho_{SA}^{\rm ss, 2}$, in the $A-B$ coupling strength. Injecting in the dynamics we obtain
\bea
{\cal L}\rho_{SA}^{\rm ss} &=& -i[H_{SA}(t),\rho_{SA}^{\rm ss, 0}] \label{1stline}\\
&& -i[H_{SA}(t), \rho_{SA}^{\rm ss, 2}] -i[H_{LS}(t),\rho_{SA}^{\rm ss, 0}] + {\cal D}\rho_{SA}^{\rm ss, 0} \label{2dline}\\
&&+ {\cal O}(\lambda^4)\label{3dline}
\eea
where $\lambda$ stands for the magnitude of the coupling between $A$ and $B$. The term in the first line \eqref{1stline} is of order 0, the terms in the second line \eqref{2dline} are of order 2, while the remaining terms \eqref{3dline} are of order at least 4. Then, $\rho_{SA}^{\rm ss}$ is a steady state, ${\cal L}\rho_{SA}^{\rm ss} =0$, when terms of each order cancel out, implying 
\bea
&&-i[H_{SA}(t),\rho_{SA}^{\rm ss, 0}] = 0 ,\label{cond0}\\
 &&-i[H_{SA}(t), \rho_{SA}^{\rm ss, 2}] -i[H_{LS}(t),\rho_{SA}^{\rm ss, 0}] + {\cal D}\rho_{SA}^{\rm ss, 0} = 0\label{cond2}.
\eea
The condition \eqref{cond0} implies that $\rho_{SA}^{\rm ss,0}$ is diagonal in $\{|e_k\ket|n\ket\}_{k,n}$, the $SA$ energy eigenbasis. We then write $\rho_{SA}^{\rm ss,0}$ as 
\be
\rho_{SA}^{\rm ss,0} = \sum_{k,n} p_{k,n}^0 \pi_k^S \pi_n^A.
\ee
The second condition \eqref{cond2} allows to obtain the higher order corrections in terms of the zeroth order term,
\be\label{B9}
[H_{SA}(t), \rho_{SA}^{\rm ss, 2}]  = -[H_{LS}(t),\rho_{SA}^{\rm ss, 0}] -i {\cal D}\rho_{SA}^{\rm ss, 0} .
\ee
We can re-write ${\cal D}\rho_{SA}^{\rm ss, 0}$ as
\bea
{\cal D}\rho_{SA}^{\rm ss, 0} &=& \sum_{k,n}p_{k,n}^0 \sum_{l,l'}\gamma_{\nu_l^t,\nu_{l'}^t}\left[A_{l}  \pi_k^S \pi_n^A A_{l'}^{\dag} -\frac{1}{2}\{A_{l'}^{\dag}A_{l}, \pi_k^S \pi_n^A\}\right]\nn\\
&=& \sum_{k,n}p_{k,n}^0 \pi_k^S\sum_{l,l' \in {\cal E}_k}\gamma_{\nu_l^t,\nu_{l'}^t}\left[a_{l}  \pi_n^A a_{l'}^{\dag} -\frac{1}{2}\{a_{l'}^{\dag}a_{l}, \pi_n^A\}\right]\nn\\
&=& \sum_{k,n}p_{k,n}^0 \pi_k^S\sum_{n',m',n'',m''}\gamma_{\nu_{k,n',m'}^t,\nu_{k,n'',m''}^t}\left[a_{n',m'}  \pi_n^A a_{n'',m''}^{\dag} -\frac{1}{2}\{a_{n'',m''}^{\dag}a_{n',m'}, \pi_n^A\}\right]\nn\\
&=&\sum_{k,n}p_{k,n}^0 \pi_k^S\sum_{n',n''}\gamma_{\nu_{k,n',n}^t,\nu_{k,n'',n}^t} a_{n',n}\pi_n^Aa_{n'',n}^\dag\nn\\
&&-\frac{1}{2}\sum_{k,n}p_{k,n}^0 \pi_k^S\sum_{n',m''}\gamma_{\nu_{k,n',n}^t,\nu_{k,n',m''}^t} a_{n',m''}^\dag a_{n',n}\pi_n^A\nn\\
&&-\frac{1}{2}\sum_{k,n}p_{k,n}^0 \pi_k^S\sum_{n',m'}\gamma_{\nu_{k,n',m'}^t,\nu_{k,n',n}^t} \pi_n^S a_{n',n}^\dag a_{n',m'}\nn\\
&=& \sum_k \pi_k^S \sum_{n,n',n''} \Big(p_{k,n'}^0\gamma_{\nu_{k,n,n'}^t,\nu_{k,n'',n'}^t}-\frac{1}{2}(p_{k,n''}^0+p_{k,n}^0)\gamma_{\nu_{k,n',n''}^t,\nu_{k,n',n}^t}\Big)\pi_n^A A\pi_{n'}^AA\pi_{n''}^A.\label{term2}
\eea
%
Similarly, we can express $[H_{LS}(t),\rho_{SA}^{\rm ss}]$ as,
\bea\label{term1}
[H_{LS}(t),\rho_{SA}^{\rm ss}] = \sum_{k} \pi_k^S \sum_{n,n',n''}(p_{k,n''}^0-p_{k,n}^0)S(\nu_{k,n',n''}^t,\nu_{k,n',n}^t)\pi_n^AA\pi_{n'}^AA\pi_{n''}^A.
\eea
Injecting \eqref{term1} and \eqref{term2} in \eqref{B9}, we finally obtain
\bea\label{finalcond}
[H_{SA}(t),\rho_{SA}^{ \rm ss,2}] =  \sum_k \pi_k^S \sum_{n,n',n''} \Big((p_{k,n}^0-p_{k,n''}^0)S(\nu_{k,n',n''}^t,\nu_{k,n',n}^t) -i p_{k,n'}^0\gamma_{\nu_{k,n,n'}^t,\nu_{k,n'',n'}^t} &+& i\frac{p_{k,n''}^0+p_{k,n}^0}{2}\gamma_{\nu_{k,n',n''}^t,\nu_{k,n',n}^t}\Big)\nn\\
&&\times\pi_n^A A\pi_{n'}^AA\pi_{n''}^A.\nn\\
\eea

Taking the diagonal element associated to $|e_k\ket|n\ket$ in the above equation \eqref{finalcond}, we obtain
\be
0 = \sum_{n'} \Big( -i p_{k,n'}^0\gamma_{\nu_{k,n,n'}^t,\nu_{k,n,n'}^t} + i\frac{p_{k,n}^0+p_{k,n}^0}{2}\gamma_{\nu_{k,n',n}^t,\nu_{k,n',n}^t}\Big)\bra n|A\pi_{n'}^A A|n\ket
\ee
leading to
\be
 \sum_{n'} \Big( p_{k,n'}^0\gamma(\nu_{k,n,n'}^t) - p_{k,n}^0\gamma(-\nu_{k,n,n'}^t)\Big)|\bra n|A|n'\ket|^2 = 0.
\ee
This implies that for all $n \leftrightarrow n'$ transitions induced by the bath (meaning $\bra n |A|n'\ket \ne 0$), the associated populations are linked by the Boltzmann factor,
\be
\frac{p_{k,n}^0}{p_{k,n'}^0} = e^{ - \beta[E_{n,k}(t)-E_{n',k}(t)]}.
\ee
The off-diagonal terms of eq. \eqref{finalcond} allow to determine the off-diagonal elements of $\rho_{SA}^\text{ss,2}$, but in the main text we neglect this second order contribution to the steady state of $SA$.   
The normalisation of the populations $p_{k,n}^0$ is done remembering that the population of $S$ is conserved by the dynamics since $[Q_S,H_S]=0$. This implies that 
\be
p_{k,n}^0 = \frac{p_k(0)}{Z_{A,k}}e^{-\beta E_{n,k}(t)}
\ee
with $Z_{A,k}:=\sum_n e^{-\beta E_{n,k}(t)}$. Finally, we obtain
\be
\rho_{SA}^\text{ss} = \sum_k p_k(0)\pi_k^S \rho_{A,k}^\text{th} + {\cal O}(\lambda^2)
\ee
with 
\be\label{appssak}
\rho_{A,k}^\text{th} := Z_{A,k}^{-1}e^{-\beta H_{A,k}},
\ee
which is the result used in the main text.

\section{Details of the time evolution of the average cavity photon number}\label{Appdetailn}
We have the following coupled dynamics,
 \be
\frac{d}{dt}\bra a^{\dag}a\ket_k = - [\gamma(\nu_k^t) - \gamma(-\nu_k^t)]\bra a^{\dag}a\ket_k +\gamma(-\nu_k^t) + 2i S(\nu_k^t,-\nu_k^t) \bra a^2\ket_k -2i S(-\nu_k^t,\nu_k^t)\bra a^{\dag 2}\ket_k
\ee
with $k=e,g$, and
\be
\frac{d}{dt}\bra a^2\ket_k = -[\gamma(\nu_k^t)-\gamma(-\nu_k^t) + 2i\nu_k^t +2i S(\nu_k^t,\nu_k^t) + 2iS(-\nu_k^t,-\nu_k^t)] \bra a^2\ket_k - 4iS(-\nu_k^t,\nu_k^t)\bra a^{\dag}a\ket_k -2i S(-\nu_k^t,\nu_k^t) - \gamma_{-\nu_k^t,\nu_k^t}.
\ee
 Integrating formally, we obtain,
 \be
 \bra a^{\dag}a \ket_k(t) = e^{-\int_0^t du \kappa_{k}}\bra a^{\dag}a\ket_k (0) + \int_0^t du e^{-\int_u^t ds \kappa_k} \gamma(-\nu_k^u) +  \int_0^t du e^{-\int_u^t ds \kappa_k} [2i S(\nu_k^u,-\nu_k^u)\bra a^2\ket_k(u) + {\rm c.c.}],
 \ee
where we introduced the notation $\kappa_k := \gamma(\nu_k^t) - \gamma(-\nu_k^t) = 2\pi J(\nu_k^t)$, $k=e,g$. Integrating formally the dynamics of $\bra a^2\ket$, 
\bea
\bra a^2\ket_k(t) &=& e^{-\int_0^t du [\kappa_k + 2i \tilde \nu_k]}\bra a^2\ket_k(0) \nn\\
&& -\int_0^t du e^{-\int_u^t ds [\kappa_k +2i\tilde \nu_k]}[ 4iS(-\nu_k^u,\nu_k^u)\bra a^{\dag}a\ket_k +2i S(-\nu_k^u,\nu_k^u) + \gamma_{-\nu_k^u,\nu_k^u}]
\eea
where we introduced $\tilde \nu_k := \nu_k^t + S(\nu_k^t,\nu_k^t) + S(-\nu_k^t,-\nu_k^t)$, omitting in the notation the explicit time dependence. Then, injecting in the above equation and retaining only terms up to order 2 in the system bath coupling strength, we obtain
\bea
 \bra a^{\dag}a \ket_k(t) &=& e^{-\int_0^t du \kappa_{k}}\bra a^{\dag}a\ket_k (0) + \int_0^t du e^{-\int_u^t ds \kappa_k} \gamma(-\nu_k^u)\nn\\
 && +  \int_0^t du e^{-\int_u^t ds \kappa_k} [2i S(\nu_k^u,-\nu_k^u) e^{-\int_0^u ds (\kappa_k + 2i \tilde \nu_k)}\bra a^2\ket_k(0) + {\rm c.c.}]\nn\\
 &&\hspace{-2cm} -  \int_0^t du e^{-\int_u^t ds \kappa_k} \left[2i S(\nu_k^u,-\nu_k^u)
 \int_0^u ds e^{-\int_s^u dv [\kappa_k +2i\tilde \nu_k]}[ 4iS(-\nu_k^s,\nu_k^s)\bra a^{\dag}a\ket_k +2i S(-\nu_k^s,\nu_k^s) + \gamma_{-\nu_k^s,\nu_k^s}]  + {\rm c.c.}\right] \nn\\
 &=& e^{-\int_0^t du \kappa_{k}}\bra a^{\dag}a\ket_k (0) + \int_0^t du e^{-\int_u^t ds \kappa_k} \gamma(-\nu_k^u)\nn\\
 && +  \int_0^t du e^{-\int_u^t ds \kappa_k} [2i S(\nu_k^u,-\nu_k^u) e^{-\int_0^u ds (\kappa_k + 2i \tilde \nu_k)}\bra a^2\ket_k(0) + {\rm c.c.}]\nn\\
 &&\hspace{-2cm} -  \int_0^t du e^{-\int_u^t ds \kappa_k} \left[2i S(\nu_k^u,-\nu_k^u)
 \int_0^u ds e^{-\int_s^u dv [\kappa_k +2i\tilde \nu_k]}[2i S(-\nu_k^s,\nu_k^s) + \gamma_{-\nu_k^s,\nu_k^s}]  + {\rm c.c.}\right] \nn\\
  &&+ {\cal O}(g_k^4)\nn\\
  &=& e^{-\int_0^t du \kappa_{k}}\bra a^{\dag}a\ket_k (0) + \int_0^t du e^{-\int_u^t ds \kappa_k} \gamma(-\nu_k^u)\nn\\
 && +  e^{-\int_0^t ds \kappa_k} \int_0^t du [2i S(\nu_k^u,-\nu_k^u) e^{-2i\int_0^u ds   \tilde \nu_k}\bra a^2\ket_k(0) + {\rm c.c.}] \nn\\
  &&\hspace{-2cm} - e^{-\int_0^t ds \kappa_k} \int_0^t du  \left[2i S(\nu_k^u,-\nu_k^u)
e^{-2i\int_0^u dv \tilde \nu_k} \int_0^u ds e^{\int_0^s dv [\kappa_k +2i\tilde \nu_k]}[2i S(-\nu_k^s,\nu_k^s) + \gamma_{-\nu_k^s,\nu_k^s}]  + {\rm c.c.}\right] \nn\\
  &&+ {\cal O}(g_k^4)\nn\\
    &=& e^{-\int_0^t du \kappa_{k}}\bra a^{\dag}a\ket_k (0) + \int_0^t du e^{-\int_u^t ds \kappa_k} \gamma(-\nu_k^u)    +  T^{(2)}(t) + {\cal O}(g_k^4),
 \eea
 where $T^{(2)}(t)$ denotes the terms of order 2 in the system-bath coupling, namely,
 \bea
 T^{(2)}(t) &:=& e^{-\int_0^t ds \kappa_k} \int_0^t du [2i S(\nu_k^u,-\nu_k^u) e^{-2i\int_0^u ds   \tilde \nu_k}\bra a^2\ket_k(0) + {\rm c.c.}] \nn\\
  &&\hspace{-2cm} - e^{-\int_0^t ds \kappa_k} \int_0^t du  \left[2i S(\nu_k^u,-\nu_k^u)
e^{-2i\int_0^u dv \tilde \nu_k} \int_0^u ds e^{\int_0^s dv [\kappa_k +2i\tilde \nu_k]}[2i S(-\nu_k^s,\nu_k^s) + \gamma_{-\nu_k^s,\nu_k^s}]  + {\rm c.c.}\right].
\eea
 The term $\int_0^t du e^{-\int_u^t ds \kappa_k} \gamma(-\nu_k^u)$ can be integrated by part, leading to
\be
\int_0^t du e^{-\int_u^t ds \kappa_k} \gamma(-\nu_k^u) = n_{\nu_k^t} - e^{-\int_0^t du \kappa_k}n_{\nu_k^0} - \int_0^t du e^{-\int_u^t ds \kappa_k}\frac{\partial n_{\nu_k^u}}{\partial u}
\ee
so that

\bea
\bra a^{\dag}a \ket_k(t) &=& n_{\nu_k^t}  -\int_0^t du e^{-\int_u^t ds \kappa_k}\frac{\partial n_{\nu_k^u}}{\partial u}+ e^{-\int_0^t du \kappa_k}[\bra a^{\dag}a\ket_k (0)-n_{\nu_k^0}] + T^{(2)}(t),
\eea
as announced in eq. \eqref{eq:A27} of Appendix \ref{app:saturatingprotocol}, with $n_{\nu_k^t}:=(e^{\beta\nu_k^t}-1)^{-1}$.\\

\section{Detail on the work cost for a linear swithcing on/off ramp}\label{applinearramp}
We assume $\chi(t) = R t$, for $t \in [0;t_M]$ (with $t_M = \chi_M/R$), and $\chi(t) =\chi_M - R (t-t_M)$, for $t \in [t_M;t_F]$ (with $t_F - t_M = \chi_M/R$). 
Note that in order to fulfill the assumption used for the derivation of the time-dependent master equation, we must have that the variation of $\chi(t)$ is much slower than the bath correlation time $\tau_c$, namely $R\tau_c \ll \chi_M $. Remembering that in order to have an efficient measurement we need $\chi_M \sim \omega_c$, this typically imposes $R \ll \omega_c/\tau_c$.

We have, remembering that $\kappa_k:= 2\pi J(\nu_k^t)$,
\bea
&&-\sum_{k=e,g} kp_k(0)\int_{0}^{t_M} du \dot\chi(u)\int_{0}^u ds e^{-2\pi\int_s^u dv J(\nu_k^v)}\frac{\partial n_{\nu_k^s}}{\partial s} \nn\\
&&~~~~~~~ = \sum_k p_k(0)q_k^2R^2\beta\int_0^{t_{M}}du \int_0^u ds e^{-2\pi\int_s^u dv J(\omega_c +q_kRv)}\frac{e^{(\omega_c+q_kRs)\beta}}{(e^{(\omega_c+q_kRs)\beta}-1)^2}\nn\\
  &&~~~~~~~ \leq \sum_k p_k(0)  R^2\beta\int_0^{t_{M}}du \int_0^u ds e^{-2\pi (u-s) J_m}\frac{e^{(\omega_c+q_kRs)\beta}}{(e^{(\omega_c+q_kRs)\beta}-1)^2}\nn\\
 &&~~~~~~~ = \sum_k p_k(0) R^2\beta\int_0^{t_{M}}ds \int_s^{t_M} du e^{-2\pi (u-s) J_m}\frac{e^{(\omega_c+q_kRs)\beta}}{(e^{(\omega_c+q_kRs)\beta}-1)^2}\nn\\
 &&~~~~~~~ = \sum_k p_k(0)  R^2\beta\int_0^{t_{M}}ds \frac{1- e^{2\pi (s-t_M) J_m}}{2\pi J_m}\frac{e^{(\omega_c+q_kRs)\beta}}{(e^{(\omega_c+q_kRs)\beta}-1)^2}\nn\\
 \eea
remembering that $q_e= 1 $, $q_g = -1$, and we introduced $J_m:= {\rm Min}_{\omega \in [\omega_c;\omega_c+l\chi_M]}J(\omega)$. Note that the above expression is always positive, even for $q_kR<0$ (and $q_k\chi_M$ is of the same sign as $q_kR$).

Similarly, 
\bea
 &-&\sum_{k=e,g} q_kp_k(0)\frac{p(r|k)}{p_r}\int_{t_M}^{t_F} du \dot\chi(u)\int_{t_M}^u ds e^{-2\pi\int_s^u dv J(\nu_k^v)}\frac{\partial n_{\nu_k^s}}{\partial s} \nn\\
 &&=\sum_k p_k(0)\frac{p(r|k)}{p_r} q_k^2R^2\beta\int_{t_M}^{t_F}du \int_{t_M}^u ds e^{-2\pi\int_s^u dv J[\omega_c +q_k\chi_M - q_kR(v-t_M)]}\frac{e^{[\omega_c+q_k\chi_M -q_kR(s-t_M)]\beta}}{(e^{[\omega_c+q_k\chi_M-q_kR(s-t_M)]\beta}-1)^2}\nn\\
 &&= \sum_k p_k(0)\frac{p(r|k)}{p_r} R^2\beta\int_{0}^{t_M}du \int_{0}^u ds e^{-2\pi\int_s^u dv J(\omega_c + q_kRv)}\frac{e^{(\omega_c+q_kRu)\beta}}{(e^{(\omega_c+q_kRu)\beta}-1)^2}
 \eea
also always positive for any sign of $q_kR$ (since $q_k\chi_M$ has the same sign as $q_kR$). The second equality was obtained doing the change of variable $v \rightarrow 2 t_M -v$, $s \rightarrow 2 t_M -s$, $u \rightarrow 2 t_M -u$, followed by interverting the integrals on $s$ and $u$ (and then swapping the name of $s$ and $u$). We also use the fact that for the ramp protocol, $t_F = 2 t_M$. Then, if additionally we assume that the bath spectral density is approximatively constant $J(\omega) \simeq J$, this simplifies to,
\bea
 &-&\sum_{k=e,g} q_kp_k(0)\frac{p(r|k)}{p_r}\int_{t_M}^{t_F} du \dot\chi(u)\int_{t_M}^u ds e^{-2\pi\int_s^u dv J(\nu_k^v)}\frac{\partial n_{\nu_k^s}}{\partial s}\nn\\
 &&=\sum_k p_k(0)\frac{p(r|k)}{p_r} R^2\beta \int_0^{t_M}du \int_0^u ds e^{-2\pi(u-s) J}\frac{e^{(\omega_c+q_kRu)\beta}}{(e^{(\omega_c+q_kRu)\beta}-1)^2} \nn\\
&&=\sum_k p_k(0)\frac{p(r|k)}{p_r} R^2\beta \int_0^{t_M}du \frac{1-e^{-2\pi Ju}}{2\pi J}\frac{e^{(\omega_c+q_kRu)\beta}}{(e^{(\omega_c+q_kRu)\beta}-1)^2}.
 \eea
Averaging over the observations $r$, this becomes 
\bea
&&\sum_r p_r \left[  \sum_{k=e,g} p_k(0)\frac{p(r|k)}{p_r} R^2\beta \int_0^{t_M}du \frac{1-e^{-2\pi Ju}}{2\pi J}\frac{e^{(\omega_c+q_kRu)\beta}}{(e^{(\omega_c+q_kRu)\beta}-1)^2}\right] \nn\\
&&= \sum_{k=e,g} p_k(0) R^2\beta \int_0^{t_M}du \frac{1-e^{-2\pi Ju}}{2\pi J}\frac{e^{(\omega_c+q_kRu)\beta}}{(e^{(\omega_c+q_kRu)\beta}-1)^2}.
\eea

Finally, the average work is,
\bea 
W_\text{dr} &=& \sum_{k=e,g} p_k(0) R^2\beta\left\{ \int_0^{t_M} du \int_0^u ds e^{-2\pi\int_s^u dv J(\omega_c +q_kRv)}\left[\frac{e^{(\omega_c+q_kRs)\beta}}{(e^{(\omega_c+q_kRs)\beta}-1)^2} + \frac{e^{(\omega_c+q_kRu)\beta}}{(e^{(\omega_c+q_kRu)\beta}-1)^2}\right]\right\}\nn\\
&\simeq&\sum_{k=e,g} p_k(0) R^2\beta\Bigg\{ \int_0^{t_M}du \int_0^u ds e^{-2\pi(u-s) J}\left[\frac{e^{(\omega_c+q_kRs)\beta}}{(e^{(\omega_c+q_kRs)\beta}-1)^2} +\frac{e^{(\omega_c+q_kRu)\beta}}{(e^{(\omega_c+q_kRu)\beta}-1)^2}\right]\Bigg\}  \nn\\
&=&\sum_{k=e,g} p_k(0)R^2\beta\Bigg\{ \int_0^{t_M}du \int_0^u ds e^{-2\pi(u-s) J}\left[\frac{e^{(\omega_c+q_kRs)\beta}}{(e^{(\omega_c+q_kRs)\beta}-1)^2} +\frac{e^{(\omega_c+q_kRu)\beta}}{(e^{(\omega_c+q_kRu)\beta}-1)^2}\right]\Bigg\}  \nn\\
&&=\sum_{k=e,g} p_k(0)\frac{ R^2\beta}{2\pi J}\Bigg\{ \int_0^{t_M}du\frac{e^{(\omega_c+q_kRu)\beta}}{(e^{(\omega_c+q_kRu)\beta}-1)^2} \left[1-e^{-2\pi J(t_M-u)} +1-e^{-2\pi J u}\right]\Bigg\}. 
\eea


\bibliography{biblio}

\begin{thebibliography}{58}%
\makeatletter
\providecommand \@ifxundefined [1]{%
 \@ifx{#1\undefined}
}%
\providecommand \@ifnum [1]{%
 \ifnum #1\expandafter \@firstoftwo
 \else \expandafter \@secondoftwo
 \fi
}%
\providecommand \@ifx [1]{%
 \ifx #1\expandafter \@firstoftwo
 \else \expandafter \@secondoftwo
 \fi
}%
\providecommand \natexlab [1]{#1}%
\providecommand \enquote  [1]{``#1''}%
\providecommand \bibnamefont  [1]{#1}%
\providecommand \bibfnamefont [1]{#1}%
\providecommand \citenamefont [1]{#1}%
\providecommand \href@noop [0]{\@secondoftwo}%
\providecommand \href [0]{\begingroup \@sanitize@url \@href}%
\providecommand \@href[1]{\@@startlink{#1}\@@href}%
\providecommand \@@href[1]{\endgroup#1\@@endlink}%
\providecommand \@sanitize@url [0]{\catcode `\\12\catcode `\$12\catcode
  `\&12\catcode `\#12\catcode `\^12\catcode `\_12\catcode `\%12\relax}%
\providecommand \@@startlink[1]{}%
\providecommand \@@endlink[0]{}%
\providecommand \url  [0]{\begingroup\@sanitize@url \@url }%
\providecommand \@url [1]{\endgroup\@href {#1}{\urlprefix }}%
\providecommand \urlprefix  [0]{URL }%
\providecommand \Eprint [0]{\href }%
\providecommand \doibase [0]{https://doi.org/}%
\providecommand \selectlanguage [0]{\@gobble}%
\providecommand \bibinfo  [0]{\@secondoftwo}%
\providecommand \bibfield  [0]{\@secondoftwo}%
\providecommand \translation [1]{[#1]}%
\providecommand \BibitemOpen [0]{}%
\providecommand \bibitemStop [0]{}%
\providecommand \bibitemNoStop [0]{.\EOS\space}%
\providecommand \EOS [0]{\spacefactor3000\relax}%
\providecommand \BibitemShut  [1]{\csname bibitem#1\endcsname}%
\let\auto@bib@innerbib\@empty
\bibitem [{\citenamefont {Wiseman}\ and\ \citenamefont
  {Milburn}(2009)}]{WisemanBook}%
  \BibitemOpen
  \bibfield  {author} {\bibinfo {author} {\bibfnamefont {H.~M.}\ \bibnamefont
  {Wiseman}}\ and\ \bibinfo {author} {\bibfnamefont {G.~J.}\ \bibnamefont
  {Milburn}},\ }\href {https://doi.org/10.1017/CBO9780511813948} {\emph
  {\bibinfo {title} {{Quantum Measurement and Control}}}}\ (\bibinfo
  {publisher} {Cambridge University Press},\ \bibinfo {address} {Cambridge,
  England, UK},\ \bibinfo {year} {2009})\BibitemShut {NoStop}%
\bibitem [{\citenamefont {Wigner}(1984)}]{Wigner1984}%
  \BibitemOpen
  \bibfield  {author} {\bibinfo {author} {\bibfnamefont {E.~P.}\ \bibnamefont
  {Wigner}},\ }\bibfield  {title} {\bibinfo {title} {{Review of the
  Quantum-Mechanical Measurement Problem}},\ }in\ \href
  {https://doi.org/10.1016/B978-0-12-404970-3.50011-2} {\emph {\bibinfo
  {booktitle} {{Science, Computers, and the Information Onslaught}}}}\
  (\bibinfo  {publisher} {Academic Press},\ \bibinfo {address} {Cambridge, MA,
  USA},\ \bibinfo {year} {1984})\ pp.\ \bibinfo {pages} {63--82}\BibitemShut
  {NoStop}%
\bibitem [{\citenamefont {Zurek}(2009)}]{Zurek09}%
  \BibitemOpen
  \bibfield  {author} {\bibinfo {author} {\bibfnamefont {W.~H.}\ \bibnamefont
  {Zurek}},\ }\bibfield  {title} {\bibinfo {title} {{Quantum Darwinism}},\
  }\href {https://doi.org/10.1038/nphys1202} {\bibfield  {journal} {\bibinfo
  {journal} {Nature Physics}\ }\textbf {\bibinfo {volume} {5}},\ \bibinfo
  {pages} {181} (\bibinfo {year} {2009})}\BibitemShut {NoStop}%
\bibitem [{\citenamefont {Korbicz}\ \emph {et~al.}(2017)\citenamefont
  {Korbicz}, \citenamefont {Aguilar}, \citenamefont
  {{\ifmmode\acute{C}\else\'{C}\fi}wikli{\ifmmode\acute{n}\else\'{n}\fi}ski},\
  and\ \citenamefont {Horodecki}}]{Korbicz17}%
  \BibitemOpen
  \bibfield  {author} {\bibinfo {author} {\bibfnamefont {J.~K.}\ \bibnamefont
  {Korbicz}}, \bibinfo {author} {\bibfnamefont {E.~A.}\ \bibnamefont
  {Aguilar}}, \bibinfo {author} {\bibfnamefont {P.}~\bibnamefont
  {{\ifmmode\acute{C}\else\'{C}\fi}wikli{\ifmmode\acute{n}\else\'{n}\fi}ski}},\
  and\ \bibinfo {author} {\bibfnamefont {P.}~\bibnamefont {Horodecki}},\
  }\bibfield  {title} {\bibinfo {title} {{Generic appearance of objective
  results in quantum measurements}},\ }\href
  {https://doi.org/10.1103/PhysRevA.96.032124} {\bibfield  {journal} {\bibinfo
  {journal} {Physical Review A}\ }\textbf {\bibinfo {volume} {96}},\ \bibinfo
  {pages} {032124} (\bibinfo {year} {2017})}\BibitemShut {NoStop}%
\bibitem [{\citenamefont {Elouard}\ \emph
  {et~al.}(2017{\natexlab{a}})\citenamefont {Elouard}, \citenamefont
  {Herrera-Mart{\ifmmode\acute{\imath}\else\'{\i}\fi}}, \citenamefont
  {Clusel},\ and\ \citenamefont
  {Auff{\ifmmode\grave{e}\else\`{e}\fi}ves}}]{Elouard17Role}%
  \BibitemOpen
  \bibfield  {author} {\bibinfo {author} {\bibfnamefont {C.}~\bibnamefont
  {Elouard}}, \bibinfo {author} {\bibfnamefont {D.~A.}\ \bibnamefont
  {Herrera-Mart{\ifmmode\acute{\imath}\else\'{\i}\fi}}}, \bibinfo {author}
  {\bibfnamefont {M.}~\bibnamefont {Clusel}},\ and\ \bibinfo {author}
  {\bibfnamefont {A.}~\bibnamefont {Auff{\ifmmode\grave{e}\else\`{e}\fi}ves}},\
  }\bibfield  {title} {\bibinfo {title} {{The role of quantum measurement in
  stochastic thermodynamics}},\ }\href
  {https://doi.org/10.1038/s41534-017-0008-4} {\bibfield  {journal} {\bibinfo
  {journal} {npj Quantum Inf.}\ }\textbf {\bibinfo {volume} {3}},\ \bibinfo
  {pages} {1} (\bibinfo {year} {2017}{\natexlab{a}})}\BibitemShut {NoStop}%
\bibitem [{\citenamefont {Yi}\ \emph {et~al.}(2017)\citenamefont {Yi},
  \citenamefont {Talkner},\ and\ \citenamefont {Kim}}]{Yi17}%
  \BibitemOpen
  \bibfield  {author} {\bibinfo {author} {\bibfnamefont {J.}~\bibnamefont
  {Yi}}, \bibinfo {author} {\bibfnamefont {P.}~\bibnamefont {Talkner}},\ and\
  \bibinfo {author} {\bibfnamefont {Y.~W.}\ \bibnamefont {Kim}},\ }\bibfield
  {title} {\bibinfo {title} {{Single-temperature quantum engine without
  feedback control}},\ }\href {https://doi.org/10.1103/PhysRevE.96.022108}
  {\bibfield  {journal} {\bibinfo  {journal} {Physical Review E}\ }\textbf
  {\bibinfo {volume} {96}},\ \bibinfo {pages} {022108} (\bibinfo {year}
  {2017})}\BibitemShut {NoStop}%
\bibitem [{\citenamefont {Ding}\ \emph {et~al.}(2018)\citenamefont {Ding},
  \citenamefont {Yi}, \citenamefont {Kim},\ and\ \citenamefont
  {Talkner}}]{Ding18}%
  \BibitemOpen
  \bibfield  {author} {\bibinfo {author} {\bibfnamefont {X.}~\bibnamefont
  {Ding}}, \bibinfo {author} {\bibfnamefont {J.}~\bibnamefont {Yi}}, \bibinfo
  {author} {\bibfnamefont {Y.~W.}\ \bibnamefont {Kim}},\ and\ \bibinfo {author}
  {\bibfnamefont {P.}~\bibnamefont {Talkner}},\ }\bibfield  {title} {\bibinfo
  {title} {{Measurement-driven single temperature engine}},\ }\href
  {https://doi.org/10.1103/PhysRevE.98.042122} {\bibfield  {journal} {\bibinfo
  {journal} {Phys. Rev. E}\ }\textbf {\bibinfo {volume} {98}},\ \bibinfo
  {pages} {042122} (\bibinfo {year} {2018})}\BibitemShut {NoStop}%
\bibitem [{\citenamefont {Elouard}\ \emph
  {et~al.}(2017{\natexlab{b}})\citenamefont {Elouard}, \citenamefont
  {Herrera-Mart{\ifmmode\acute{\imath}\else\'{\i}\fi}}, \citenamefont {Huard},\
  and\ \citenamefont {Auff{\ifmmode\grave{e}\else\`{e}\fi}ves}}]{Elouard17}%
  \BibitemOpen
  \bibfield  {author} {\bibinfo {author} {\bibfnamefont {C.}~\bibnamefont
  {Elouard}}, \bibinfo {author} {\bibfnamefont {D.}~\bibnamefont
  {Herrera-Mart{\ifmmode\acute{\imath}\else\'{\i}\fi}}}, \bibinfo {author}
  {\bibfnamefont {B.}~\bibnamefont {Huard}},\ and\ \bibinfo {author}
  {\bibfnamefont {A.}~\bibnamefont {Auff{\ifmmode\grave{e}\else\`{e}\fi}ves}},\
  }\bibfield  {title} {\bibinfo {title} {{Extracting Work from Quantum
  Measurement in Maxwell's Demon Engines}},\ }\href
  {https://doi.org/10.1103/PhysRevLett.118.260603} {\bibfield  {journal}
  {\bibinfo  {journal} {Phys. Rev. Lett.}\ }\textbf {\bibinfo {volume} {118}},\
  \bibinfo {pages} {260603} (\bibinfo {year} {2017}{\natexlab{b}})}\BibitemShut
  {NoStop}%
\bibitem [{\citenamefont {Manikandan}\ \emph {et~al.}(2022)\citenamefont
  {Manikandan}, \citenamefont {Elouard}, \citenamefont {Murch}, \citenamefont
  {Auff{\ifmmode\grave{e}\else\`{e}\fi}ves},\ and\ \citenamefont
  {Jordan}}]{Manikandan22}%
  \BibitemOpen
  \bibfield  {author} {\bibinfo {author} {\bibfnamefont {S.~K.}\ \bibnamefont
  {Manikandan}}, \bibinfo {author} {\bibfnamefont {C.}~\bibnamefont {Elouard}},
  \bibinfo {author} {\bibfnamefont {K.~W.}\ \bibnamefont {Murch}}, \bibinfo
  {author} {\bibfnamefont {A.}~\bibnamefont
  {Auff{\ifmmode\grave{e}\else\`{e}\fi}ves}},\ and\ \bibinfo {author}
  {\bibfnamefont {A.~N.}\ \bibnamefont {Jordan}},\ }\bibfield  {title}
  {\bibinfo {title} {{Efficiently fueling a quantum engine with incompatible
  measurements}},\ }\href {https://doi.org/10.1103/PhysRevE.105.044137}
  {\bibfield  {journal} {\bibinfo  {journal} {Physical Review E}\ }\textbf
  {\bibinfo {volume} {105}},\ \bibinfo {pages} {044137} (\bibinfo {year}
  {2022})}\BibitemShut {NoStop}%
\bibitem [{\citenamefont {Bresque}\ \emph {et~al.}(2021)\citenamefont
  {Bresque}, \citenamefont {Camati}, \citenamefont {Rogers}, \citenamefont
  {Murch}, \citenamefont {Jordan},\ and\ \citenamefont
  {Auff{\ifmmode\grave{e}\else\`{e}\fi}ves}}]{Bresque21}%
  \BibitemOpen
  \bibfield  {author} {\bibinfo {author} {\bibfnamefont {L.}~\bibnamefont
  {Bresque}}, \bibinfo {author} {\bibfnamefont {P.~A.}\ \bibnamefont {Camati}},
  \bibinfo {author} {\bibfnamefont {S.}~\bibnamefont {Rogers}}, \bibinfo
  {author} {\bibfnamefont {K.}~\bibnamefont {Murch}}, \bibinfo {author}
  {\bibfnamefont {A.~N.}\ \bibnamefont {Jordan}},\ and\ \bibinfo {author}
  {\bibfnamefont {A.}~\bibnamefont {Auff{\ifmmode\grave{e}\else\`{e}\fi}ves}},\
  }\bibfield  {title} {\bibinfo {title} {{Two-Qubit Engine Fueled by
  Entanglement and Local Measurements}},\ }\href
  {https://doi.org/10.1103/PhysRevLett.126.120605} {\bibfield  {journal}
  {\bibinfo  {journal} {Physical Review Letters}\ }\textbf {\bibinfo {volume}
  {126}},\ \bibinfo {pages} {120605} (\bibinfo {year} {2021})}\BibitemShut
  {NoStop}%
\bibitem [{\citenamefont {Fellous-Asiani}\ \emph {et~al.}(2023)\citenamefont
  {Fellous-Asiani}, \citenamefont {Chai}, \citenamefont {Thonnart},
  \citenamefont {Ng}, \citenamefont {Whitney},\ and\ \citenamefont
  {Auff{\ifmmode\grave{e}\else\`{e}\fi}ves}}]{Fellous-Asiani23}%
  \BibitemOpen
  \bibfield  {author} {\bibinfo {author} {\bibfnamefont {M.}~\bibnamefont
  {Fellous-Asiani}}, \bibinfo {author} {\bibfnamefont {J.~H.}\ \bibnamefont
  {Chai}}, \bibinfo {author} {\bibfnamefont {Y.}~\bibnamefont {Thonnart}},
  \bibinfo {author} {\bibfnamefont {H.~K.}\ \bibnamefont {Ng}}, \bibinfo
  {author} {\bibfnamefont {R.~S.}\ \bibnamefont {Whitney}},\ and\ \bibinfo
  {author} {\bibfnamefont {A.}~\bibnamefont
  {Auff{\ifmmode\grave{e}\else\`{e}\fi}ves}},\ }\bibfield  {title} {\bibinfo
  {title} {{Optimizing Resource Efficiencies for Scalable Full-Stack Quantum
  Computers}},\ }\href {https://doi.org/10.1103/PRXQuantum.4.040319} {\bibfield
   {journal} {\bibinfo  {journal} {PRX Quantum}\ }\textbf {\bibinfo {volume}
  {4}},\ \bibinfo {pages} {040319} (\bibinfo {year} {2023})}\BibitemShut
  {NoStop}%
\bibitem [{\citenamefont {Jacobs}(2012)}]{Jacobs12}%
  \BibitemOpen
  \bibfield  {author} {\bibinfo {author} {\bibfnamefont {K.}~\bibnamefont
  {Jacobs}},\ }\bibfield  {title} {\bibinfo {title} {{Quantum measurement and
  the first law of thermodynamics: The energy cost of measurement is the work
  value of the acquired information}},\ }\href
  {https://doi.org/10.1103/PhysRevE.86.040106} {\bibfield  {journal} {\bibinfo
  {journal} {Physical Review E}\ }\textbf {\bibinfo {volume} {86}},\ \bibinfo
  {pages} {040106} (\bibinfo {year} {2012})}\BibitemShut {NoStop}%
\bibitem [{\citenamefont {Deffner}\ \emph {et~al.}(2016)\citenamefont
  {Deffner}, \citenamefont {Paz},\ and\ \citenamefont {Zurek}}]{Deffner16}%
  \BibitemOpen
  \bibfield  {author} {\bibinfo {author} {\bibfnamefont {S.}~\bibnamefont
  {Deffner}}, \bibinfo {author} {\bibfnamefont {J.~P.}\ \bibnamefont {Paz}},\
  and\ \bibinfo {author} {\bibfnamefont {W.~H.}\ \bibnamefont {Zurek}},\
  }\bibfield  {title} {\bibinfo {title} {{Quantum work and the thermodynamic
  cost of quantum measurements}},\ }\href
  {https://doi.org/10.1103/PhysRevE.94.010103} {\bibfield  {journal} {\bibinfo
  {journal} {Physical Review E}\ }\textbf {\bibinfo {volume} {94}},\ \bibinfo
  {pages} {010103} (\bibinfo {year} {2016})}\BibitemShut {NoStop}%
\bibitem [{\citenamefont {Sagawa}\ and\ \citenamefont {Ueda}(2009)}]{Sagawa09}%
  \BibitemOpen
  \bibfield  {author} {\bibinfo {author} {\bibfnamefont {T.}~\bibnamefont
  {Sagawa}}\ and\ \bibinfo {author} {\bibfnamefont {M.}~\bibnamefont {Ueda}},\
  }\bibfield  {title} {\bibinfo {title} {{Minimal Energy Cost for Thermodynamic
  Information Processing: Measurement and Information Erasure}},\ }\href
  {https://doi.org/10.1103/PhysRevLett.102.250602} {\bibfield  {journal}
  {\bibinfo  {journal} {Phys. Rev. Lett.}\ }\textbf {\bibinfo {volume} {102}},\
  \bibinfo {pages} {250602} (\bibinfo {year} {2009})}\BibitemShut {NoStop}%
\bibitem [{\citenamefont {Jacobs}(2009)}]{Jacobs09}%
  \BibitemOpen
  \bibfield  {author} {\bibinfo {author} {\bibfnamefont {K.}~\bibnamefont
  {Jacobs}},\ }\bibfield  {title} {\bibinfo {title} {{Second law of
  thermodynamics and quantum feedback control: Maxwell's demon with weak
  measurements}},\ }\href {https://doi.org/10.1103/PhysRevA.80.012322}
  {\bibfield  {journal} {\bibinfo  {journal} {Phys. Rev. A}\ }\textbf {\bibinfo
  {volume} {80}},\ \bibinfo {pages} {012322} (\bibinfo {year}
  {2009})}\BibitemShut {NoStop}%
\bibitem [{\citenamefont {Funo}\ \emph {et~al.}(2013)\citenamefont {Funo},
  \citenamefont {Watanabe},\ and\ \citenamefont {Ueda}}]{Funo13}%
  \BibitemOpen
  \bibfield  {author} {\bibinfo {author} {\bibfnamefont {K.}~\bibnamefont
  {Funo}}, \bibinfo {author} {\bibfnamefont {Y.}~\bibnamefont {Watanabe}},\
  and\ \bibinfo {author} {\bibfnamefont {M.}~\bibnamefont {Ueda}},\ }\bibfield
  {title} {\bibinfo {title} {{Integral quantum fluctuation theorems under
  measurement and feedback control}},\ }\href
  {https://doi.org/10.1103/PhysRevE.88.052121} {\bibfield  {journal} {\bibinfo
  {journal} {Phys. Rev. E}\ }\textbf {\bibinfo {volume} {88}},\ \bibinfo
  {pages} {052121} (\bibinfo {year} {2013})}\BibitemShut {NoStop}%
\bibitem [{\citenamefont {Watanabe}\ \emph {et~al.}(2014)\citenamefont
  {Watanabe}, \citenamefont {Venkatesh}, \citenamefont {Talkner}, \citenamefont
  {Campisi},\ and\ \citenamefont
  {H{\ifmmode\ddot{a}\else\"{a}\fi}nggi}}]{Watanabe14}%
  \BibitemOpen
  \bibfield  {author} {\bibinfo {author} {\bibfnamefont {G.}~\bibnamefont
  {Watanabe}}, \bibinfo {author} {\bibfnamefont {B.~P.}\ \bibnamefont
  {Venkatesh}}, \bibinfo {author} {\bibfnamefont {P.}~\bibnamefont {Talkner}},
  \bibinfo {author} {\bibfnamefont {M.}~\bibnamefont {Campisi}},\ and\ \bibinfo
  {author} {\bibfnamefont {P.}~\bibnamefont
  {H{\ifmmode\ddot{a}\else\"{a}\fi}nggi}},\ }\bibfield  {title} {\bibinfo
  {title} {{Quantum fluctuation theorems and generalized measurements during
  the force protocol}},\ }\href {https://doi.org/10.1103/PhysRevE.89.032114}
  {\bibfield  {journal} {\bibinfo  {journal} {Phys. Rev. E}\ }\textbf {\bibinfo
  {volume} {89}},\ \bibinfo {pages} {032114} (\bibinfo {year}
  {2014})}\BibitemShut {NoStop}%
\bibitem [{\citenamefont {Manikandan}\ \emph {et~al.}(2019)\citenamefont
  {Manikandan}, \citenamefont {Elouard},\ and\ \citenamefont
  {Jordan}}]{Manikandan19}%
  \BibitemOpen
  \bibfield  {author} {\bibinfo {author} {\bibfnamefont {S.~K.}\ \bibnamefont
  {Manikandan}}, \bibinfo {author} {\bibfnamefont {C.}~\bibnamefont
  {Elouard}},\ and\ \bibinfo {author} {\bibfnamefont {A.~N.}\ \bibnamefont
  {Jordan}},\ }\bibfield  {title} {\bibinfo {title} {{Fluctuation theorems for
  continuous quantum measurements and absolute irreversibility}},\ }\href
  {https://doi.org/10.1103/PhysRevA.99.022117} {\bibfield  {journal} {\bibinfo
  {journal} {Phys. Rev. A}\ }\textbf {\bibinfo {volume} {99}},\ \bibinfo
  {pages} {022117} (\bibinfo {year} {2019})}\BibitemShut {NoStop}%
\bibitem [{\citenamefont {Belenchia}\ \emph {et~al.}(2020)\citenamefont
  {Belenchia}, \citenamefont {Mancino}, \citenamefont {Landi},\ and\
  \citenamefont {Paternostro}}]{Belenchia20}%
  \BibitemOpen
  \bibfield  {author} {\bibinfo {author} {\bibfnamefont {A.}~\bibnamefont
  {Belenchia}}, \bibinfo {author} {\bibfnamefont {L.}~\bibnamefont {Mancino}},
  \bibinfo {author} {\bibfnamefont {G.~T.}\ \bibnamefont {Landi}},\ and\
  \bibinfo {author} {\bibfnamefont {M.}~\bibnamefont {Paternostro}},\
  }\bibfield  {title} {\bibinfo {title} {{Entropy production in continuously
  measured Gaussian quantum systems}},\ }\href
  {https://doi.org/10.1038/s41534-020-00334-6} {\bibfield  {journal} {\bibinfo
  {journal} {npj Quantum Inf.}\ }\textbf {\bibinfo {volume} {6}},\ \bibinfo
  {pages} {1} (\bibinfo {year} {2020})}\BibitemShut {NoStop}%
\bibitem [{\citenamefont {Mancino}\ \emph {et~al.}(2018)\citenamefont
  {Mancino}, \citenamefont {Sbroscia}, \citenamefont {Roccia}, \citenamefont
  {Gianani}, \citenamefont {Somma}, \citenamefont {Mataloni}, \citenamefont
  {Paternostro},\ and\ \citenamefont {Barbieri}}]{Mancino18}%
  \BibitemOpen
  \bibfield  {author} {\bibinfo {author} {\bibfnamefont {L.}~\bibnamefont
  {Mancino}}, \bibinfo {author} {\bibfnamefont {M.}~\bibnamefont {Sbroscia}},
  \bibinfo {author} {\bibfnamefont {E.}~\bibnamefont {Roccia}}, \bibinfo
  {author} {\bibfnamefont {I.}~\bibnamefont {Gianani}}, \bibinfo {author}
  {\bibfnamefont {F.}~\bibnamefont {Somma}}, \bibinfo {author} {\bibfnamefont
  {P.}~\bibnamefont {Mataloni}}, \bibinfo {author} {\bibfnamefont
  {M.}~\bibnamefont {Paternostro}},\ and\ \bibinfo {author} {\bibfnamefont
  {M.}~\bibnamefont {Barbieri}},\ }\bibfield  {title} {\bibinfo {title} {{The
  entropic cost of quantum generalized measurements}},\ }\href
  {https://doi.org/10.1038/s41534-018-0069-z} {\bibfield  {journal} {\bibinfo
  {journal} {npj Quantum Inf.}\ }\textbf {\bibinfo {volume} {4}},\ \bibinfo
  {pages} {1} (\bibinfo {year} {2018})}\BibitemShut {NoStop}%
\bibitem [{\citenamefont {Linpeng}\ \emph {et~al.}(2022)\citenamefont
  {Linpeng}, \citenamefont {Bresque}, \citenamefont {Maffei}, \citenamefont
  {Jordan}, \citenamefont {Auff{\ifmmode\grave{e}\else\`{e}\fi}ves},\ and\
  \citenamefont {Murch}}]{Linpeng22}%
  \BibitemOpen
  \bibfield  {author} {\bibinfo {author} {\bibfnamefont {X.}~\bibnamefont
  {Linpeng}}, \bibinfo {author} {\bibfnamefont {L.}~\bibnamefont {Bresque}},
  \bibinfo {author} {\bibfnamefont {M.}~\bibnamefont {Maffei}}, \bibinfo
  {author} {\bibfnamefont {A.~N.}\ \bibnamefont {Jordan}}, \bibinfo {author}
  {\bibfnamefont {A.}~\bibnamefont {Auff{\ifmmode\grave{e}\else\`{e}\fi}ves}},\
  and\ \bibinfo {author} {\bibfnamefont {K.~W.}\ \bibnamefont {Murch}},\
  }\bibfield  {title} {\bibinfo {title} {{Energetic Cost of Measurements Using
  Quantum, Coherent, and Thermal Light}},\ }\href
  {https://doi.org/10.1103/PhysRevLett.128.220506} {\bibfield  {journal}
  {\bibinfo  {journal} {Physical Review Letters}\ }\textbf {\bibinfo {volume}
  {128}},\ \bibinfo {pages} {220506} (\bibinfo {year} {2022})}\BibitemShut
  {NoStop}%
\bibitem [{\citenamefont {Elouard}\ and\ \citenamefont
  {Jordan}(2018)}]{Elouard18}%
  \BibitemOpen
  \bibfield  {author} {\bibinfo {author} {\bibfnamefont {C.}~\bibnamefont
  {Elouard}}\ and\ \bibinfo {author} {\bibfnamefont {A.~N.}\ \bibnamefont
  {Jordan}},\ }\bibfield  {title} {\bibinfo {title} {{Efficient Quantum
  Measurement Engines}},\ }\href
  {https://doi.org/10.1103/PhysRevLett.120.260601} {\bibfield  {journal}
  {\bibinfo  {journal} {Phys. Rev. Lett.}\ }\textbf {\bibinfo {volume} {120}},\
  \bibinfo {pages} {260601} (\bibinfo {year} {2018})}\BibitemShut {NoStop}%
\bibitem [{\citenamefont {Guryanova}\ \emph {et~al.}(2020)\citenamefont
  {Guryanova}, \citenamefont {Friis},\ and\ \citenamefont
  {Huber}}]{Guryanova20}%
  \BibitemOpen
  \bibfield  {author} {\bibinfo {author} {\bibfnamefont {Y.}~\bibnamefont
  {Guryanova}}, \bibinfo {author} {\bibfnamefont {N.}~\bibnamefont {Friis}},\
  and\ \bibinfo {author} {\bibfnamefont {M.}~\bibnamefont {Huber}},\ }\bibfield
   {title} {\bibinfo {title} {{Ideal Projective Measurements Have Infinite
  Resource Costs}},\ }\href {https://doi.org/10.22331/q-2020-01-13-222}
  {\bibfield  {journal} {\bibinfo  {journal} {Quantum}\ }\textbf {\bibinfo
  {volume} {4}},\ \bibinfo {pages} {222} (\bibinfo {year} {2020})},\ \Eprint
  {https://arxiv.org/abs/1805.11899v3} {1805.11899v3} \BibitemShut {NoStop}%
\bibitem [{\citenamefont {Elouard}\ \emph {et~al.}(2021)\citenamefont
  {Elouard}, \citenamefont {Lewalle}, \citenamefont {Manikandan}, \citenamefont
  {Rogers}, \citenamefont {Frank},\ and\ \citenamefont {Jordan}}]{Elouard21}%
  \BibitemOpen
  \bibfield  {author} {\bibinfo {author} {\bibfnamefont {C.}~\bibnamefont
  {Elouard}}, \bibinfo {author} {\bibfnamefont {P.}~\bibnamefont {Lewalle}},
  \bibinfo {author} {\bibfnamefont {S.~K.}\ \bibnamefont {Manikandan}},
  \bibinfo {author} {\bibfnamefont {S.}~\bibnamefont {Rogers}}, \bibinfo
  {author} {\bibfnamefont {A.}~\bibnamefont {Frank}},\ and\ \bibinfo {author}
  {\bibfnamefont {A.~N.}\ \bibnamefont {Jordan}},\ }\bibfield  {title}
  {\bibinfo {title} {{Quantum erasing the memory of Wigner's friend}},\ }\href
  {https://doi.org/10.22331/q-2021-07-08-498} {\bibfield  {journal} {\bibinfo
  {journal} {Quantum}\ }\textbf {\bibinfo {volume} {5}},\ \bibinfo {pages}
  {498} (\bibinfo {year} {2021})},\ \Eprint
  {https://arxiv.org/abs/2009.09905v4} {2009.09905v4} \BibitemShut {NoStop}%
\bibitem [{\citenamefont {Sokolovski}\ and\ \citenamefont
  {Matzkin}(2021)}]{Sokolovski21}%
  \BibitemOpen
  \bibfield  {author} {\bibinfo {author} {\bibfnamefont {D.}~\bibnamefont
  {Sokolovski}}\ and\ \bibinfo {author} {\bibfnamefont {A.}~\bibnamefont
  {Matzkin}},\ }\bibfield  {title} {\bibinfo {title} {{Wigner{'}s Friend
  Scenarios and the Internal Consistency of Standard Quantum Mechanics}},\
  }\bibfield  {journal} {\bibinfo  {journal} {Entropy}\ }\textbf {\bibinfo
  {volume} {23}},\ \href {https://doi.org/10.3390/e23091186}
  {10.3390/e23091186} (\bibinfo {year} {2021})\BibitemShut {NoStop}%
\bibitem [{\citenamefont {Mohammady}(2023)}]{Mohammady23}%
  \BibitemOpen
  \bibfield  {author} {\bibinfo {author} {\bibfnamefont {M.~H.}\ \bibnamefont
  {Mohammady}},\ }\bibfield  {title} {\bibinfo {title} {{Thermodynamically free
  quantum measurements}},\ }\href {https://doi.org/10.1088/1751-8121/acad4a}
  {\bibfield  {journal} {\bibinfo  {journal} {Journal of Physics A:
  Mathematical and Theoretical}\ }\textbf {\bibinfo {volume} {55}},\ \bibinfo
  {pages} {505304} (\bibinfo {year} {2023})}\BibitemShut {NoStop}%
\bibitem [{\citenamefont {Allahverdyan}\ \emph {et~al.}(2013)\citenamefont
  {Allahverdyan}, \citenamefont {Balian},\ and\ \citenamefont
  {Nieuwenhuizen}}]{Allahverdyan13}%
  \BibitemOpen
  \bibfield  {author} {\bibinfo {author} {\bibfnamefont {A.~E.}\ \bibnamefont
  {Allahverdyan}}, \bibinfo {author} {\bibfnamefont {R.}~\bibnamefont
  {Balian}},\ and\ \bibinfo {author} {\bibfnamefont {T.~M.}\ \bibnamefont
  {Nieuwenhuizen}},\ }\bibfield  {title} {\bibinfo {title} {{Understanding
  quantum measurement from the solution of dynamical models}},\ }\href
  {https://doi.org/10.1016/j.physrep.2012.11.001} {\bibfield  {journal}
  {\bibinfo  {journal} {Phys. Rep.}\ }\textbf {\bibinfo {volume} {525}},\
  \bibinfo {pages} {1} (\bibinfo {year} {2013})}\BibitemShut {NoStop}%
\bibitem [{\citenamefont {Goan}\ \emph {et~al.}(2001)\citenamefont {Goan},
  \citenamefont {Milburn}, \citenamefont {Wiseman},\ and\ \citenamefont
  {Bi~Sun}}]{Goan01}%
  \BibitemOpen
  \bibfield  {author} {\bibinfo {author} {\bibfnamefont {H.-S.}\ \bibnamefont
  {Goan}}, \bibinfo {author} {\bibfnamefont {G.~J.}\ \bibnamefont {Milburn}},
  \bibinfo {author} {\bibfnamefont {H.~M.}\ \bibnamefont {Wiseman}},\ and\
  \bibinfo {author} {\bibfnamefont {H.}~\bibnamefont {Bi~Sun}},\ }\bibfield
  {title} {\bibinfo {title} {{Continuous quantum measurement of two coupled
  quantum dots using a point contact: A quantum trajectory approach}},\ }\href
  {https://doi.org/10.1103/PhysRevB.63.125326} {\bibfield  {journal} {\bibinfo
  {journal} {Phys. Rev. B}\ }\textbf {\bibinfo {volume} {63}},\ \bibinfo
  {pages} {125326} (\bibinfo {year} {2001})}\BibitemShut {NoStop}%
\bibitem [{\citenamefont {Breuer}\ and\ \citenamefont
  {Petruccione}(2007)}]{BreuerPetruccione}%
  \BibitemOpen
  \bibfield  {author} {\bibinfo {author} {\bibfnamefont {H.-P.}\ \bibnamefont
  {Breuer}}\ and\ \bibinfo {author} {\bibfnamefont {F.}~\bibnamefont
  {Petruccione}},\ }\href
  {https://doi.org/10.1093/acprof:oso/9780199213900.001.0001} {\emph {\bibinfo
  {title} {{The Theory of Open Quantum Systems}}}}\ (\bibinfo  {publisher}
  {Oxford University Press},\ \bibinfo {year} {2007})\BibitemShut {NoStop}%
\bibitem [{\citenamefont {Zurek}(2003)}]{Zurek03}%
  \BibitemOpen
  \bibfield  {author} {\bibinfo {author} {\bibfnamefont {W.~H.}\ \bibnamefont
  {Zurek}},\ }\bibfield  {title} {\bibinfo {title} {{Decoherence, einselection,
  and the quantum origins of the classical}},\ }\href
  {https://doi.org/10.1103/RevModPhys.75.715} {\bibfield  {journal} {\bibinfo
  {journal} {Reviews of Modern Physics}\ }\textbf {\bibinfo {volume} {75}},\
  \bibinfo {pages} {715} (\bibinfo {year} {2003})}\BibitemShut {NoStop}%
\bibitem [{\citenamefont {Horodecki}\ \emph {et~al.}(2013)\citenamefont
  {Horodecki}, \citenamefont {Korbicz},\ and\ \citenamefont
  {Horodecki}}]{Horodecki13}%
  \BibitemOpen
  \bibfield  {author} {\bibinfo {author} {\bibfnamefont {R.}~\bibnamefont
  {Horodecki}}, \bibinfo {author} {\bibfnamefont {J.~K.}\ \bibnamefont
  {Korbicz}},\ and\ \bibinfo {author} {\bibfnamefont {P.}~\bibnamefont
  {Horodecki}},\ }\bibfield  {title} {\bibinfo {title} {{Quantum origins of
  objectivity}},\ }\bibfield  {journal} {\bibinfo  {journal} {arXiv}\ }\href
  {https://doi.org/10.1103/PhysRevA.91.032122} {10.1103/PhysRevA.91.032122}
  (\bibinfo {year} {2013}),\ \Eprint {https://arxiv.org/abs/1312.6588}
  {1312.6588} \BibitemShut {NoStop}%
\bibitem [{\citenamefont {Landauer}(1961)}]{Landauer}%
  \BibitemOpen
  \bibfield  {author} {\bibinfo {author} {\bibfnamefont {R.}~\bibnamefont
  {Landauer}},\ }\bibfield  {title} {\bibinfo {title} {{Irreversibility and
  Heat Generation in the Computing Process}},\ }\href
  {https://doi.org/10.1147/rd.53.0183} {\bibfield  {journal} {\bibinfo
  {journal} {IBM Journal of Research and Development}\ }\textbf {\bibinfo
  {volume} {5}},\ \bibinfo {pages} {183} (\bibinfo {year} {1961})}\BibitemShut
  {NoStop}%
\bibitem [{\citenamefont {Abdelkhalek}\ \emph {et~al.}(2016)\citenamefont
  {Abdelkhalek}, \citenamefont {Nakata},\ and\ \citenamefont
  {Reeb}}]{Abdelkhalek16}%
  \BibitemOpen
  \bibfield  {author} {\bibinfo {author} {\bibfnamefont {K.}~\bibnamefont
  {Abdelkhalek}}, \bibinfo {author} {\bibfnamefont {Y.}~\bibnamefont
  {Nakata}},\ and\ \bibinfo {author} {\bibfnamefont {D.}~\bibnamefont {Reeb}},\
  }\bibfield  {title} {\bibinfo {title} {{Fundamental energy cost for quantum
  measurement}},\ }\bibfield  {journal} {\bibinfo  {journal} {arXiv}\ }\href
  {https://doi.org/10.48550/arXiv.1609.06981} {10.48550/arXiv.1609.06981}
  (\bibinfo {year} {2016}),\ \Eprint {https://arxiv.org/abs/1609.06981}
  {1609.06981} \BibitemShut {NoStop}%
\bibitem [{\citenamefont {Korbicz}\ \emph {et~al.}(2014)\citenamefont
  {Korbicz}, \citenamefont {Horodecki},\ and\ \citenamefont
  {Horodecki}}]{Korbicz14}%
  \BibitemOpen
  \bibfield  {author} {\bibinfo {author} {\bibfnamefont {J.~K.}\ \bibnamefont
  {Korbicz}}, \bibinfo {author} {\bibfnamefont {P.}~\bibnamefont {Horodecki}},\
  and\ \bibinfo {author} {\bibfnamefont {R.}~\bibnamefont {Horodecki}},\
  }\bibfield  {title} {\bibinfo {title} {{Objectivity in a Noisy Photonic
  Environment through Quantum State Information Broadcasting}},\ }\href
  {https://doi.org/10.1103/PhysRevLett.112.120402} {\bibfield  {journal}
  {\bibinfo  {journal} {Phys. Rev. Lett.}\ }\textbf {\bibinfo {volume} {112}},\
  \bibinfo {pages} {120402} (\bibinfo {year} {2014})}\BibitemShut {NoStop}%
\bibitem [{\citenamefont {Engineer}\ \emph {et~al.}(2024)\citenamefont
  {Engineer}, \citenamefont {Rivlin}, \citenamefont {Wollmann}, \citenamefont
  {Malik},\ and\ \citenamefont {Lock}}]{Engineer24}%
  \BibitemOpen
  \bibfield  {author} {\bibinfo {author} {\bibfnamefont {S.}~\bibnamefont
  {Engineer}}, \bibinfo {author} {\bibfnamefont {T.}~\bibnamefont {Rivlin}},
  \bibinfo {author} {\bibfnamefont {S.}~\bibnamefont {Wollmann}}, \bibinfo
  {author} {\bibfnamefont {M.}~\bibnamefont {Malik}},\ and\ \bibinfo {author}
  {\bibfnamefont {M.~P.~E.}\ \bibnamefont {Lock}},\ }\bibfield  {title}
  {\bibinfo {title} {{Equilibration of objective observables in a dynamical
  model of quantum measurements}},\ }\bibfield  {journal} {\bibinfo  {journal}
  {arXiv}\ }\href {https://doi.org/10.48550/arXiv.2403.18016}
  {10.48550/arXiv.2403.18016} (\bibinfo {year} {2024}),\ \Eprint
  {https://arxiv.org/abs/2403.18016} {2403.18016} \BibitemShut {NoStop}%
\bibitem [{\citenamefont {Le}\ and\ \citenamefont
  {Olaya-Castro}(2019)}]{Le2019Jan}%
  \BibitemOpen
  \bibfield  {author} {\bibinfo {author} {\bibfnamefont {T.~P.}\ \bibnamefont
  {Le}}\ and\ \bibinfo {author} {\bibfnamefont {A.}~\bibnamefont
  {Olaya-Castro}},\ }\bibfield  {title} {\bibinfo {title} {{Strong Quantum
  Darwinism and Strong Independence are Equivalent to Spectrum Broadcast
  Structure}},\ }\href {https://doi.org/10.1103/PhysRevLett.122.010403}
  {\bibfield  {journal} {\bibinfo  {journal} {Phys. Rev. Lett.}\ }\textbf
  {\bibinfo {volume} {122}},\ \bibinfo {pages} {010403} (\bibinfo {year}
  {2019})}\BibitemShut {NoStop}%
\bibitem [{\citenamefont {Feller}\ \emph {et~al.}(2021)\citenamefont {Feller},
  \citenamefont {Roussel}, \citenamefont
  {Fr{\ifmmode\acute{e}\else\'{e}\fi}rot},\ and\ \citenamefont
  {Degiovanni}}]{Feller21}%
  \BibitemOpen
  \bibfield  {author} {\bibinfo {author} {\bibfnamefont {A.}~\bibnamefont
  {Feller}}, \bibinfo {author} {\bibfnamefont {B.}~\bibnamefont {Roussel}},
  \bibinfo {author} {\bibfnamefont {I.}~\bibnamefont
  {Fr{\ifmmode\acute{e}\else\'{e}\fi}rot}},\ and\ \bibinfo {author}
  {\bibfnamefont {P.}~\bibnamefont {Degiovanni}},\ }\bibfield  {title}
  {\bibinfo {title} {{Comment on {\textasciigrave}{\textasciigrave}Strong
  Quantum Darwinism and Strong Independence Are Equivalent to Spectrum
  Broadcast Structure''}},\ }\href
  {https://doi.org/10.1103/PhysRevLett.126.188901} {\bibfield  {journal}
  {\bibinfo  {journal} {Phys. Rev. Lett.}\ }\textbf {\bibinfo {volume} {126}},\
  \bibinfo {pages} {188901} (\bibinfo {year} {2021})}\BibitemShut {NoStop}%
\bibitem [{\citenamefont {Jacobs}\ and\ \citenamefont
  {Steck}(2006)}]{Jacobs06}%
  \BibitemOpen
  \bibfield  {author} {\bibinfo {author} {\bibfnamefont {K.}~\bibnamefont
  {Jacobs}}\ and\ \bibinfo {author} {\bibfnamefont {D.~A.}\ \bibnamefont
  {Steck}},\ }\bibfield  {title} {\bibinfo {title} {{A straightforward
  introduction to continuous quantum measurement}},\ }\href
  {https://doi.org/10.1080/00107510601101934} {\bibfield  {journal} {\bibinfo
  {journal} {Contemp. Phys.}\ }\textbf {\bibinfo {volume} {47}},\ \bibinfo
  {pages} {279} (\bibinfo {year} {2006})}\BibitemShut {NoStop}%
\bibitem [{\citenamefont {Esposito}\ \emph {et~al.}(2010)\citenamefont
  {Esposito}, \citenamefont {Lindenberg},\ and\ \citenamefont {Van~den
  Broeck}}]{Esposito10}%
  \BibitemOpen
  \bibfield  {author} {\bibinfo {author} {\bibfnamefont {M.}~\bibnamefont
  {Esposito}}, \bibinfo {author} {\bibfnamefont {K.}~\bibnamefont
  {Lindenberg}},\ and\ \bibinfo {author} {\bibfnamefont {C.}~\bibnamefont
  {Van~den Broeck}},\ }\bibfield  {title} {\bibinfo {title} {{Entropy
  production as correlation between system and reservoir}},\ }\href
  {https://doi.org/10.1088/1367-2630/12/1/013013} {\bibfield  {journal}
  {\bibinfo  {journal} {New J. Phys.}\ }\textbf {\bibinfo {volume} {12}},\
  \bibinfo {pages} {013013} (\bibinfo {year} {2010})}\BibitemShut {NoStop}%
\bibitem [{\citenamefont {Nielsen}\ and\ \citenamefont
  {Chuang}(2010)}]{Nielsen10}%
  \BibitemOpen
  \bibfield  {author} {\bibinfo {author} {\bibfnamefont {M.~A.}\ \bibnamefont
  {Nielsen}}\ and\ \bibinfo {author} {\bibfnamefont {I.~L.}\ \bibnamefont
  {Chuang}},\ }\href {https://doi.org/10.1017/CBO9780511976667} {\emph
  {\bibinfo {title} {{Quantum Computation and Quantum Information: 10th
  Anniversary Edition}}}}\ (\bibinfo  {publisher} {Cambridge University
  Press},\ \bibinfo {address} {Cambridge, England, UK},\ \bibinfo {year}
  {2010})\BibitemShut {NoStop}%
\bibitem [{\citenamefont {Sagawa}\ and\ \citenamefont
  {Ueda}(2011)}]{Erratum-Sagawa09}%
  \BibitemOpen
  \bibfield  {author} {\bibinfo {author} {\bibfnamefont {T.}~\bibnamefont
  {Sagawa}}\ and\ \bibinfo {author} {\bibfnamefont {M.}~\bibnamefont {Ueda}},\
  }\bibfield  {title} {\bibinfo {title} {{Erratum: Minimal Energy Cost for
  Thermodynamic Information Processing: Measurement and Information Erasure
  [Phys. Rev. Lett. 102, 250602 (2009)]}},\ }\href
  {https://doi.org/10.1103/PhysRevLett.106.189901} {\bibfield  {journal}
  {\bibinfo  {journal} {Phys. Rev. Lett.}\ }\textbf {\bibinfo {volume} {106}},\
  \bibinfo {pages} {189901} (\bibinfo {year} {2011})}\BibitemShut {NoStop}%
\bibitem [{\citenamefont {Sagawa}(2012)}]{Sagawa12}%
  \BibitemOpen
  \bibfield  {author} {\bibinfo {author} {\bibfnamefont {T.}~\bibnamefont
  {Sagawa}},\ }\bibfield  {title} {\bibinfo {title} {{Second Law-Like
  Inequalities with Quantum Relative Entropy: An Introduction}},\ }\bibfield
  {journal} {\bibinfo  {journal} {arXiv}\ }\href
  {https://doi.org/10.48550/arXiv.1202.0983} {10.48550/arXiv.1202.0983}
  (\bibinfo {year} {2012}),\ \Eprint {https://arxiv.org/abs/1202.0983}
  {1202.0983} \BibitemShut {NoStop}%
\bibitem [{\citenamefont {Sagawa}\ and\ \citenamefont {Ueda}(2008)}]{Sagawa08}%
  \BibitemOpen
  \bibfield  {author} {\bibinfo {author} {\bibfnamefont {T.}~\bibnamefont
  {Sagawa}}\ and\ \bibinfo {author} {\bibfnamefont {M.}~\bibnamefont {Ueda}},\
  }\bibfield  {title} {\bibinfo {title} {{Second Law of Thermodynamics with
  Discrete Quantum Feedback Control}},\ }\href
  {https://doi.org/10.1103/PhysRevLett.100.080403} {\bibfield  {journal}
  {\bibinfo  {journal} {Phys. Rev. Lett.}\ }\textbf {\bibinfo {volume} {100}},\
  \bibinfo {pages} {080403} (\bibinfo {year} {2008})}\BibitemShut {NoStop}%
\bibitem [{\citenamefont {Naghiloo}\ \emph {et~al.}(2018)\citenamefont
  {Naghiloo}, \citenamefont {Alonso}, \citenamefont {Romito}, \citenamefont
  {Lutz},\ and\ \citenamefont {Murch}}]{Naghiloo18}%
  \BibitemOpen
  \bibfield  {author} {\bibinfo {author} {\bibfnamefont {M.}~\bibnamefont
  {Naghiloo}}, \bibinfo {author} {\bibfnamefont {J.~J.}\ \bibnamefont
  {Alonso}}, \bibinfo {author} {\bibfnamefont {A.}~\bibnamefont {Romito}},
  \bibinfo {author} {\bibfnamefont {E.}~\bibnamefont {Lutz}},\ and\ \bibinfo
  {author} {\bibfnamefont {K.~W.}\ \bibnamefont {Murch}},\ }\bibfield  {title}
  {\bibinfo {title} {{Information Gain and Loss for a Quantum Maxwell's
  Demon}},\ }\href {https://doi.org/10.1103/PhysRevLett.121.030604} {\bibfield
  {journal} {\bibinfo  {journal} {Physical Review Letters}\ }\textbf {\bibinfo
  {volume} {121}},\ \bibinfo {pages} {030604} (\bibinfo {year}
  {2018})}\BibitemShut {NoStop}%
\bibitem [{\citenamefont {Lecamwasam}\ \emph {et~al.}(2024)\citenamefont
  {Lecamwasam}, \citenamefont {Assad}, \citenamefont {Hope}, \citenamefont
  {Lam}, \citenamefont {Thompson},\ and\ \citenamefont {Gu}}]{Lecamwasam24}%
  \BibitemOpen
  \bibfield  {author} {\bibinfo {author} {\bibfnamefont {R.}~\bibnamefont
  {Lecamwasam}}, \bibinfo {author} {\bibfnamefont {S.}~\bibnamefont {Assad}},
  \bibinfo {author} {\bibfnamefont {J.~J.}\ \bibnamefont {Hope}}, \bibinfo
  {author} {\bibfnamefont {P.~K.}\ \bibnamefont {Lam}}, \bibinfo {author}
  {\bibfnamefont {J.}~\bibnamefont {Thompson}},\ and\ \bibinfo {author}
  {\bibfnamefont {M.}~\bibnamefont {Gu}},\ }\bibfield  {title} {\bibinfo
  {title} {{Relative Entropy of Coherence Quantifies Performance in Bayesian
  Metrology}},\ }\href {https://doi.org/10.1103/PRXQuantum.5.030303} {\bibfield
   {journal} {\bibinfo  {journal} {PRX Quantum}\ }\textbf {\bibinfo {volume}
  {5}},\ \bibinfo {pages} {030303} (\bibinfo {year} {2024})}\BibitemShut
  {NoStop}%
\bibitem [{\citenamefont {Buffoni}\ \emph {et~al.}(2019)\citenamefont
  {Buffoni}, \citenamefont {Solfanelli}, \citenamefont {Verrucchi},
  \citenamefont {Cuccoli},\ and\ \citenamefont {Campisi}}]{Buffoni19}%
  \BibitemOpen
  \bibfield  {author} {\bibinfo {author} {\bibfnamefont {L.}~\bibnamefont
  {Buffoni}}, \bibinfo {author} {\bibfnamefont {A.}~\bibnamefont {Solfanelli}},
  \bibinfo {author} {\bibfnamefont {P.}~\bibnamefont {Verrucchi}}, \bibinfo
  {author} {\bibfnamefont {A.}~\bibnamefont {Cuccoli}},\ and\ \bibinfo {author}
  {\bibfnamefont {M.}~\bibnamefont {Campisi}},\ }\bibfield  {title} {\bibinfo
  {title} {{Quantum Measurement Cooling}},\ }\href
  {https://doi.org/10.1103/PhysRevLett.122.070603} {\bibfield  {journal}
  {\bibinfo  {journal} {Phys. Rev. Lett.}\ }\textbf {\bibinfo {volume} {122}},\
  \bibinfo {pages} {070603} (\bibinfo {year} {2019})}\BibitemShut {NoStop}%
\bibitem [{\citenamefont {Bresque}\ \emph {et~al.}(2020)\citenamefont
  {Bresque}, \citenamefont {Camati}, \citenamefont {Rogers}, \citenamefont
  {Murch}, \citenamefont {Jordan},\ and\ \citenamefont
  {Auff{\ifmmode\grave{e}\else\`{e}\fi}ves}}]{Bresque20}%
  \BibitemOpen
  \bibfield  {author} {\bibinfo {author} {\bibfnamefont {L.}~\bibnamefont
  {Bresque}}, \bibinfo {author} {\bibfnamefont {P.~A.}\ \bibnamefont {Camati}},
  \bibinfo {author} {\bibfnamefont {S.}~\bibnamefont {Rogers}}, \bibinfo
  {author} {\bibfnamefont {K.}~\bibnamefont {Murch}}, \bibinfo {author}
  {\bibfnamefont {A.~N.}\ \bibnamefont {Jordan}},\ and\ \bibinfo {author}
  {\bibfnamefont {A.}~\bibnamefont {Auff{\ifmmode\grave{e}\else\`{e}\fi}ves}},\
  }\bibfield  {title} {\bibinfo {title} {{A two-qubit engine fueled by
  entangling operations and local measurements}},\ }\bibfield  {journal}
  {\bibinfo  {journal} {arXiv}\ }\href
  {https://doi.org/10.1103/PhysRevLett.126.120605}
  {10.1103/PhysRevLett.126.120605} (\bibinfo {year} {2020}),\ \Eprint
  {https://arxiv.org/abs/2007.03239} {2007.03239} \BibitemShut {NoStop}%
\bibitem [{\citenamefont {Perna}\ and\ \citenamefont
  {Calzetta}(2023)}]{Perna23}%
  \BibitemOpen
  \bibfield  {author} {\bibinfo {author} {\bibfnamefont {G.}~\bibnamefont
  {Perna}}\ and\ \bibinfo {author} {\bibfnamefont {E.}~\bibnamefont
  {Calzetta}},\ }\bibfield  {title} {\bibinfo {title} {{Limits on quantum
  measurement engines}},\ }\bibfield  {journal} {\bibinfo  {journal} {ArXiv
  e-prints}\ }\href {https://doi.org/10.48550/arXiv.2312.08148}
  {10.48550/arXiv.2312.08148} (\bibinfo {year} {2023}),\ \Eprint
  {https://arxiv.org/abs/2312.08148} {2312.08148} \BibitemShut {NoStop}%
\bibitem [{\citenamefont {Braginsky}\ \emph {et~al.}(1980)\citenamefont
  {Braginsky}, \citenamefont {Vorontsov},\ and\ \citenamefont
  {Thorne}}]{Braginsky1980}%
  \BibitemOpen
  \bibfield  {author} {\bibinfo {author} {\bibfnamefont {V.~B.}\ \bibnamefont
  {Braginsky}}, \bibinfo {author} {\bibfnamefont {Y.~I.}\ \bibnamefont
  {Vorontsov}},\ and\ \bibinfo {author} {\bibfnamefont {K.~S.}\ \bibnamefont
  {Thorne}},\ }\bibfield  {title} {\bibinfo {title} {{Quantum Nondemolition
  Measurements}},\ }\href {https://doi.org/10.1126/science.209.4456.547}
  {\bibfield  {journal} {\bibinfo  {journal} {Science}\ }\textbf {\bibinfo
  {volume} {209}},\ \bibinfo {pages} {547} (\bibinfo {year}
  {1980})}\BibitemShut {NoStop}%
\bibitem [{\citenamefont {Parrondo}\ \emph {et~al.}(2015)\citenamefont
  {Parrondo}, \citenamefont {Horowitz},\ and\ \citenamefont
  {Sagawa}}]{Parrondo15}%
  \BibitemOpen
  \bibfield  {author} {\bibinfo {author} {\bibfnamefont {J.~M.~R.}\
  \bibnamefont {Parrondo}}, \bibinfo {author} {\bibfnamefont {J.~M.}\
  \bibnamefont {Horowitz}},\ and\ \bibinfo {author} {\bibfnamefont
  {T.}~\bibnamefont {Sagawa}},\ }\bibfield  {title} {\bibinfo {title}
  {{Thermodynamics of information}},\ }\href
  {https://doi.org/10.1038/nphys3230} {\bibfield  {journal} {\bibinfo
  {journal} {Nature Physics}\ }\textbf {\bibinfo {volume} {11}},\ \bibinfo
  {pages} {131} (\bibinfo {year} {2015})}\BibitemShut {NoStop}%
\bibitem [{\citenamefont {Kammerlander}\ and\ \citenamefont
  {Anders}(2016)}]{Kammerlander16}%
  \BibitemOpen
  \bibfield  {author} {\bibinfo {author} {\bibfnamefont {P.}~\bibnamefont
  {Kammerlander}}\ and\ \bibinfo {author} {\bibfnamefont {J.}~\bibnamefont
  {Anders}},\ }\bibfield  {title} {\bibinfo {title} {{Coherence and measurement
  in quantum thermodynamics}},\ }\href {https://doi.org/10.1038/srep22174}
  {\bibfield  {journal} {\bibinfo  {journal} {Sci. Rep.}\ }\textbf {\bibinfo
  {volume} {6}},\ \bibinfo {pages} {1} (\bibinfo {year} {2016})}\BibitemShut
  {NoStop}%
\bibitem [{\citenamefont {Bera}\ \emph {et~al.}(2019)\citenamefont {Bera},
  \citenamefont {Riera}, \citenamefont {Lewenstein}, \citenamefont {Khanian},\
  and\ \citenamefont {Winter}}]{Bera19}%
  \BibitemOpen
  \bibfield  {author} {\bibinfo {author} {\bibfnamefont {M.~N.}\ \bibnamefont
  {Bera}}, \bibinfo {author} {\bibfnamefont {A.}~\bibnamefont {Riera}},
  \bibinfo {author} {\bibfnamefont {M.}~\bibnamefont {Lewenstein}}, \bibinfo
  {author} {\bibfnamefont {Z.~B.}\ \bibnamefont {Khanian}},\ and\ \bibinfo
  {author} {\bibfnamefont {A.}~\bibnamefont {Winter}},\ }\bibfield  {title}
  {\bibinfo {title} {{Thermodynamics as a Consequence of Information
  Conservation}},\ }\href {https://doi.org/10.22331/q-2019-02-14-121}
  {\bibfield  {journal} {\bibinfo  {journal} {Quantum}\ }\textbf {\bibinfo
  {volume} {3}},\ \bibinfo {pages} {121} (\bibinfo {year} {2019})},\ \Eprint
  {https://arxiv.org/abs/1707.01750v3} {1707.01750v3} \BibitemShut {NoStop}%
\bibitem [{\citenamefont {Elouard}\ and\ \citenamefont
  {Lombard~Latune}(2023)}]{Elouard23}%
  \BibitemOpen
  \bibfield  {author} {\bibinfo {author} {\bibfnamefont {C.}~\bibnamefont
  {Elouard}}\ and\ \bibinfo {author} {\bibfnamefont {C.}~\bibnamefont
  {Lombard~Latune}},\ }\bibfield  {title} {\bibinfo {title} {{Extending the
  Laws of Thermodynamics for Arbitrary Autonomous Quantum Systems}},\ }\href
  {https://doi.org/10.1103/PRXQuantum.4.020309} {\bibfield  {journal} {\bibinfo
   {journal} {PRX Quantum}\ }\textbf {\bibinfo {volume} {4}},\ \bibinfo {pages}
  {020309} (\bibinfo {year} {2023})}\BibitemShut {NoStop}%
\bibitem [{\citenamefont {Fadler}\ \emph {et~al.}(2023)\citenamefont {Fadler},
  \citenamefont {Friedenberger},\ and\ \citenamefont {Lutz}}]{Fadler23}%
  \BibitemOpen
  \bibfield  {author} {\bibinfo {author} {\bibfnamefont {P.}~\bibnamefont
  {Fadler}}, \bibinfo {author} {\bibfnamefont {A.}~\bibnamefont
  {Friedenberger}},\ and\ \bibinfo {author} {\bibfnamefont {E.}~\bibnamefont
  {Lutz}},\ }\bibfield  {title} {\bibinfo {title} {{Efficiency at Maximum Power
  of a Carnot Quantum Information Engine}},\ }\href
  {https://doi.org/10.1103/PhysRevLett.130.240401} {\bibfield  {journal}
  {\bibinfo  {journal} {Physical Review Letters}\ }\textbf {\bibinfo {volume}
  {130}},\ \bibinfo {pages} {240401} (\bibinfo {year} {2023})}\BibitemShut
  {NoStop}%
\bibitem [{\citenamefont {Cottet}\ \emph {et~al.}(2017)\citenamefont {Cottet},
  \citenamefont {Jezouin}, \citenamefont {Bretheau}, \citenamefont
  {Campagne-Ibarcq}, \citenamefont {Ficheux}, \citenamefont {Anders},
  \citenamefont {Auff{\ifmmode\grave{e}\else\`{e}\fi}ves}, \citenamefont
  {Azouit}, \citenamefont {Rouchon},\ and\ \citenamefont {Huard}}]{Cottet17}%
  \BibitemOpen
  \bibfield  {author} {\bibinfo {author} {\bibfnamefont {N.}~\bibnamefont
  {Cottet}}, \bibinfo {author} {\bibfnamefont {S.}~\bibnamefont {Jezouin}},
  \bibinfo {author} {\bibfnamefont {L.}~\bibnamefont {Bretheau}}, \bibinfo
  {author} {\bibfnamefont {P.}~\bibnamefont {Campagne-Ibarcq}}, \bibinfo
  {author} {\bibfnamefont {Q.}~\bibnamefont {Ficheux}}, \bibinfo {author}
  {\bibfnamefont {J.}~\bibnamefont {Anders}}, \bibinfo {author} {\bibfnamefont
  {A.}~\bibnamefont {Auff{\ifmmode\grave{e}\else\`{e}\fi}ves}}, \bibinfo
  {author} {\bibfnamefont {R.}~\bibnamefont {Azouit}}, \bibinfo {author}
  {\bibfnamefont {P.}~\bibnamefont {Rouchon}},\ and\ \bibinfo {author}
  {\bibfnamefont {B.}~\bibnamefont {Huard}},\ }\bibfield  {title} {\bibinfo
  {title} {{Observing a quantum Maxwell demon at work}},\ }\href
  {https://doi.org/10.1073/pnas.1704827114} {\bibfield  {journal} {\bibinfo
  {journal} {Proceedings of the National Academy of Sciences}\ }\textbf
  {\bibinfo {volume} {114}},\ \bibinfo {pages} {7561} (\bibinfo {year}
  {2017})}\BibitemShut {NoStop}%
\bibitem [{\citenamefont {Manzano}\ \emph {et~al.}(2020)\citenamefont
  {Manzano}, \citenamefont {S{\ifmmode\acute{a}\else\'{a}\fi}nchez},
  \citenamefont {Silva}, \citenamefont {Haack}, \citenamefont {Brask},
  \citenamefont {Brunner},\ and\ \citenamefont {Potts}}]{Manzano20}%
  \BibitemOpen
  \bibfield  {author} {\bibinfo {author} {\bibfnamefont {G.}~\bibnamefont
  {Manzano}}, \bibinfo {author} {\bibfnamefont {R.}~\bibnamefont
  {S{\ifmmode\acute{a}\else\'{a}\fi}nchez}}, \bibinfo {author} {\bibfnamefont
  {R.}~\bibnamefont {Silva}}, \bibinfo {author} {\bibfnamefont
  {G.}~\bibnamefont {Haack}}, \bibinfo {author} {\bibfnamefont {J.~B.}\
  \bibnamefont {Brask}}, \bibinfo {author} {\bibfnamefont {N.}~\bibnamefont
  {Brunner}},\ and\ \bibinfo {author} {\bibfnamefont {P.~P.}\ \bibnamefont
  {Potts}},\ }\bibfield  {title} {\bibinfo {title} {{Hybrid thermal machines:
  Generalized thermodynamic resources for multitasking}},\ }\href
  {https://doi.org/10.1103/PhysRevResearch.2.043302} {\bibfield  {journal}
  {\bibinfo  {journal} {Physical Review Research}\ }\textbf {\bibinfo {volume}
  {2}},\ \bibinfo {pages} {043302} (\bibinfo {year} {2020})}\BibitemShut
  {NoStop}%
\bibitem [{\citenamefont {Elouard}\ \emph {et~al.}()\citenamefont {Elouard},
  \citenamefont {Jordan},\ and\ \citenamefont {Haack}}]{MeasFridgePaper}%
  \BibitemOpen
  \bibfield  {author} {\bibinfo {author} {\bibfnamefont {C.}~\bibnamefont
  {Elouard}}, \bibinfo {author} {\bibfnamefont {A.~N.}\ \bibnamefont
  {Jordan}},\ and\ \bibinfo {author} {\bibfnamefont {G.~a.}\ \bibnamefont
  {Haack}},\ }\bibfield  {title} {\bibinfo {title} {{Revealing the fuel of a
  measurement-powered refrigerator (provisional title)}},\ }\href@noop {}
  {\bibinfo  {journal} {in preparation}\ }\BibitemShut {NoStop}%
\bibitem [{\citenamefont {Minagawa}\ \emph {et~al.}(2023)\citenamefont
  {Minagawa}, \citenamefont {Mohammady}, \citenamefont {Sakai}, \citenamefont
  {Kato},\ and\ \citenamefont {Buscemi}}]{Minagawa23}%
  \BibitemOpen
\bibfield  {journal} {  }\bibfield  {author} {\bibinfo {author} {\bibfnamefont
  {S.}~\bibnamefont {Minagawa}}, \bibinfo {author} {\bibfnamefont {M.~H.}\
  \bibnamefont {Mohammady}}, \bibinfo {author} {\bibfnamefont {K.}~\bibnamefont
  {Sakai}}, \bibinfo {author} {\bibfnamefont {K.}~\bibnamefont {Kato}},\ and\
  \bibinfo {author} {\bibfnamefont {F.}~\bibnamefont {Buscemi}},\ }\bibfield
  {title} {\bibinfo {title} {{Universal validity of the second law of
  information thermodynamics}},\ }\bibfield  {journal} {\bibinfo  {journal}
  {ArXiv e-prints}\ }\href {https://doi.org/10.48550/arXiv.2308.15558}
  {10.48550/arXiv.2308.15558} (\bibinfo {year} {2023}),\ \Eprint
  {https://arxiv.org/abs/2308.15558} {2308.15558} \BibitemShut {NoStop}%
\end{thebibliography}%


\begin{thebibliography}{1} 
\bibitem{CohenBook} C. Cohen-Tannoudji, J. Dupont-Roc, and G. Grynberg, {\it Processus d'interaction entre photons et atomes}, (EDP Science/CNRS {\'E}ditions, Paris, 2001).
\bibitem{Guryanova_2020} Y. Guryanova, N. Friis, and M. Huber {\it Ideal Projective Measurements Have Infinite Resource Costs}, Quantum {\bf 4}, 222 (2020).  
\bibitem{NielsenChuangBook} Nielsen, M., \& Chuang, I. (2010). Quantum Computation and Quantum Information: 10th Anniversary Edition. Cambridge: Cambridge University Press. doi:10.1017/CBO9780511976667
\bibitem{FrancescoBook}  Heinz-Peter Breuer, Francesco Petruccione, {\it The Theory of Open Quantum Systems},  (Oxford, 2007; online edn, Oxford Academic, 1 Feb. 2010). 
\bibitem{Elouard21},
Cyril Elouard, Philippe Lewalle, Sreenath K. Manikandan, Spencer Rogers, Frank Adam and Andrew N. Jordan, Quantum erasing the memory of Wigner's friend, Quantum, 5, 498 (2021).
\bibitem{WisemanBook} H.M. Wiseman and G.J. Milburn, Quantum Measurement and Control (Cambridge University Press, Cambridge, 2009).
\bibitem{Bera_2019} M. N. Bera, A. Riera, M. Lewenstein, Z. B. Khanian, and A. Winter, {\it Thermodynamics as a Consequence of Information Conservation}, Quantum {\bf3}, 121 (2019).
\bibitem{Esposito_2010} Massimiliano Esposito, Katja Lindenberg, Christian Van den Broeck, {\it Entropy production as correlation between system and reservoir}, New J. Phys. {\bf12}, 013013 (2010).
\bibitem{Elouard_2023} C. Elouard and C. L. Latune, arXiv:2207.04850. 
\end{thebibliography}

\end{document}